\documentclass[a4paper,twocolumn,showpacs,superscriptaddress,floatfix]{iopart}

\usepackage{iopams}
\expandafter\let\csname equation*\endcsname\relax
\expandafter\let\csname endequation*\endcsname\relax
\usepackage{amsmath}
\usepackage{soul}

\usepackage{graphicx}
\usepackage{epsf}
\usepackage{epsfig}
\usepackage[usenames]{color}
\usepackage{times}

\usepackage{amssymb}
\usepackage{subfigure}
\usepackage{slashed}
\usepackage{comment}

\newcommand{\be}{\begin{equation}}
\newcommand{\ee}{\end{equation}}
\newcommand{\bea}{\begin{eqnarray}}
\newcommand{\eea}{\end{eqnarray}}
\newcommand{\bal}{\begin{align}}
\newcommand{\eal}{\end{align}}
\newcommand{\nn}{\nonumber}

\font\tenscr=rsfs10 scaled1100
\font\sevenscr=rsfs7 % scaled \magstep1
\font\fivescr=rsfs5 % scaled \magstep1
\skewchar\tenscr='177
\skewchar\sevenscr='177
\skewchar\fivescr='177
\newfam\scrfam
\textfont\scrfam=\tenscr
\scriptfont\scrfam=\sevenscr
\scriptscriptfont\scrfam=\fivescr

\def\scri{{\fam\scrfam I}}

\usepackage{xcolor}
\usepackage{color}

\begin{document}

%\title{Physical scales in black hole  scattering  pseudospectra: the role of the scalar product}
        
\title{Energy scales and black hole pseudospectra: the structural role of the scalar product% in quasinormal mode stability
}

%%%%%%%%%%%%%%%%%%%%%%

\author{E. Gasper\'in and J.L. Jaramillo}
%%%%%%%%%%%%%%%%%%%%%%%%%
\address{Institut  de  Math\'ematiques  de  Bourgogne  (IMB),  UMR  5584,  CNRS, Universit\'e  de  Bourgogne  Franche-Comt\'e,  F-21000  Dijon,  France.}
%%%%%%%%%%%%%%%%%%%%%%%%%%%%%%

\begin{abstract}

  %%%%%% Shorter abstract for CQG %%%%%%%%

  A pseudospectrum analysis has recently provided evidence of a
  potential generic instability of black hole (BH) quasinormal mode
  (QNM) overtones under high-frequency perturbations. Such instability
  analysis depends on the assessment of the size of perturbations. The
  latter is encoded in the scalar product and its choice is not
  unique. Here, we address the impact of the scalar product choice,
  advocating for founding it on the physical energy scales of the
  problem. The article is organized in three parts: basics,
  applications and heuristic proposals. In the first part, we revisit
  the energy scalar product used in the hyperboloidal approach to
  QNMs, extending previous effective analyses and placing them on
  solid spacetime basis. The second part focuses on systematic
  applications of the scalar product in the QNM problem: i) we
  demonstrate that the QNM instability is not an artifact of previous
  spectral numerical schemes, by implementing a finite elements
  calculation from a weak formulation; ii) using Keldysh’s asymptotic
  expansion of the resolvent, we provide QNM resonant expansions for
  the gravitational waveform, with explicit expressions of the
  expansion coefficients; iii) we propose the notion of `epsilon-dual
  QNM expansions' to exploit BH QNM instability in BH spectroscopy,
  complementarily exploiting both non-perturbed and perturbed QNMs,
  the former informing on large scales and the latter probing small
  scales. The third part enlarges the conceptual scope of BH QNM
  instability proposing: a) spiked perturbations are more efficient in
  triggering BH QNM instabilities than smooth ones, b) a general
  picture of the BH QNM instability problem is given, supporting the
  conjecture (built on Burnett’s conjecture on the spacetime
  high-frequency limit) that Nollert-Price branches converge
  universally to logarithmic Regge branches in the high-frequency
  limit and c) aiming at a fully geometric description of QNMs, BMS
  states are hinted as possible asymptotic/boundary degrees of freedom
  for an inverse scattering problem.

\end{abstract}
\tableofcontents


%\maketitle
\section{Introduction: scales and QNM instability}\label{sec:Introduction}
The goal of the present article is to address some conceptual and
technical issues concerning the role of the scalar product in the
analysis of BH QNM instability, namely in the setting of a recently
introduced pseudospectrum framework~\cite{Jaramillo:2020tuu}.
Spectral instability of QNM overtones under high-frequency
perturbations is a phenomenon of potential relevance in the study of
basic properties of BHs, in particular probing their small scale
physics.  Specifically, it may be of interest for fundamental gravity
at the quantum scale and, given current and future prospects
regarding interferometric antennae, for gravitational wave (GW)
physics and the emerging field of BH
spectroscopy~\cite{Berti:2005ys,Dreyer:2003bv,Baibhav_2018,Ota:2019bzl,Isi:2019aib,Giesler:2019uxc,Isi:2020tac,Cabero:2019zyt,Ota:2021ypb}.

QNMs play a central role in the modeling of (scattering) dissipative
systems with applications in research areas so diverse as
spectroscopy, optics, oceanography or general relativity, to cite some
examples. In the latter gravitational case, QNMs appear naturally in
the study of linear perturbations of BHs and in the study of
propagating fields on black hole backgrounds ---see
\cite{Kokkotas:1999bd,Nollert:1999ji,BerCarSta09,Konoplya:2011qq} for
comprehensive reviews (in particular also addressing QNMs of other
classes of compact objects, e.g. neutron stars).  The
dissipative nature of scattering by black holes follows
  from the presence of the event horizon and null infinity. In the
case of, for example, optics, QNMs appear in the study of the
interaction of light and plasmonic nanoresonators
\cite{sauvan2013theory,LalYanVynSauHug18} and in the study of optical
leaky cavities \cite{LeuLiuYou94,LeuLiuTon94,ChiLeuMaa98,BouShe20}.

Regardless of the specific physical setup in consideration, the
dissipative behavior is encoded, at the mathematical level, in the
lack of selfadjointness of the operator governing the dynamics of
perturbations to the physical system. In the scattering setting, this
follows in particular from the imposed outgoing boundary conditions.
QNMs in all these contexts can be shown to correspond to the
eigenvalues of such non-selfadjoint operators. This statement can be
made precise, in particular, by using a hyperboloidal formulation of
the equations
\cite{Warnick:2013hba,Ansorg:2016ztf,HinVas16,Jaramillo:2020tuu}.  The
non-selfadjoint nature of the operator makes these eigenvalues complex
and alter profoundly the properties of the set of eigenfunctions, in
particular their completeness and orthogonality.  Hence the prefix
quasi is used to distinguish them from the usual normal modes of
selfadjoint operators which have real eigenvalues with eigenfunctions
forming orthonormal bases.  Thus, in such a setting, the study of QNMs
can be thought condensedly as an application of the spectral analysis
theory for non-selfadjoint operators.

A natural question that arises is the robustness ---or structural
stability--- of QNMs in such physical systems, namely, how much the
QNM spectrum changes upon (small) perturbations of the operator. In
other words, an assessment of the \emph{spectral QNM
  (in)stability}. In the case of normal operators (in particular
selfadjoint ones) spectral stability is ensured by the spectral
theorem.  In contrast, non-normal operators can potentially be
spectrally unstable.  A systematic tool to investigate the spectral
instability of a given operator is the analysis of its pseudospectrum
---see \cite{TreEmb05Book,Davie07,Sjo19Book} for precise definitions
and an exhaustive discussion.  Heuristically, the pseudospectrum of an
operator can be thought as a cartographic map of its analytic
structure (actually, of its resolvent) providing, in particular, a
tool that can be used to qualitatively read off ``the sensitivity'' of
an operator upon perturbations.  The analysis of the pseudospectrum
has been exploited in several areas of physics where eigenvalues of
non-selfadjoint operators arise such as hydrodynamics and turbulence
or non-Hermitian quantum mechanics ---see
\cite{TreRedDri93,KreSieTatVio15}, and has been recently introduced in
gravitational physics in \cite{Jaramillo:2020tuu} in the context of BH
QNMs (cf. \cite{EmbTre_webpage} for extensive discussion of
pseudospectrum applications).

One of the equivalent characterizations of the pseudospectrum, which
is suitable in the present setting, is that the pseudospectrum of an
operator $L$ consists of the set of points in the complex plane that
are realized as eigenvalues of some perturbation $L + \delta L$ of the
original operator $L$.  To address possible spectral instabilities,
one is interested in estimating how large the change in the
eigenvalues, produced by small perturbations of the operator, can
be. Such variation range defines the so-called
$\epsilon$-pseudospectra sets associated with an $\epsilon$-sized
operator perturbations. We will say that the system is spectrally
stable if, under operator perturbations of 'size'
$\epsilon$, the change in the eigenvalues is also of order
$\epsilon$. On the contrary, if small changes of order $\epsilon$ in
the operator give rise to variations in the spectrum
(orders-of-magnitude) much larger than $\epsilon$, then we say that
$L$ is spectrally unstable.  Therefore the assessment of spectral
instability critically depends on the ``rule'' we choose to determine
the notion of 'big' and 'small' in the size of operator perturbations,
namely the employed operator norm.  Thus, in contrast with the notion
of spectrum which is an intrinsic property of the operator, the
notion of pseudospectrum ---and therefore the notion of
(in)stability--- depends also on the choice of norm.  This is crucial,
since perturbations considered as small in one norm could correspond
to large perturbations in another in another norm.  In sum, the
appropriate choice of scalar product ---with its associated norm---
becomes a key aspect in the study of QNM (in)stability.

From a mathematical perspective, the study of the spectral stability
of a non-normal operator can be linked to assessing the
existence of a scalar product permitting to keep the control of
 eigenvalue perturbations in tiny sets around the spectrum, i.e
with  $\epsilon$-pseudospectra sets tighly packed around the spectrum.
However, from a physical perspective, the notion of large and small
can be actually fixed by the physics of the problem, therefore not
having necessarily the freedom of making up a norm to mathematically
``control'' the instabilities.

The analysis of this question defines the problem addressed in this work.
In particular, we pursue further the program initiated in
\cite{Jaramillo:2020tuu} and extended in \cite{Jaramillo:2021tmt}
---with a first analysis on the implications on GW physics: assessment
of the presence of perturbed QNMs in the GW signal, isospectrality
loss and the introduction effective parameters for data analysis---
and in \cite{Destounis:2021lum} ---with a first exploration of asymptotic
universality and the geometric character of the
pseudospectrum, namely the slicing independence inside the hyperboloidal approach.
In a further step, in this article we discuss the role of the
inner product and its associated norm for assessing the spectral (in)stability of
QNMs and some of its implications.  In the following, we summarize the
main results to be developed later in the text.

\section{Hyperboloidal approach to QNMs: the energy scalar product in a nutshell}
\label{s:intro_QNM_instability}
The present work is framed in a compactified hyperboloidal approach to
the scattering problem of a field $\Phi$ in stationary asymptotically
flat black hole (BH) spacetimes
(cf. \cite{Jaramillo:2020tuu,Jaramillo:2021tmt} for details; see also
\cite{Ansorg:2016ztf,PanossoMacedo:2018hab,PanossoMacedo:2020biw}).
Specifically, we deal with linear wave equations in a first-order
reduction in time,
with spatial slices $\Sigma_\tau$ intersecting the
BH horizon and null infinity.  As a methodological simplification, we
focus on spherically symmetric spacetimes, leading to reduced
$1+1$-dimensional problems for harmonic modes $\phi_{\ell m}$.
Dropping the mode labels $(\ell m)$ and writing $\psi =
\partial_\tau\phi$, the wave equation is written
as
\be
\label{e:wave_eq_1storder_intro}
 \partial_\tau u = i L u \ \ , \ \ u =
\begin{pmatrix}
  \phi \\ \psi
\end{pmatrix}  \ ,
\ee
with the operator $L$ given by
\be
\label{e:L_operator_intro}
L =\frac{1}{i}\!  \left(
  \begin{array}{c|c}
    0 & 1 \\ \hline L_1 & L_2
  \end{array}
  \right) \ ,
  \ee
For scalar field potentials we have (here $x$ denotes a coordinate parametrizing spheres
  in $\Sigma_\tau$)
\be \label{e:L_1-L_2_intro}
 L_1 = \frac{1}{w(x)}\big(\partial_x\left(p(x)\partial_x\right) -
 \tilde{V}(x)\big), \;\; 
 L_2 =
\frac{1}{w(x)}\big(2\gamma(x)\partial_x + \partial_x\gamma(x)\big) \ ,
\ee where functions $p(x)$ and $\tilde{V}(x)$ depend only on the
intrinsic geometry of the slice $\Sigma_\tau$, whereas the function
$\gamma(x)$ and the weight function $w(x)$ depend also on the
extrinsic geometry of $\Sigma_\tau$ in the spacetime (see details in
\cite{Jaramillo:2020tuu}).

A key point in the discussion concerns the choice of a scalar product,
defining the Hilbert space in which solutions $u$ to equation
\eqref{e:wave_eq_1storder_intro} live and, crucially for our
discussion, defining the adjoint operator $L^\dagger$. This article
focuses on different questions around such scalar product
choice. Given the broad range of discussed aspects, the rest of this
section summarises the main points later developed in the text, with
the aim of acquiring an overall picture of the involved problems and
indicating the sections where corresponding results are presented.

\subsection{Energy scalar product: basics}
\label{s:scalar_product_basics}
This subsection focuses on structural issues regarding the scalar
product and its associated norm, namely its role in the construction
of the pseudospectrum. Then, upon the specific choice of an ``energy
scalar product'', we adress the relation between the total spacetime
problem and the reduced spherically-symmetric problem that was studied
in \cite{Jaramillo:2020tuu}.

\subsubsection{Assessing big and small in spectral stability problems: the relevance of the norm.}
\label{s:spectral_inst_intro}
Taking the Fourier transform in $\tau$ of equation
(\ref{e:wave_eq_1storder_intro}), the QNM problem can be cast as an
eigenvalue problem for a non-selfadjoint operator $L$ \be
\label{e:intro_L_eigen}
L v_n = \omega_n v_n \ . \ee If one considers, for $\epsilon>0$, a
perturbation $\delta L(\epsilon)$ of norm $\epsilon$, the perturbed
(right-)eigenvalue problem is written as \be
\label{e:intro_perturbed_spectral_problem}
\left(L + \delta L\right) v_n^\epsilon = \omega_n^\epsilon
v_n^\epsilon \ \ , \ \ ||\delta L||= \epsilon \ .  \ee Then it holds
(cf. e.g. \cite{Kat76,trefethen2005spectra}) \be
\label{e:intro_eigenvalue_perturbation}
|\omega_n^\epsilon - \omega_n| \leq \epsilon \kappa_n + O(\epsilon^2)
\ , \ee where $\kappa_n$ is the so-called condition number of
$\omega_n$ (see definition later in equation
(\ref{e:condition_number})).  We say that $\omega_n$ is unstable if
perturbations with small $||\delta L||= \epsilon$ produce large
changes $|\omega_n^\epsilon - \omega_n|$, associated with large values
of $\kappa_n$.  Therefore, we need to control the notion of `big' and
`small' perturbations, encoded in the operator norm $||\cdot||$
  for $\delta L$, constructed from a vector norm $||\cdot||$ (we abuse
  the repeated notation here) associated with a scalar product
$\langle \cdot, \cdot\rangle$.

A complementary approach to spectral instability is given by the
notion of
pseudospectrum~\cite{trefethen2005spectra,Davie07,Sjostrand2019}.
Such notion captures how far perturbed eigenvalues can migrate in the
complex plane under an operator perturbation with $||\delta L||=
\epsilon$ (see \cite{Jaramillo:2020tuu} for a detailed discussion in
the present context). Specifically, the notion of
$\epsilon$-pseudospectrum set $\sigma^\epsilon(L)$ can be equivalently
characterized as (here $\sigma(L)$ is the spectrum\footnote{We dwell
  in a finite-dimensional setting. See \cite{Davie07,Sjostrand2019}
  for the discussion in the infinite-dimensional case. Also, in this
  abstract discussion of the pseudospectrum, we use $\lambda$ to
  denote eigenvalues, reserving $\omega$ for the case where they are
  actually frequencies in the spectrum of a time generator $L$, as in
  eq. \eqref{e:intro_L_eigen} obtained by Fourier transform of
  eq. \eqref{e:wave_eq_1storder_intro}.}
   of $L$) \begin{eqnarray}
\label{e:pseudospectrum_def}
\sigma^\epsilon(L)&=& \{\lambda\in\mathbb{C}, \exists \; \delta
L\!\in\! M_n(\mathbb{C}), ||\delta L||<\epsilon:
\lambda\!\in\!\sigma(L+\delta L) \} \nn \\ &=& \{\lambda\in\mathbb{C}:
||(\lambda \mathrm{Id}- L)^{-1}||>1/\epsilon\} \nn \\ &=&
\{\lambda\in\mathbb{C}, \exists v\in\mathbb{C}^n: ||L v-\lambda
v||<\epsilon\} \ .  \end{eqnarray} As above in equations
   (\ref{e:intro_perturbed_spectral_problem})-(\ref{e:intro_eigenvalue_perturbation}),
   norms appear explicitly in the construction of the sets
   $\sigma^\epsilon(L)$, inside which eigenvalues can migrate under
   operator perturbations. Therefore, the control of notions of `big'
   and `small' is in-built in the notion of spectral stability.

In section \ref{sec:ABC} we discuss a simple example that dramatically
illustrates the impact of the choice of scalar product and its
norm in the assessment of spectral instability in non-selfadjoint
operators. In particular, it neatly highlights the impact of the norm
choice in the construction of the pseudospectrum.

\subsubsection{Physical scales: energy norm and scalar product
in the hyperboloidal approach.}
\label{s:intro_energy_norm}
A key element in the pseudospectrum discussion in
\cite{Jaramillo:2020tuu} is the use of a so-called ``energy scalar
product'', encoding the geometric and analytical structure of the
problem. In particular, in the assessment of big/small perturbations,
this translates into an operator norm aiming at capturing the relevant
scales in terms of the energy.  Such ``energy (operator) norm'' (see
also \cite{Driscoll:1996,Schmi07,TreTreRed93}) is induced from a
vector norm defined directly from the energy of perturbations. Given a
perturbation $\Phi$ with associated stress-energy tensor $T_{ab}$ (in
a stationary spacetime with timelike Killing $t^a$), its energy
$E(\Phi)$ \cite{Wald84} in a time slice $\Sigma_\tau$ defines the norm
$||\Phi||_{_E}$
 \be
\label{e:intro_E_norm}
||\Phi||_{_E} := E(\Phi) = \int_{\Sigma_\tau} T_{ab}(\Phi) t^a n^b d\Sigma \ ,
\ee with $n^a$ the timelike normal to $\Sigma$.  Two caveats are
associated with the use of such a norm. First, the operator norm
$||\delta L||_{_E}$ of a perturbation $\delta L$ is induced from the
vector energy norm (\ref{e:intro_E_norm}), but it is not itself an
energy.  Therefore, the relation between the energy in field
perturbations and the assessment of scales of QNM instabilities
is a point needing elucidation, something that will be
discussed in section \ref{sec:physical_interpretation_energy}.
Second, and more technically, the energy norm constructed in
\cite{Jaramillo:2020tuu} for spherically symmetric problems is not
built on the full stress-energy tensor $T_{ab}$ of $\Phi$, but
actually on the effective stress-energy tensor $\mathring{T}_{ab}$
associated with the $1+1$ dimensional problem for a given mode
$\phi_{\ell m}$ (in a spherical harmonic decomposition). A key
question is then: how does the energy $\mathring{E}_{\ell
  m}:=\mathring{E}(\phi_{\ell m})$ relates to the total energy
$E(\Phi)$? Section \ref{sec:Energy_associated_inner_products} answers
this question for a scalar field, giving the expected (but
significant) result
\begin{equation}\label{eq:intro_TotalEnergy}
  E = \sum_{\ell m} \mathring{E}_{\ell m}
  -\Bigg(\frac{f}{2r}\phi_{\ell m}\bar{\phi}_{\ell
    m}\Bigg)\Bigg|_{a}^{b},
\end{equation}
where $a$ and $b$ represent the (compactified) boundaries in our
problem, $f$ is the function in the line element in Schwarzschild
coordinates vanishing at the horizon and $r$ is the (Schwarzschild)
radial coordinate (diverging at infinity). From this it follows
  that the energy of the total problem is exactly given by the sum of
the energies of the effective $1+1$ problems. Section
\ref{sec:Energy_associated_inner_products} completes the analysis with
the study of energy fluxes through the (null) boundaries
\begin{equation}\label{eq:totalFluxExpression_reduced:intro}
  F = \sum_{\ell m} \gamma \partial_\tau\phi_{\ell m} \partial_\tau
  \bar{\phi}_{\ell m} \ ,
\end{equation}
accounted in terms of the function $\gamma$ in the $L_2$
  operator in \eqref{e:L_1-L_2_intro}.  Such operator $L_2$ is
  therefore neatly identified with the non-conservative character of
the system and its non-selfadjointness.

The discussion of the energy scalar product and energy norm is
  presented in section \ref{sec:Energy_associated_inner_products}.

\subsection{Energy scalar product: applications}
Once the basic elements of the scalar product and its associated
  norm have been presented in sections \ref{sec:ABC} and
  \ref{sec:Energy_associated_inner_products}, we make use of them in
  some applications in sections \ref{sec:Weak_Formulation} and
  \ref{s:resonant_expansions}.

\subsubsection{QNM weak formulation and finite element calculations.}

Using the scalar product, the eigenvalue problem
(\ref{e:intro_L_eigen}) can be cast in a weak formulation (cf. section
\ref{sec:Weak_Formulation}, equation \eqref{e:weak_problem_v0})
\be
\label{eq:Weak_Formulation_NoBoundary_intro} \int_{a}^{b}
W_{L}(u_T,u) \; dx = i \omega \int_{a}^{b} W_{R}(u_T,u) \;dx \ ,
\ee
in terms of appropriate bilinear forms $W_{L}(u_T,u)$ and
$W_{R}(u_T,u)$ and the $u_T$ test functions.  This is by itself an
interesting perspective on the QNM problem, in particular when aiming
at identifying the appropriate functional space where QNMs are defined
or when systematically studying the effect of random perturbations on
the distribution of resonant frequencies, seen as eigenvalue values
\cite{Sjostrand2019}.  As a practical application, such a weak
formulation is employed here to study the possibility of a numerical
artifact behind the observed QNM instability, namely arising from the
employed numerical scheme in
\cite{Jaramillo:2020tuu,Jaramillo:2021tmt,Destounis:2021lum}, based on
Chebyshev spectral methods.  This point specifically addresses one of
the caveats raised in \cite{Jaramillo:2020tuu}, in relation to the
adopted (numerical) analysis scheme.  In concrete terms, the weak
formulation \eqref{eq:Weak_Formulation_NoBoundary_intro} permits to
implement a completely independent numerical setting, namely a finite
elements scheme. The results presented here confirm the presence of
Nollert-Price-like QNM branches described in \cite{Jaramillo:2020tuu},
demonstrating the independence of the latter from the Chebyshev's
spectral scheme.

\subsubsection{QNM resonant expansions: a Keldysh's projection expression}
A second application, presented in section
\ref{s:resonant_expansions}, concerns resonant expansions.  One of the
key features of the hyperbolic formulation is that QNM are
normalizable, in particular eigenfunctions live in a Banach (actually
a Hilbert) space. This permits to make use of Keldysh's theorem for
the expansion for the resolvent of an operator in a Banach space
\cite{Keldy51,Keldy71,MenMol03,BeyLatRot12,Beyn12}.  Specifically,
associated with the introduced energy scalar product, we have the
adjoint $L^\dagger$ of $L$, so we can complete (\ref{e:intro_L_eigen})
with the left-eigenvalue problem (namely the right-eigenvalue problem
of $L^\dagger$)
\be
\label{e:intro_LR_eigen}
L \hat{v}_n = \omega_n \hat{v}_n \ \ , \ \ L^\dagger \hat{w}_n =
\overline{\omega}_n \hat{w}_n \ ,
\ee where $\hat{v}_n$ and
$\hat{w}_n$ are, respectively right- and left-eigenvectors normalized
in the energy norm. 
The application of Keldysh's theorem
(under the assumption of a spectrum given only by eigenvalues) then leads
to the following expansion of the scattered field $u$ in terms of
QNMs
\be
   \label{e:Keldysh_QNM_expansion_u_intro}
   u(\tau,x) \sim \sum_n e^{i\omega_n\tau} a_n \hat{v}_n(x) \ , \ee
   with coefficients $a_n$ obtained by projection on initial data
   $u_0$ for the equation \eqref{e:wave_eq_1storder_intro} \be
\label{e:coeffient_QNM_expansion}
a_n = \kappa_n \langle \hat{w}_n|u_0\rangle_{_{E}} \ , \ee with
$\kappa_n$ the associated condition numbers (see their general
definition in equation \eqref{e:condition_number})
%%%%
%in terms (here, $w_n$ and $v_n$ are not normalized)
%%%
\be
\label{e:condition_number_intro}
\kappa_n = \frac{1}{\langle \hat{w}_n, \hat{v}_n\rangle_{_E}} \ .
%%%
%=\frac{||w_n||_{_E} ||v_n||_{_E} }{\langle w_n, v_n\rangle_{_E}}
%%
\ee
This expression presents the following features:
\begin{itemize}
\item[i)] The QNM expansion is, in the generic case, a (non-convergent)
  asymptotic series (cf. discussion around equation (\ref{e:u_Keldysh_v7}) for details).

\item[ii)] In a hyperboloidal slicing scheme, it provides an explicit
  expression for the QNM (asymptotic) expansion coefficients in terms of initial
  data, even if $\hat{v}_n$ do not form a complete set, providing a
  prescription for a frequency-domain solution to equation \eqref{e:wave_eq_1storder_intro}.

\item[iii)] It reduces to the standard form in the selfadjoint (in
  general, normal) case, where $\kappa_n=1$ and
  $\hat{w}_n=\hat{v}_n$. In this case, the series is actually
  convergent and the set $\{\hat{v}_n\}$ is a (Hilbert) orthonormal
  basis. In particular, it recovers the explicit expression for normal
  modes.
\end{itemize}

\subsection{Energy scalar product: general picture, heuristics and  geometry.}
Sections \ref{e:QNM_instability_general_picture} and
\ref{s:Geometry_QNMs} have a more heuristic flavor, aiming at framing
on a geometric setting some of the open conceptual problems in the
approach to QNM instability, built on the hyperboloidal approach to
scattering and the pseudospectrum, as introduced in reference
\cite{Jaramillo:2020tuu}.

\subsubsection{Building a general picture of QNM ultraviolet instability.}
\label{s:general_picture_intro}

Section \ref{e:QNM_instability_general_picture} presents a description
of the generic picture of ultraviolet QNM instability as emerging from
the results in \cite{Jaramillo:2020tuu} and \cite{Jaramillo:2021tmt}.
Specifically, QNM overtones of non-perturbed black holes are
structurally unstable under high-frequency perturbations, migrating to
new QNM branches when perturbed. The resulting perturbed branches, on
the contrary, are themselves structurally stable, that is, QNM stay in
the perturbed branches if further perturbed.

Such new QNM branches, referred to generically as Nollert-Price QNM
branches, display open patterns in the complex plane pushed by
perturbations closer to the real axial (increase of damping) and
without upper bounds in the real part (therefore exploring small
spacetime structures). Nollert-Price branches are constrained to lay
in complex plane regions bounded below by QNM-free regions ---
determined from the boundaries of $\epsilon$-pseudospectra--- that
present asymptotic logarithmic boundaries for large real parts
\be
\label{e:log_pseudospectrum_lines_intro}
\mathrm{Im}(\omega) \sim C_1 + C_2 \ln \big(|\mathrm{Re}(\omega)| +
C_3\big) \ \ , \  \ |\mathrm{Re}(\omega)|\gg 1 .
\ee
Qualitative different Nollert-Price-like branches appear under
perturbations, possibly associated with different underlying resonance
mechanisms. Transition between such distinct regimes are marked by the
presence of ``inner QNMs'', much closer to the imaginary axis.
Finally, the slowest decaying QNM is stable under the considered
ultraviolet perturbations.

\subsubsection{BH spectroscopy: $\epsilon$-dual QNM expansions.}
As a by-product of combining QNM instability and QNM resonant
expansions, section \ref{s:epsilon_dual_QNMexpansions} introduces the
notion of $\epsilon$-dual QNM expansions of a (time-domain) scattered
signal, of potential use in BH spectroscopy.

BH
spectroscopy~\cite{Berti:2005ys,Dreyer:2003bv,Baibhav_2018,Ota:2019bzl,Isi:2019aib,Giesler:2019uxc,Isi:2020tac,Cabero:2019zyt,Maggio:2020jml,Ota:2021ypb}
aims at retrieving the physical information of a BH from the expansion
of the ringdown gravitational wave signal (GW) in terms of QNMs. In
the present setting, this amounts to expand the scattered field
$u(\tau,x)$ in equation (\ref{e:wave_eq_1storder_intro}) in terms of
QNMs, according to equation (\ref{e:Keldysh_QNM_expansion_u_intro}).
If we consider now an astrophysical BH subject to small perturbations,
modeled by operator perturbations with $||\delta L||=\epsilon$, the
resulting perturbed scattered field $u^\epsilon(\tau,x) $ that solves
equation (\ref{e:wave_eq_1storder_intro}) with $L + \delta L$, admits
rather a QNM expansion in terms of the corresponding perturbed QNMs in
(\ref{e:intro_perturbed_spectral_problem}) \be
   \label{e:Keldysh_QNM_expansion_u_pert_intro}
   u^\epsilon(\tau,x) \sim \sum_n e^{i\omega^\epsilon_n\tau}
   a^\epsilon_n \hat{v}^\epsilon_n(x) \ . \ee Given the QNM spectral
   instability under ultraviolet perturbations described in
   \cite{Jaramillo:2020tuu,Jaramillo:2021tmt}, perturbed QNM
   frequencies $\omega^\epsilon_n$ can strongly differ from QNM
   frequencies $\omega_n$, corresponding to the non-perturbed BH. The
   presence of such perturbed QNMs in the perturbed scattered signal,
   as well as the need to expand in terms of perturbed QNMs in order
   to reach arbitrary accuracy of the resonant expansion, have been
   demonstrated in \cite{Jaramillo:2021tmt}.  This raises the
   following question: {\em are QNM expansions of GW signals in terms
     of non-perturbed QNMs meaningful?}

Remarkably, the answer is in the affirmative, leading to the notion of
$\epsilon$-dual QNM expansions. Such notion follows from combining: i)
the instability of the (frequency-domain) spectral QNM problem
(\ref{e:intro_L_eigen}), and ii) the stability of the (time-domain)
evolution problem (\ref{e:wave_eq_1storder_intro}).  Such
$\epsilon$-dual QNM expansions are very different resonant expansions
of ``essentially the same scattered waveform''. More precisely, they
are either resonant expansions of a time-domain signal $u(\tau,x)$ in
terms of the standard non-perturbed QNMs $\omega_n$ ---namely
expansion (\ref{e:Keldysh_QNM_expansion_u_intro})--- or,
alternatively, a expansion of $u^\epsilon(\tau,x)$ in terms of the
(very different) perturbed QNMs $\omega^\epsilon_n$ ---namely
expansion \eqref{e:Keldysh_QNM_expansion_u_pert_intro}. From the
dynamical stability of the evolution problem
\eqref{e:wave_eq_1storder_intro}, one has $u^\epsilon(\tau,x)\sim
u(\tau,x) +O(\epsilon)$, so that the respective QNM expansions cannot
be distinguished at the order $\epsilon$. In other words, they are
equivalent at this order $\epsilon$, something we note as
\be
   \label{e:epsilon-dual_expansions}
   \sum_n e^{i\omega^\epsilon_n\tau} a^\epsilon_n
   \hat{v}^\epsilon_n(x) \stackrel{\epsilon}{\sim} \sum_n
   e^{i\omega_n\tau} a_n \hat{v}_n(x) \ .  \ee Both expansions
   approach the actual astrophysical (i.e. the perturbed) waveform, the one on the
   left with arbitrary accuracy, the one on the right to order
   $O(\epsilon)$. But for all practical purposes (the errors in the
   observational GW data are much bigger than $\epsilon$) both are
   valid.

Which expansion should then be chosen? From our perspective, and
  if technically possible, both of them. Indeed, they contain
  complementary information.  Whereas the standard expansion in terms
  of non-perturbed $\omega_n$ QNMs encodes information of the large
  scale structures of the (averaged) BH, the expansion in terms of
  perturbed $\omega^\epsilon_n$ informs on the small perturbation
  scales \cite{Jaramillo:2021tmt}.  Because of this complementarity we
  refer to them as ``$\epsilon$-dual''.

\subsubsection{Operator norm and metric fluctuations.}
The previous emerging picture is of descriptive nature, without
discussing the underlying physical and geometrical phenomena.  A first
key point is discussed in section
\ref{sec:physical_interpretation_energy}, namely the fact that, when
considering perturbations with a given ``energy norm'' $\epsilon$ in
the QNM instability problem, such $\epsilon$ is {\em not} the actual
energy of a spacetime perturbation $\delta g_{ab}$. It
corresponds rather to the ``size'' $||\delta L||_{_E}$ of the
  perturbation of the operator $L$ generating evolution in equation
  (\ref{e:wave_eq_1storder_intro}).  Such a quantity is indeed related
  to the "energy" (norm) of spacetime perturbations $\delta g_{ab}$,
  but such a relation is not a simple one.

The discussion in section \ref{sec:physical_interpretation_energy}, leads to a ``pointwise''
estimation of $||\delta L||_{_E}$ in terms of the energy density of
$\delta g_{ab}$ (understood as a field perturbation on a fixed
background).  More specifically, $||\delta L||_{_E}$ is estimated by
the maximum of (compact support) perturbations $\delta \tilde{V}$ to
the effective potential in equation \eqref{VtildeDef}, namely \be ||\delta
L||_{_E} \sim \max_{\mathrm{supp}(\delta \tilde{V})}{|\delta
  \tilde{V}|} \ .  \ee On the other side, the variations $\delta
\tilde{V}$ are of the order of second derivatives of the metric
perturbations, $\partial^2\delta g_{ab}$, this leading to $\delta
\tilde{V}$ being estimated by an energy density of the perturbations
$\rho_{E, \delta g}(x)\sim |\partial \delta g_{ab}|^2$, so we can
crudely write \be
\label{e:delta_L_max_rho}
|\delta g_{ab}| \; ||\delta L||_{_E} \sim \max_{\mathrm{supp}(\delta g)} |\partial \delta
g_{ab}|^2 \sim \max_{\mathrm{supp}(\delta g)}\rho_E(\delta g; x) \ . \ee
This pointwise ($L^\infty$-like) expression indicates that when
estimating the perturbations $\delta L$ triggering
QNM instability, the right quantity to consider is not the total
energy of the perturbations $\delta g_{ab}$, but the {\em peaks}
of their spacetime energy distribution.  In other words: given two
spacetime perturbations $\delta g_{ab}$ with the same energy $E
= \int_{\mathrm{supp}(\delta g)} \rho_{E, \delta g}(x)dx$, the one
impacting more strongly the QNM instability is the one with a larger
maximum in $\rho_{E, \delta g}(x)$.

From a physical perspective the conclusion is that {\em spiked}
distributions of spacetime perturbations are more efficient in
triggering QNM instabilities than smooth
distributions. Astrophysically, clouds of compact objects would be
more efficient than clouds of gas.

\subsubsection{ QNM instability and compact object QNM universality: from high-frequency
Burnett's conjecture to a Regge-QNM's conjecture.} 
As discussed in section \ref{e:QNM_instability_general_picture}
  (cf. also \ref{s:general_picture_intro}
above), ultraviolet perturbations push
QNMs toward logarithmic curves \eqref{e:log_pseudospectrum_lines_intro}, getting closer
and closer as the norm or the frequency of the perturbation increase, but always laying
above them as long as the spacetime perturbations are smooth.  A question is posed:
{\em do Nollert-Price branches reach logarithmic pseudospectra boundaries in
  the limit of infinite frequency?}

Reference \cite{Jaramillo:2021tmt} conjectures that this is the case,
based on numerical considerations. In section \ref{s:Burnett}, a
heuristic avenue to support this conjecture and to systematically
address this question is proposed based on the so-called Burnett's
conjecture \cite{Burne89} on the high-frequency limit of spacetimes.
Burnett's conjecture states that the limit of high-frequency
oscillations of a vacuum spacetime is effectively described by an
effective matter spacetime described by massless
Einstein-Vlasov. However, Burnett's original approach does not allow
for spacetime ``concentrations'' (spikes), the key feature of
spacetime perturbations for efficiently triggering instabilities,
according to expression (\ref{e:delta_L_max_rho}). A refinement of the
high-frequency limit scheme, allowing for concentrations, has been
presented by Luk \& Rodnianski \cite{Luk:2020pyn}.  The main outcome
of such analysis, for the considerations our present setting, is that
the high-frequency spacetime limit to $g_\infty$ is of low-regularity
nature, namely
\be
\label{e:conditions_Luk-Rodnianski_intro}
g_n\to g_\infty \ \ \hbox{in } C^0 \ \ , \ \ \partial g_n\to \partial
g_\infty \ \ \hbox{in } L^2 \ .
\ee Putting this together with the
results \cite{Regge58,Zwors87} (see also \cite{Berry82}), stating
that QNMs of low-regularity $C^p$ potentials have a universal
structure in terms of so-called Regge (broad) resonances, 
we formulate the QNM high-frequency limit conjecture in \cite{Jaramillo:2021tmt}  as
:\\ {\em Under generic ultraviolet
  perturbations and in the limit of infinite frequency, BH
  Nollert-Price QNM branches become Regge QNM branches, presenting the
  $n\gg1$ asymptotics} \begin{eqnarray}
\label{e:Regge_branches_intro}
\mathrm{Re}(\omega_n) &\sim& \pm\Big(\frac{\pi}{L} n + \frac{\pi
  \gamma_p}{2L}\Big) \\ \mathrm{Im}(\omega_n) &\sim& \frac{1}{L}
\bigg[\gamma\ln\bigg(\Big(\frac{\pi}{L} n + \frac{\pi
    \gamma_p}{2L}\Big) + \frac{\pi \gamma_\Delta}{2L}\bigg) - \ln
  S\bigg] \ . \nn \end{eqnarray}  This conjecture states not only that
Nollert-Price QNMs asymptote to logarithmic lines in the infinite
high-frequency limit, but also proposes the specific pattern that QNMs
follow on those logarithmic curves.  This connects with the
asymptotics of $w$-modes of matter compact objects
\cite{Kokkotas:1999bd,ZhaWuLeu11}, enforcing the idea proposed in
\cite{Jaramillo:2020tuu} that ($w$-)QNM asymptotics of {\em generic} (vacuum
or matter) astrophysical compact objects are universal in the
high-frequency limit.

\subsubsection{Towards a geometric formulation of QNM instability.}
In their current form, all discussions above depend strongly on a
particular choice of coordinates in the hyperboloidal approach. This
could be an obstacle when trying to extend such analyses to
generic situations (e.g. beyond spherical symmetry), but also
can obscure their underlying invariant content.

Section \ref{s:geometric_approach} aims at laying the
basic elements to build a genuinely geometric construction of the
approach.  The main elements ingredients can be summarized in the
following points:
\begin{itemize}
\item[i)] {\em Foliations of the null boundaries}.  The basic starting
  point would be the choice of respective foliations $\{\cal  S^\infty_\tau\}$
  of null infinity $\scri^+$ and $\{\cal S^{\cal H}_\tau\}$
  of the BH horizon ${\cal H}$, together with a function
  $\gamma$ defined on sections $\cal S^\infty_\tau$ and $\cal S^{\cal
    H}_\tau$ that controls the energy flux in
  equation (\ref{eq:totalFluxExpression_reduced:intro}).  These objects are
  arbitrary, but are kept fixed. Then, the boundary foliations
  are extended to the bulk in a spacetime (hyperboloidal) foliation $\{\Sigma_\tau\}$
  that ``interpolates'' between $\{\cal S^\infty_\tau\}$ and $\{\cal S^{\cal  H}_\tau\}$.
  The resulting slices $\Sigma_{\tau}$ are then compactified.
  Such bulk extension and the subsequent compactification are (essentially) arbitrary.
   
\item[ii)] {\em QNMs in a geometric approach.} Expressions for $L_1$
  and $L_2$ in equations \eqref{e:L_1-L_2_intro} are promoted to a
  geometric form (\ref{e:L_1-L_2}) on $\Sigma_\tau$, and boundaries
  $\cal S^\infty_\tau$ and $\cal S^{\cal H}_\tau$ are ``shrinked'' to
  points $i^\infty$ and $i^{\cal H}$, respectively.  Slices
  $\Sigma_\tau$ become $3$-spheres $\mathbb{S}^3$ ``pinched'' at
  $i^\infty$ and $i^{\cal H}$.  In this setting, $L_1$ is promoted to
  an operator $\mathring{L}_1$ acting on a ``completed'' 3-sphere,
  whereas $L_2$ acts on the pinched $3$-sphere. No boundary conditions
  enter through $\mathring{L}_1$ (that acts on a closed manifold) and,
  regarding $L_2$, only the value of $\gamma$ at the pinching holes is
  relevant. Under these conditions, the QNM spectrum would become
  automatically discrete.

\item[iii)] {\em ``Missing'' degrees of freedom and BMS symmetry.}
  Denoting by $\ell^a$ the null generators of $\scri^+$ and ${\cal
    H}$, vector fields on such null boundaries satisfying \be
\label{e:xi_BMS_intro}
\xi^a = \gamma \ell^a \ \ , \ \ {\cal L}_\ell \gamma = 0 \ , \ee can
be associated with BMS supertranslations (in ${\cal H}$ this is an
abuse of language). Such symmetries would relate the different
possible choices of boundary foliations in the point $i)$ above,
restoring invariance to the construction. On the other hand,
understanding BMS as a dynamical symmetry generating a phase space of
degrees of freedom at the null boundaries ---or, in another language,
understanding BMS as (boundary residual) broken symmetry enlarging the
physical phase space--- outgoing bulk degrees of freedom $\phi$ would
interact with ``pinching-BMS'' degrees of freedom $\gamma$ at
$i^\infty$ and $i^{\cal H}$, with a coupling fixed in terms of the
flux term \eqref{eq:totalFluxExpression_reduced:intro}.  In
particular, $\mathring{L}_1$ would be a bulk operator oblivious to
pinching holes, whereas $L_2$ would be essentially a ``pinching-BMS''
operator.

The extension of the phase space of degrees of freedom with such
``boundary data'' could provide clues into QNM instability from an
``inverse scattering'' perspective and, in parallel, could offer an
avenue to restore selfadjointness into the full problem, leading to a
notion of global spacetime {\em normal modes} in the classical
formulation of the problem.

\end{itemize}

\section{QNM spectral instability, pseudospectrum and scalar product choice}
\label{s:QNM_stability_norm_choice}
In this section we discuss the question of the choice of scalar
product ---and its associated norm--- in the study of the spectral
stability properties of an operator. First, we illustrate the impact
of such a choice with a simple example.  Second, we present the
spacetime construction of the energy scalar product proposed in
\cite{Jaramillo:2020tuu} for the hyperboloidal approach to QNMs.

\subsection{Spectral instability as a norm-dependent notion: a study case}\label{sec:ABC}
We consider a simple linear second order differential operator ---with constant
  coefficients--- and study the role of the choice of scalar product in the assessment
  of selfadjointness and normality. Once this is done, we display the consequence
  of such choices in its eigenvalue stability problem, by using the pseudospectrum.

%%%%%%%%%%%%
%do this by detailing the calculation of the formal
%adjoint of a simple \blue{linear elliptic} second order differential operator and discussing its
%normality properties.
%
%BWe start by motiIn this section we motivate and stress the importance of the choice
%of inner product when assessing the properties of linear (differential)
%operators, \blue{in particular spectral stability}.
%We do this by detailing the calculation of the formal
%adjoint of a simple \blue{linear elliptic} second order differential operator and discussing its
%normality properties.
%%%%%%%%%%%

Although this calculation is elementary, we find it illustrative for
its explicit character.
%%%%%%%%%%%
%since it shows in an explicit
%form, the impact of choosing an appropriate inner product in arriving to
%conclusions about the properties of a given operator.
%%%%%%%%%%%
Additionally, it exhibits, in a very simple setting, the difference
between a formally normal operator and an actual normal operator, key
in the functional context of QNM instability.

\subsubsection{Scalar product, selfadjointness and normality.}
\label{s:example_Q_selfadjointness}
Let $Q$ be the operator \be \label{e:Q}Q = a \frac{d^2}{dx^2}+ b
\frac{d}{dx} + c \ ,\ee where $a, b, c \in \mathbb{R}$, with $a
  \neq 0$, acting on functions defined on the interval $x\in
       [x_0,x_1]$.  To complete the definition of $Q$ we must specify
is domain ${\cal D}$ , i.e. the functional space in which its
acts. Let us illustrate the involved issues in different points:
\begin{itemize}
\item[i)] {\em Square-integral functions: assessing self-adjointness}.
  First we consider the standard space of square-integrable functions
  $L^2([x_0,x_1],dx)$, with scalar product
  \be \label{eq:id_inner_product} \langle \varphi, \phi\rangle_{_2} =
  \int_{x_0}^{x_1} \bar{\varphi}\phi\; dx \ .  \ee A direct
  calculation using integration by parts shows that
  \be \label{e:scalar_product_2}\langle \varphi, Q \phi \rangle_{_2} =
  \bigg( a (\bar{\varphi}\phi' - \phi\bar{\varphi}') +
  b\bar{\varphi}\phi \bigg)\bigg|_{x_0}^{x_1} + \langle
  Q^\dagger_{_2}\varphi, \phi \rangle_{_2} \ ,\ee where
  $Q^\dagger_{_2}$ is given by \be\label{e:Q_dagger} Q^\dagger_{_2} =
  a \frac{d^2}{dx^2} - b \frac{d}{dx} + c \ .\ee If we now impose
  homogeneous Dirichlet boundary conditions, we can write
  \be \label{e:formal_adjoint_Q}\langle \varphi, Q \phi \rangle_{_2} =
  \langle Q^\dagger \varphi, \phi \rangle_{_2} \ ,\ee and
  $Q^\dagger_{_2}$ is the ``formal adjoint'' of $Q$ ---the subscript
  $2$ stresses that this result depends on the employed scalar product
  $\langle\cdot,\cdot\rangle_{_2}$, as it will be further discussed
  below.  A symmetric operator is characterised by
\be\langle \varphi, Q \phi \rangle = \langle Q \varphi, \phi \rangle
\ ,\ee and, from equations \eqref{e:Q} and \eqref{e:Q_dagger}, this
happens for $Q$ when $b=0$. If $b\neq 0$ the operator $Q$ is not
symmetric and, therefore, cannot be selfadjoint. For $b=0$, $Q$ is
selfadjoint if, in addition, the domains of $Q$ and $Q^\dagger_{_2}$
coincide, something we can achieve by imposing $\mathcal{D}(Q)
=\mathcal{D}(Q^\dagger_{_2}) = \mathcal{D}$ with
\be \mathcal{D} = \{
\phi \in L^2([x_0,x_1],dx) \mathrel{}\mid\mathrel{}
\phi(x_0)=\phi(x_1)=0 \} \ ,
\ee the $Q$ is a selfadjoint
operator\footnote{As pointed out above, if the domains
    $\mathcal{D}(Q)$ and $\mathcal{D}(Q^\dagger_{_2})$ are not
    specified one refers to $Q^\dagger_{_2}$ as the {\em{formal}}
    adjoint of $Q$ and, if \eqref{e:formal_adjoint_Q} is satisfied,
    then $Q$ is formally selfadjoint.}  In sum, for homogeneous
  Dirichlet (more generally, homogeneous Robin) boundary conditions,
  $Q$ is selfadjoint if $b=0$. In our setting, this translates in the
  fact that $Q$ is spectrally stable for $b=0$ and there is no
  guarantee of spectral stability for $b\neq 0$. This is indeed
  confirmed numerically: under small randon perturbations $\delta Q$
  of small size $||\delta Q||_{_2} = \epsilon$ the eigenvalues for
  $b=0$ changes linearly in $\epsilon$, whereas in the non-selfadjoint
  case $b\neq 0$ we find indeed spectral instability.

%%%%%%%%%
%Although specifying the domains of $Q$ and
%$Q^\dagger_{_2}$ may seem just ludic, other properties such as normality
%are less obvious and the appropriate domains should be taken into
%account.
%%%%%%%
  
\item[ii)] {\em Normality and ``formal normality''}. The
  non-selfadjoint case $b\neq 0$ presents an apparent catch. Spectral
  stability is guaranteed for a larger class than selfadjoint
  operators. Normal operators ---namely operators that commute with
  their adjoints--- also satisfy a spectral theorem guaranteeing
  spectrally stability. In our case, because of the constancy of $a$,
  $b$ and $c$ in equations \eqref{e:Q} and \eqref{e:formal_adjoint_Q},
  operators $Q$ and $Q^\dagger_{_2}$ actually satisfy
  \be\label{e:formal_normal}[Q,Q^\dagger_{_2}] = QQ^\dagger_{_2} -
  Q^\dagger_{_2}Q =0 \ \ \Longleftrightarrow \ \ QQ^\dagger_{_2} =
  Q^\dagger_{_2}Q \ .\ee In spite of this, the operator $Q$ for $b\neq
  0$ is indeed spectrally unstable (in the $||\cdot||_{_2}$ norm, see
  later): how does this reconcile with $Q$ satifying
  \eqref{e:formal_normal} for any $b$?
  
  The way out of this apparent contradiction is that $Q$ satisfying
  \eqref{e:formal_normal} only qualifies it as ``formally normal''. In
  order to conclude actual normality (for which spectral stability
  holds) for $Q$, one needs to verify that the domains
  $\mathcal{D}(QQ^\dagger_{_2})$ and $\mathcal{D}(Q^\dagger_{_2}Q)$
  coincide, so that operators $QQ^\dagger_{_2}$ and $Q^\dagger_{_2}Q$
  are indeed the same.  Such domains are characterised as
\begin{eqnarray}
\qquad\quad \mathcal{D}(QQ^\dagger_{_2}) & =& \{ \phi \in \mathcal{D}(Q^\dagger_{_2})
\mathrel{}\mid\mathrel{} Q^\dagger_{_2} \phi \in \mathcal{D}(Q)
\},  \nonumber \\
\qquad\quad \mathcal{D}(Q^\dagger_{_2}Q) & =& \{ \phi \in \mathcal{D}(Q)
\mathrel{}\mid\mathrel{} Q \phi \in \mathcal{D}(Q^\dagger_{_2}) \}
\end{eqnarray}
Since $\mathcal{D}(Q)=\mathcal{D}(Q^\dagger_{_2})=\mathcal{D}$,
unwrapping the definitions leads to
\begin{align}
\mathcal{D}(QQ^\dagger_{_2}) &= 
\{ \phi \in \mathcal{H}_{_2} \mathrel{}\mid\mathrel{}
\phi(x_0)=\phi(x_1)=0, %\qquad\qquad\quad \nonumber \\&
\phi''(x_0)=
a^{-1} b\phi'(x_0), \; \phi''(x_1)=a^{-1}b\phi'(x_1)\} %\nonumber
\\
\mathcal{D}(Q^\dagger_{_2} Q) &=  \{ \phi \in \mathcal{H}_{_2}
\mathrel{}\mid\mathrel{} \phi(x_0) =\phi(x_1)=0, % \qquad\qquad\quad \nonumber \\ & \hspace{-1.0cm}
\phi''(x_0)=- a^{-1} b\phi'(x_0), \;
\phi''(x_1)=-a^{-1}b\phi'(x_1)\} \ . %\nn
\end{align} Hence, if $b \neq 0$, we can conclude  
\be
\mathcal{D}(QQ^\dagger_{_2}) \neq \mathcal{D}(Q^\dagger_{_2}Q) \ . \ee
Therefore, if $b \neq 0$, then $Q$ is not normal 
due to the mismatch of the domains: it is only formally normal.
This restores consistency with the observed spectral instability when using $||\cdot||_{_2}$
to measure the size of perturbations.

\setcounter{footnote}{0}

\item[iii)] {\em The crucial role of the scalar product: a full
  reassessment of selfadjointness and spectral stability}. But there
  is still a further twist with the $Q$ operator in equation
  \eqref{e:Q}: this operator is actually selfadjoint for all values
  $a$, $b$ and $c$, if the scalar product is chosen in the appropriate
  way. In particular, for $b\neq 0$, there is a choice of scalar
  product for which $Q$ is not only normal ---and therefore normal and
  spectrally stable--- but actually selfadjoint\footnote{We thank
    Graham Cox for pointing out this fact, that makes the example of
    the operator $Q$ in equation \eqref{e:Q} even richer than we had
    originally noticed.}.
%%%%%%%%%%%
%To appreciate the role of the inner product in the above
%  discussion we recall that any
%%%%%%%%%

  This follows from the fact that any linear differential second-order
  operator in one dimension can be written in Sturm-Liouville form by
  using an appropriate ``integrating factor'' $w(x)$.  The latter can
  the be used to build the scalar product $\langle\cdot{}, \cdot{}
  \rangle_{_w}$, that makes the original operator actually
  selfadjoint.  To see this in our particular case, let us consider
  the function $ w(x)= a^{-1} \exp( a^{-1}bx)$, with $x\in [x_0,x_1]$.
  Then using this integrating factor, $Q$ can be rewritten in the
  Sturm-Liouville form
  \be Q = \frac{1}{w(x)}\bigg(\frac{d}{dx}
  \bigg(p(x) \frac{d}{dx}\bigg) - q(x) \bigg) \ ,
  \ee where $p(x) =
  aw(x)$ and $q(x) = - cw(x)$.  Then, defining the scalar product
  \be \label{eq:SL-inner_product} \langle \varphi, \phi\rangle_{_w} =
  \int_{x_0}^{x_1} \bar{\varphi}\phi \; w(x) dx \ , \ee and
  integrating by parts one obtains \be\label{eq:QSelfadjoint} \langle
  \varphi, Q \phi \rangle_{_w} = \bigg( p (\bar{\varphi}\phi' -
  \phi\bar{\varphi}') \bigg)\bigg|_{x_0}^{x_1} + \langle
  Q^\dagger_{_w}\varphi, \phi \rangle_{_w} \ , \ee where
  $Q^\dagger_{_w} = Q$. Thus, if we impose homogeneous Dirichlet
  (again, more generally, homogeneous Robin) boundary conditions
  \be
  \mathcal{D} = \{ \phi \in (L^2([x_0,x_1],w(x)dx)
  \mathrel{}\mid\mathrel{} \phi(x_0)=\phi(x_1)=0 \} \ ,
  \ee one
  concludes that $Q$ is a selfadjoint operator.

\end{itemize}
If we focus now in the case $b \neq 0$, how can we reconcile the fact
that $Q$ is non-normal and spectrally unstable when using the norm
$||\cdot||_{_2}$, while selfadjoint and therefore spectrally stable
consider in the norm $||\cdot||_{_w}$?  Then answer is that a given
perturbation $\delta Q$ can have a small $2$-norm $||\delta Q||_{_2}$
while having a large $w$-norm $||\delta Q||_{_w}$. In the former case,
a significant change in the eigenvalues is interpreted as a spectral
instability, whereas in the latter case the large change in the
spectrum is consistent with stability, since the operator perturbation
is also large. This provides a neat illustration of the important of
assessing ``big'' and ``small'' through the norm, when discussing
spectral instability. In the next subsection we address this point
from the point of view of the pseudspectrum.

\subsubsection{QNM stability through the pseudospectrum: the role of the scalar product.}
\label{s:pseudospectru_ABC}
As discussed in section \ref{s:spectral_inst_intro}, the notion of pseudospectrum
provides an avenue to assess spectral instability. Here we discuss how the choice
of the scalar product impacts the structure of the pseudospectrum, illustrating
it with the operator $Q$ and scalars products dicussed above.

%%%%%%%%%%%%%%%
%As we have just seen, the choice of inner product is crucial for the study of
%perturbations and stability of an operator via its pseudospectrum.
%%%%%%%%%%%%%%%

Following the systematic presentation in \cite{TreEmb05Book} (cf. also
the discussion in \cite{Jaramillo:2020tuu} in the present
gravitational setting) the pseudospectrum consists, by definition, of
nested sets in the complex plane around the eigenvalues of the
operator.  The pseudospectrum allows to measure how much a
perturbations to the operator affect its eigenvalues.  As a
consequence of the Bauer-Fike theorem~\cite{TreEmb05Book}, the
eigenvalues of a selfadjoint operator are stable in the sense that a
change $\delta Q$ of ``size'' $\epsilon$ in the operator result in a
change of the same order $\epsilon$ in the eigenvalues. As a
consequence, graphically, selfadjoint operators have a ``flat
pseudospectrum'' (see discussion in \cite{Jaramillo:2020tuu}).  Notice
that although the eigenvalues of an operator are independent of the
chosen norm $||\cdot||$, the notion of pseudospectrum is not, since
determining the size of the perturbation $||\delta Q||$ depends
intimately on the choice for $||\cdot||$. This is particularly
important for physical applications where one is interested in small
perturbations where the norm should correspond to a physically
relevant notion of energy for the problem.

\begin{figure}[h!]
\centering \includegraphics[width=8.5cm]{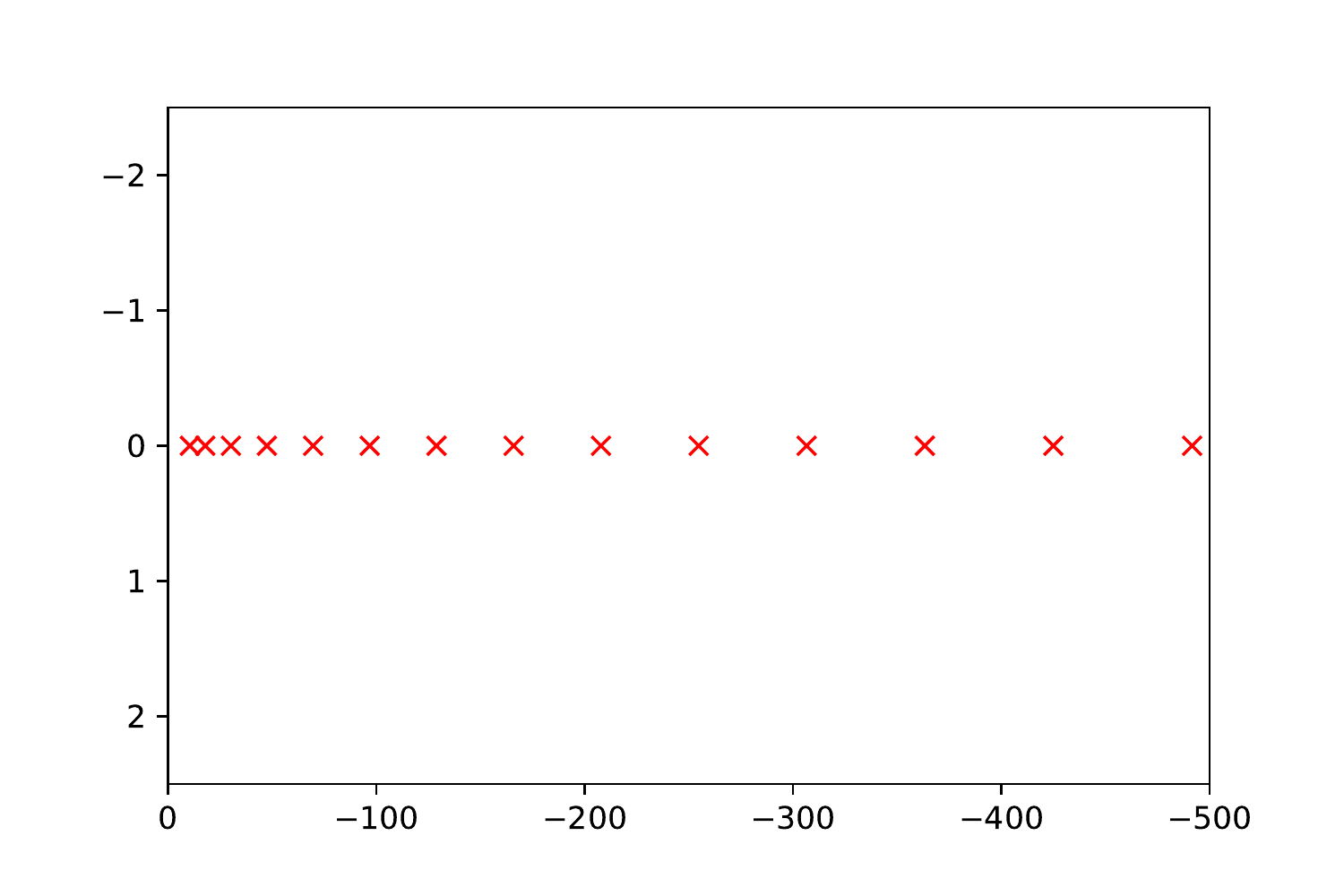}
\caption{Numerically computed eigenvalues for the operator $\hat{Q}$
  with $x_0=-1,x_1=1$ and $a=1,b=6,c=1$ with a spatial discretization
  $N=100$ points}
\label{fig:SpectrumQ}
\end{figure}

The latter point is illustrated in the following through the
construction of the pseudospectrum of the operator $Q$, computed
numerically using spectral methods, by employing the two norms induced
by the inner products $\langle\cdot, \cdot\rangle_{_2}$ and
$\langle\cdot, \cdot\rangle_{_w}$, respectively introduced in
equations \eqref{eq:id_inner_product} and
\eqref{eq:SL-inner_product}. To be more precise, we consider the
spectral stability of a related operator $\hat{Q}$, whose eigenvalue
problem is closer to the ones studied in \cite{Jaramillo:2020tuu} in
QNM context.  Specifically, starting from the eigenvalue for $Q$
\begin{equation}\label{eq:Simple_EigenProblem}
Q\Phi = \lambda \Phi \ ,
\end{equation}
we define the rescaled variable $\Phi = -(x-x_0)(x-x_1)\phi$.  This
incorporates the boundary conditions into the operator~\footnote{This
  is in the same spirit that outgoing boundary conditions in the
  hyperboloidal
  approach~\cite{Ansorg:2016ztf,PanossoMacedo:2018hab,PanossoMacedo:2020biw,Jaramillo:2020tuu}
  are incorporated into the operator by choosing a slicing intersects
  the BH horizon and null infinity.}.  Specifically, requiring
boundedness for $\phi$ at $x_0$ and $x_1$, encodes imposing
homogeneous Dirichlet boundary conditions for $\Phi$.  Then, we can
write the eigenvalue problem \eqref{eq:Simple_EigenProblem}
as~\footnote{ From the discussion in \ref{s:example_Q_selfadjointness}
  we know that there exists a scalar product ---that associated to the
  Sturm-Liouville form of $Q$--- for which the operator $Q$ is
  selfadjoint ---a completely analogous calculation can be done for
  $\hat{Q}$--- hence we know $Q$ must have real eigenvalues.  One can
  can numerically corroborate this fact. A numerical implementation in
  {\tt{python}} using the {\tt{linalg}} package for linear algebra in
  {\tt{numpy}}, and using Chebyshev spectral methods to approximate
  the derivatives ---see \cite{Tre00,Jaramillo:2020tuu} for further
  discussion on spectral methods--- renders Figure
  \ref{fig:SpectrumQ}.}
\begin{equation}
\hat{Q}\phi = \lambda \phi \ ,
\end{equation}
where,
\begin{equation}
\hat{Q} = A(x) \frac{d^2}{dx^2}+ B(x) \frac{d}{dx} + C(x) \ ,
\end{equation}
with
\begin{flalign}\label{eq:abcfunctions}
A(x) & = -a (x-x_0) (x-x_1), \nn\\ B(x) & = -b (x-x_0) (x-x_1)+2 a (-2
x+x_0+x_1),\nn\\ C(x) & = -2 a- c (x-x_0) (x-x_1)+b (-2 x+x_0+x_1) \ .
\end{flalign}

We illustrate the impact of the choice scalar product and its
associated norm in the pseudospectrum of the operator $\hat{Q}$, by
computating it numerically.  The starting point is its second
characterization in equation \eqref{e:pseudospectrum_def}, in a
discretized version of the operators and scalar product (details are
given in Appendices B and C in \cite{Jaramillo:2020tuu}). In
particular, given a scalar product $\langle \cdot, \cdot \rangle_{_G}$
and two functions $\varphi, \phi$, the discretized version of their
scalar product can be written as $\langle \varphi, \phi \rangle_{_G} =
(\varphi^*)^i G_{ij}\phi^j$ where $G_{ij}$ is the Gram matrix
associated with the scalar product and $*$ is the conjugate-transpose
(note the abuse of notation with $\varphi, \phi \in \mathbb{C}^n$ also
denoting the discretized version of the functions). The discretised
adjoint $\hat{Q}^\dagger$ is then written in terms of the Gram matrix
as $\hat{Q}^\dagger = G^{-1}\hat{Q}^*G$. With these elements, it can
be shown \cite{trefethen2005spectra,Jaramillo:2020tuu} that the
$\sigma^{\epsilon}_{_G}(\hat{Q})$ pseudospectrum associated with the
scalar product $\langle \cdot, \cdot \rangle_{_G}$ is characterised as
\be
\label{e:pseudospectrum_carac_G}
\sigma^\epsilon_{_G} (\hat{Q}) = \{\lambda\in\mathbb{C}:
s_{_G}^\mathrm{min}(\lambda \mathrm{Id}-\hat{Q})<\epsilon\} \ , \ee
where $s_{_G}^\mathrm{min}(M) = \min \{\sqrt{\lambda}: \lambda\in
\sigma(M^\dagger M) \}$, is the minimum of the singular values of the
matrix $M$ (in a generalised version associated adjoints respect to
$\langle \cdot, \cdot \rangle_{_G}$).

\begin{figure}[ht!]
\centering
\includegraphics[width=8.5cm]{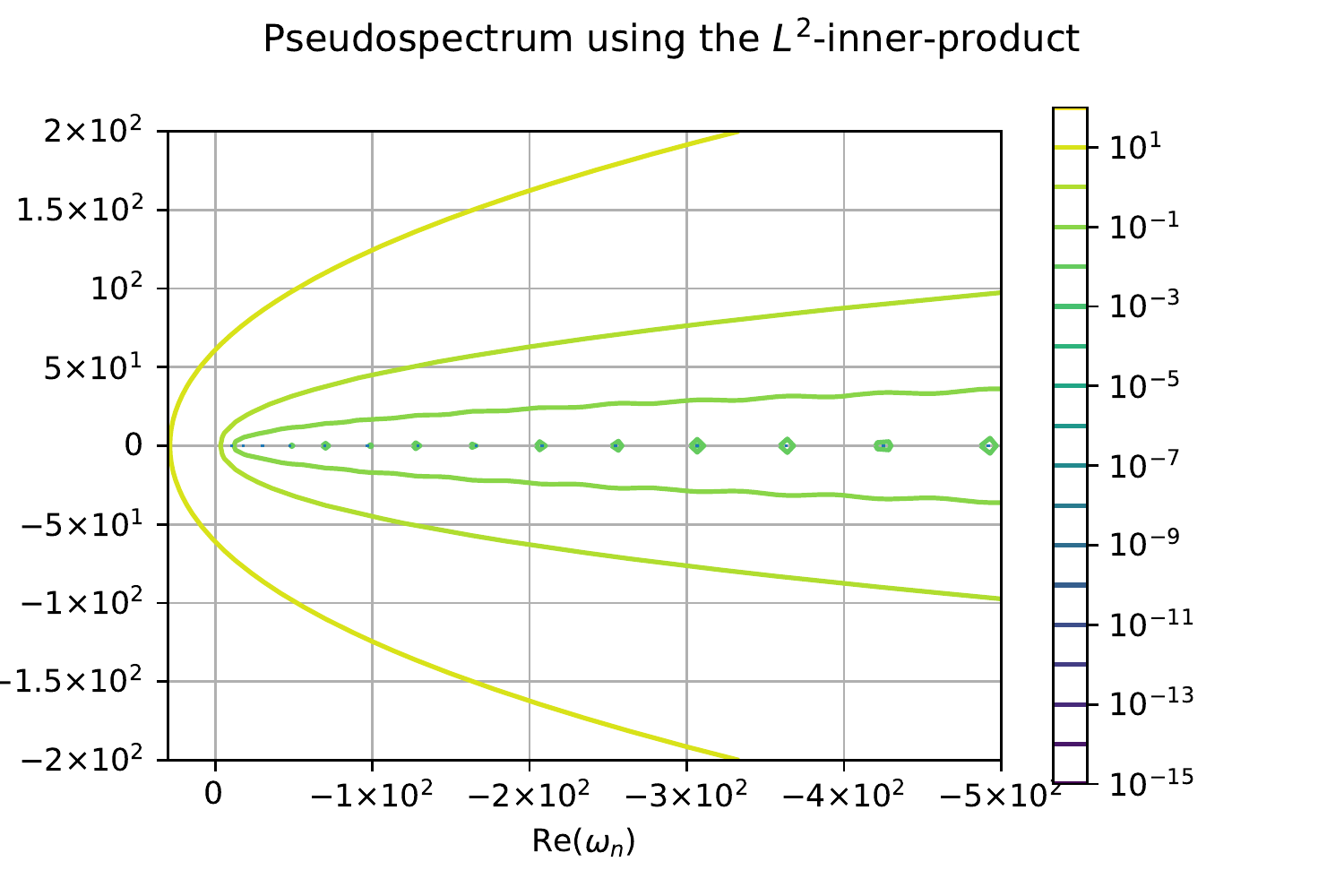}
\caption{Numerically computed pseudospectrum for the operator
  $\hat{Q}$ with $x_0=-1,x_1=1$ and $a=1,b=6,c=1$ with a spatial
  discretization $N=100$ points.  The standard matrix norm inherited
  from the standard $L^2$-inner product was used.  One can observe
  that although the pseudospectrum consists of nested sets around the
  eigenvalues of $\hat{Q}$ it is not flat: the $\epsilon$ contour
  lines open up as one considers eigenvalues with larger modulus
  (overtones).}
\label{fig:PsedudospectrumQStandardInnerProduct}
\end{figure}

We are now in condition of calculating the pseudospectra
$\sigma^{\epsilon}_{_2}(\hat{Q})$ and
$\sigma^{\epsilon}_{_w}(\hat{Q})$ associated, respectively to the
standard $L^2$ scalar product \eqref{eq:id_inner_product} and the
Sturm-Liouville-like \eqref{eq:SL-inner_product}.  The former produces
the pseudospectrum shown in Figure
\ref{fig:PsedudospectrumQStandardInnerProduct} for the operator
$\hat{Q}$. As one can observe in Figure
\ref{fig:PsedudospectrumQStandardInnerProduct}, using the $L^2$-inner
product renders a ``non-flat'' pseudospectrum for $\hat{Q}$. Note the
range of values of $\epsilon$: the borad extension of pseudospectra
sets with small $\epsilon$ indicates spectral instability.

\begin{figure}[ht!]
  \centering
  \includegraphics[width=8.5cm]{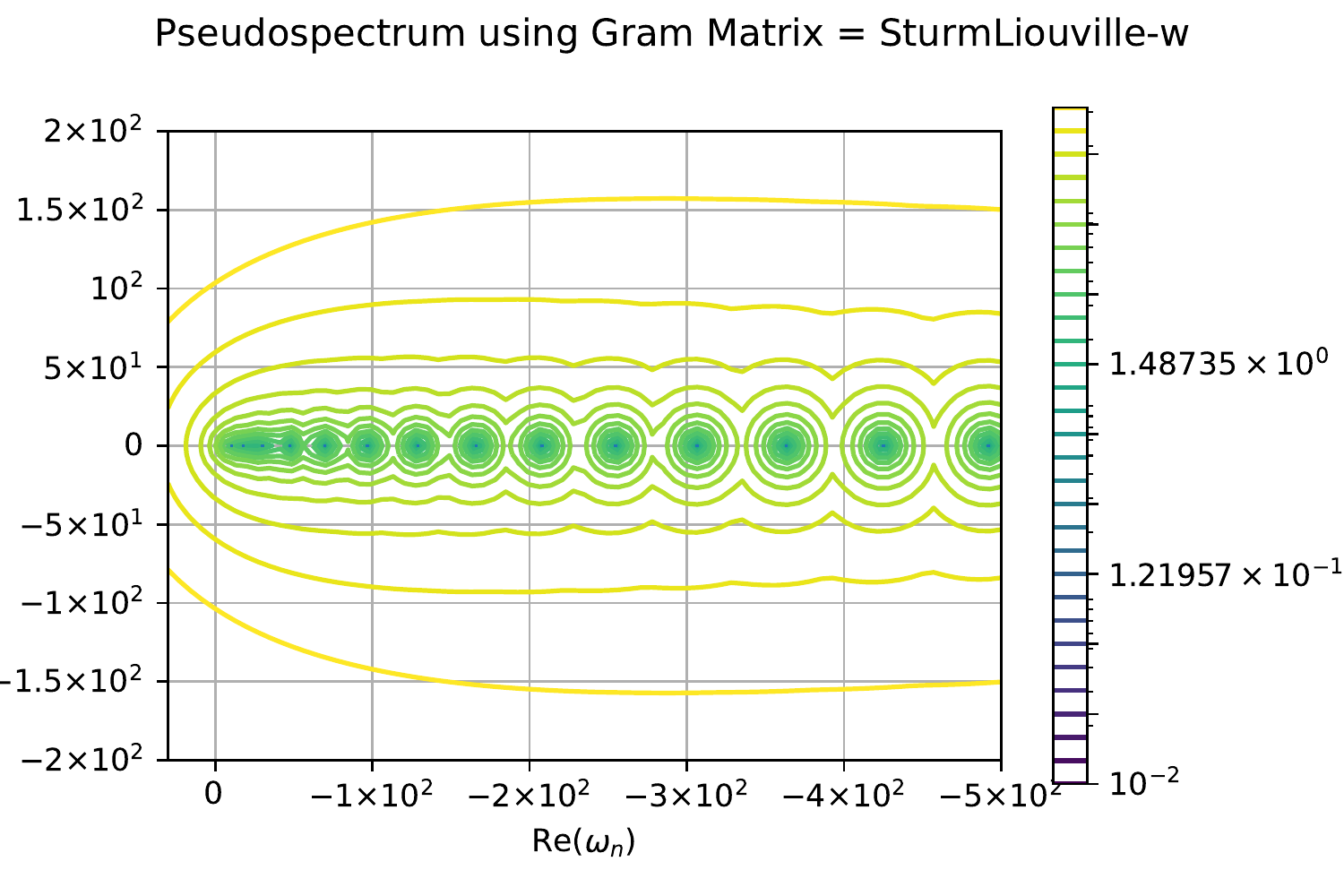}
\caption{Numerically computed pseudospectrum for the operator
  $\hat{Q}$ with $x_0=-1,x_1=1$ and $a=1,b=6,c=1$ with a spatial
  discretization $N=100$ points.  The Gram matrix associated to the
  Sturm-Liouville inner product was used.  One can observe that with
  this inner product, the pseudospectrum of $\hat{Q}$ is flat,
  indicating stability which is consistent with the selfadjointness of
  this operator respect to the inner product
  \eqref{eq:SL-inner_product}}
\label{fig:PsedudospectrumQSturmLiouvilleInnerProduct}
\end{figure}

Regarding $\sigma^{\epsilon}_{_w}(\hat{Q})$, the key information is encoded
in the integration factor $w(x)$ associated to the Sturm-Liouville 
form of the $\hat{Q}$ operator, namely
\begin{equation}\label{eq:w_sturm_Liouville_general}
  w(x) = A(x)^{-1} \exp\Bigg(\int_{x} A(x')^{-1}B(x')dx'\Bigg) \ .
\end{equation}
Explicitly for the present case, substituting the polynomials of
equation \eqref{eq:abcfunctions} into equation
\eqref{eq:w_sturm_Liouville_general} and integrating, renders \be
w(x)=- \frac{(x-x_0)(x-x_1)}{a} \exp\Big(\frac{b(x-x_0)}{a}\Big)
\ . \ee The numerical Chebyshev spectral methods implementation of the
Gram matrix associated to the scalar product in $\langle\cdot, \cdot
\rangle_{_w}$ in equation \eqref{eq:SL-inner_product}, following the
discussion in Appendix C of \cite{Jaramillo:2020tuu}, gives the
pseudospectrum shown in Figure
\ref{fig:PsedudospectrumQSturmLiouvilleInnerProduct}.  As one can
observe in Figure \ref{fig:PsedudospectrumQSturmLiouvilleInnerProduct}
the pseudospectrum of $\hat{Q}$ is ``flat'' as expected, with
concentric circles around eigenvalues, and a completely different
range of $\epsilon$'s with large $\epsilon$' in pseudospectra
extending far from the eigenvalues: this indicates stability as it
corresponds to a selfadjoitn operator.  The contrast between Figures
\ref{fig:PsedudospectrumQStandardInnerProduct} and
\ref{fig:PsedudospectrumQSturmLiouvilleInnerProduct} shows, in a
controlled set up, that given an operator, the choice of inner product
and associated norm ---with their corresponding (different) notions of
big and small--- play a fundamental role for the assessment of
spectral stability: an operator can be stable in one norm and unstable
in another. The proper choice of scalar product is therefore critical.

\subsection{Hyperboloidal slicing framework: the energy scalar product choice}
\label{sec:Energy_associated_inner_products}
Under the light of the previous discussion, the choice of scalar product
  becomes fundamental for the physical assessment of QNM stability.
  Following the proposal in \cite{Jaramillo:2020tuu}, we propose a scalar
product based on the energy of the propagating field, as the proper measure
of big and small in physical scenarios involving QNM perturbations.

In this section we revisit and extend the discussion in
\cite{Jaramillo:2020tuu}, to place it on a sounder spacetime
ground. We first discuss and motivate such choice of scalar product in
our problem.  Then we derive the relation between the physical energy
norm associated to a scalar field propagating on a spherically
symmetric spacetime background and the ``effective'' stress energy
energy used in \cite{Jaramillo:2020tuu} to define energy and an
``effective'' energy scalar product.  At the end of this section the
energy flux at the boundary is obtained.

\subsubsection{Elements of the hyperboloidal framework in spherical symmetry.}
Let $(\mathcal{M}, g_{ab}, \nabla)$ denote
a manifold equipped with is a Lorentzian metric $g$ and associated
Levi-Civita connection $\nabla$.  Consider a complex scalar field
$\Phi$ satisfying the wave equation
\begin{equation} \label{eq:Wave_eq_Phi}
\Box{}_{g} \Phi = g^{ab}\nabla_a\nabla_b \Phi = 0.
\end{equation}

The energy momentum tensor associated to this wave equation is given
by \be
\label{eq:EnergyMomemtumComplexScalar}
T_{ab} = \frac{1}{2} \Bigg(\nabla_a \bar{\Phi} \nabla_b \Phi -
\frac{1}{2}g_{ab}g^{cd}\nabla_c \bar{\Phi}\nabla_d{\Phi} + c.c. \Bigg)
\ee This definition can be motivated from stress-enegy conservation, namely
by the fact that
\begin{equation}
g^{ac}\nabla_cT_{ab} = \Box \Phi \nabla_b \bar{\Phi} + c.c. \ .
\end{equation}
Thus
\begin{equation}
\Box_g \Phi = 0 \implies g^{ac}\nabla_cT_{ab} = 0 \ .
\end{equation}
The expression \eqref{eq:EnergyMomemtumComplexScalar} corresponds to
the energy momentum associated to the action \begin{equation} S =
\int_{\mathcal{M}} g^{ab}\nabla_a\Phi\nabla_b\bar{\Phi} dV_g \ , \end{equation}
where $dV_g$ is the volume element in $(\mathcal{M},g_{ab},\nabla)$.  To
motivate the energy scalar product, to be discussed in this section for
the QNM problem, we use the notation of the vector field method to put
this discussion in a wider context. An introductory discussion of the
vector field method can be found in \cite{Are13}.  Let $X^a$ be any
vector field and let the associated deformation tensor
${}^{[X]}\pi_{ab}$, the vector and scalar currents, denoted as
${}^{[X]}J_a$ and $K^{[X]}$ respectively, be defined by
\begin{equation}
    {}^{[X]}J_a = T_{ab}X^b, \qquad {}^{[X]}\pi_{ab} =
    \nabla_{(a}X_{b)}, \qquad K^{[X]} = T_{ab}\pi^{ab} \ .
\end{equation}
A straightforward calculation shows that
\begin{equation}\label{eq:key_energy_identity}
\nabla^a {}^{[X]}J_a = {}^{[X]}K \ .
\end{equation}
Integration over a spacetime region $\mathcal{R}$ with boundary
$\partial \mathcal{R}$ gives the identity
\begin{equation}\label{eq:key_energy_identity_integrated}
\int_{\partial \mathcal{R}} {}^{[X]}J_a n^a dS = \int_{\mathcal{R}}
    {}^{[X]}K dV \ ,
\end{equation}
where $n^a$ is the normal to $\partial \mathcal{R}$ and $d\Sigma$ and $dV$
are the associated volume elements in $\partial \mathcal{R}$ and
$\mathcal{R}$ respectively.  Thus, if $(\mathcal{M}, g_{ab}, \nabla)$
admits a timelike Killing vector $t^a$ then, its associated
deformation tensor vanishes and hence, motivates the definition
\begin{equation}
{}^{[t]}E = \int_{\Sigma} {}^{[t]}J_a n^a d\Sigma \ ,
\end{equation}
where $\Sigma$ is a spacelike hypersurface with normal $n^a$, recovering
the expression in \cite{Wald84} for stationary spacetimes.  Observe
that, in general ${}^{[t]}E$ is not conserved quantity ---see
\cite{DafRod08} for a comprehensive discussion on the vector field
method and the construction of energy estimates.

\medskip

Aiming at revisiting the discussion of the scalar product in
\cite{Jaramillo:2020tuu}, and focusing on the scalar case, we consider
the wave equation \eqref{eq:Wave_eq_Phi} on a spherically symmetric
spacetime $(\mathcal{M}, g_{ab}, \nabla)$.  In particular, we adopt
local coordinates $x^\mu =\{t,r,\theta, \phi\}$ in which the line
element $ds^2 = g_{\mu\nu}dx^\mu dx^\nu$ reads
\be \label{eq:line_element_basic} ds^2 = -f(r)dt^2 + f(r)^{-1}dr^2 +
r^2 \Omega_{AB}dx^Adx^B, \ee where $\Omega_{ab}$ is the standard
metric on $\mathbb{S}^2$ with coordinates $x^A =\{\theta, \phi\}$.  By
considering the following Ansatz for the field
\begin{equation}\label{eq:Ansatz_Phi}
  \Phi = \frac{1}{r} \sum_{\ell m} \phi_{\ell m}(t,r) Y_{\ell m}(x^A)
  \ ,
\end{equation}
where $ \sum_{\ell m}$ is a shorthand for $\sum_{\ell=0}^{\infty}
\sum_{m=-\ell}^{l}$, the wave equation \eqref{eq:Wave_eq_Phi} reduces
to the mode analysis of
\begin{equation}\label{eq_static_spherically_symetric}
\Big(\frac{\partial^2}{\partial t^2} - \frac{\partial^2}{\partial
  r_\star^2} + V_\ell\Big)\phi_{\ell m} = 0 \ ,
\end{equation}
where $r_\star$ is a tortoise coordinate defined by
$\frac{dr}{dr_\star} = f(r)$ and $V_{\ell}=V_{\ell}(r)$.  In other
words, the spherically symmetric problem is reduced 
(for each $\ell m$) to the analysis of a (1+1)-dimensional wave
equation with an effective potential $V$ in Minkowski spacetime
$(\mathring{\mathcal{M}}, \mathring{\eta}_{ij}, \mathring{\nabla})$,
\be\label{eq:reducedeq}
\Box_{\mathring{\eta}}\phi - V\phi =0 \ .  \ee This equation has an
associated effective energy momentum tensor given by
%%%%%%%%%%%%
\begin{eqnarray}
\label{eq:effective_EM_tensor}
\hspace{-1.5mm}\mathring{T}_{ab} \hspace{-0.5mm}=\hspace{-0.5mm}
\frac{1}{2}\bigg(\hspace{-1mm}\mathring{\nabla}_a \bar{\phi}
\mathring{\nabla}_b \phi \hspace{-0.5mm}-\hspace{-0.5mm}
\frac{1}{2}\mathring{\eta}_{ab}\hspace{-0.5mm}\left(\hspace{-0.5mm}
\mathring{\eta}^{cd}\mathring{\nabla}_c
\bar{\phi}\mathring{\nabla}_d{\phi}\hspace{-0.5mm} + \hspace{-0.5mm} V
\phi \bar{\phi}\right)\hspace{-0.5mm} +\hspace{-0.5mm}
c.c.\hspace{-1mm} \bigg) \ ,
\end{eqnarray}
%%%%%%%%%%
(when discussing the effective approach the mode indices
$\ell m$ are omitted).  From this equation, and given a spacelike hypersurface
$\Sigma$ and a static Killing vector, one can derive an energy
associated to the field $\phi$. This was the strategy followed in \cite{Jaramillo:2020tuu}.
The main purpose of this section is to clarify the relation between
the energy associated to the field $\Phi$ and that of the field modes
$\phi$ in a  hyperboloidal foliation of the spacetime.

Following \cite{Jaramillo:2020tuu}, let us introduce coordinates $(\tau, x)$ adapted to
the compactified hyperboloidal foliation via \begin{equation}\label{eq:HypCoords}  t  =
\tau -h(x), \qquad  r_\star = g(x) \ .  \end{equation} Here, in order to simplify the
notation, we have absorbed the length scalar $\lambda$ discussed in
\cite{Jaramillo:2020tuu} into the coordinates $(t, r_{\star})$.  If one
needs to bring the length scalar back, one simply has to identify the
coordinates $(t,r_{\star})$ with the dimensionless coordinates
$(\bar{t},\bar{x})$ in the discussion of \cite{Jaramillo:2020tuu}.  The
function $h(x)$ is known as the height function while the function
$g(x)$ is usually chosen to order to compactify the radial coordinate,
hence bringing $\scri^+$ to a finite coordinate distance.  The line
element \eqref{eq:line_element_basic} in these coordinate reads
\be
\label{eq:line_element} ds^2 = f(r)\mathring{\eta}_{ij}dx^i dx^j +
r^2 \Omega_{AB}dx^Adx^B, \ee
with $x^i = \{\tau, x\}$ and the
2-dimensional line element, \be \label{eq:Mink_Hyp}
\mathring{\eta}_{ij}dx^i dx^j = -d\tau^2 + ((g')^2-(h')^2)dx^2
+2h'd\tau dx \ .  \ee where we have used $'$ to denote a derivative
respect to $x$.  Let $\Sigma_\tau$ be given by,
%%%%%%%%%%%%%%%
%\begin{equation}
%\red{\label{eq:HyperboloidalFoliation_extension} \Sigma_\tau = \{x^\mu
%\in \tilde{\mathcal{M}} \;|\; \tau =0, \; x \in [a,b]\},}
%\end{equation}
%%%%%%%%%%%%%%
\begin{equation}
\label{eq:HyperboloidalFoliation_extension} \Sigma_\tau = \{x^\mu
\in \tilde{\mathcal{M}} \;|\; \tau =0, \; x \in [a,b], \; x^A \in \mathbb{S}^2\} \ ,
\end{equation}
where
$(\tilde{\mathcal{M}},\tilde{g}_{ab})$ is the conformal extension of
the physical spacetime $(\mathcal{M},g_{ab})$.  In contrast with other
approaches to include null infinity ---for instance that of the
conformal Einstein field equations introduced by H. Friedrich; see
\cite{Fri81a, CFEBook}--- in the present hyperboloidal approach, the
conformal factor $\Xi$, relating the metrics $\tilde{g}_{ab}=\Xi^2
g_{ab}$ is not a dynamical quantity but it is fixed a priori and its
specific form depends on the choice of hyperboloidal coordinates
---see \cite{VanHusHil15,PanossoMacedo:2020biw, PanossoMacedo:2018hab}.  The function $g(x)$
implements the mapping from $]-\infty, \infty[$ to $]a,b[$, that is
    then compactified to $[a,b]$. Then, $a$ and $b$ correspond to the
    location of the BH horizon and future null infinity (notice that,
    depending on the actual implementation, a reflection can happen,
    so $a$ is actually null infinity, whereas $b$ corresponds to the
    BH horizon; this is indeed the case for the Schwarzschild
    implementation in
    \cite{Ansorg:2016ztf,PanossoMacedo:2018hab,Jaramillo:2020tuu}).

\subsubsection{Total energy versus effective mode energy.}
\label{e:total_energy}
We can now relate the proper energy of the full spacetime field $\Phi$ with that of the modes
$\phi$, namely the one employed in \cite{Jaramillo:2020tuu} to define
the energy norm. A direct calculation shows that the timelike
unit normal $n^a$ to $\Sigma_\tau$ is given by
\be n^a =
\frac{1}{\sqrt{f}\sqrt{{g'}^2-{h'}^2}}\left( w (\partial_\tau)^a
-\gamma (\partial_x)^a\right) \ , \ee and the volume element is given by
\be\label{eq:vol_sigma} d \Sigma_\tau =
r^2\sqrt{f}\sqrt{{g'}^2-{h'}^2}dx d\Omega \ ,
\ee where
$d\Omega=\sqrt{|\Omega|}d{\theta} d{\phi}$ with $|\Omega|=
\det[\Omega_{AB}]$, is the standard area element on $\mathbb{S}^2$,
and we have introduced the following notation \cite{Jaramillo:2020tuu}
\be
\label{wpdefinitions}
w = \frac{{g'}^2-{h'}^2}{g'}, \qquad \gamma =\frac{h'}{g'}, \qquad p =
\frac{1}{g'} \ .
\ee
Since we are working on the 4-dimensional spacetime
$(\mathcal{M},g_{ab})$ instead of the effective 2-dimensional
Minkowski spacetime $(\mathring{\mathcal{M}}, \mathring{\eta}_{ab})$,
the expressions for normal vector $n^a$ and volume element
$d\Sigma_\tau$ reported here and those given in \cite{Jaramillo:2020tuu}
differ by a factor of $\sqrt{f}$.  Observe that
$\partial_t=\partial_\tau$ and hence $\partial_\tau$ is a Killing
vector. Thus, ${}^{[\tau]}J_a$ is divergence free,
$\nabla^a{}^{[\tau]}J_a=0$, and we have the associated energy
associated with (the full) $\Phi$ 
\begin{equation}\label{eq:Physical_Energy}
{}^{[\tau]}E = \int_{\Sigma_{\tau}} {}^{[\tau]}J_a[\Phi] n^a
d\Sigma_\tau \ .
\end{equation}
A direct calculation using the energy momentum tensor $T_{ab}$ as
given in equation \eqref{eq:EnergyMomemtumComplexScalar} renders
\begin{flalign}\label{eq:Jn_current}
  {}^{[\tau]}J_a[\Phi] n^a &= \frac{1}{\sqrt{f}\sqrt{{g'}^2-{h'}^2}}(w
  T_{\tau \tau}-\gamma T_{\tau x}) \nn \\ &=
  \frac{1}{2\sqrt{f}\sqrt{{g'}^2-{h'}^2}} \bigg(
  \frac{{g'}^2-{h'}^2}{g'}\partial_\tau\Phi \partial_\tau \bar{\Phi}
   + \frac{1}{g'}\partial_x\Phi\partial_x\bar{\Phi} +
  \frac{g'f}{r^2}\Omega_{AB}\partial_A\Phi\partial_B\bar{\Phi}\bigg) \ .
\end{flalign}
Notice in particular that the factor $\sqrt{f}\sqrt{{g'}^2-{h'}^2}$
coming from the normal $n^a$ will combine well with that of
the volume element $d\Sigma_\tau$ when computing the energy
\eqref{eq:Physical_Energy}. 
Substituting equations \eqref{eq:Jn_current}
and \eqref{eq:vol_sigma} into \eqref{eq:Physical_Energy}, gives
\begin{multline}\label{eq:Physical_energy_expanded}
  {}^{[\tau]}E = \frac{1}{2}\int_{a}^{b} \int_{\mathbb{S}^2} r^2
  \bigg( w\partial_\tau\Phi \partial_\tau \bar{\Phi}  + p
  \partial_x\Phi\partial_x\bar{\Phi} +
  \frac{g'f}{r^2}\Omega^{AB}\partial_A\Phi\partial_B\bar{\Phi}\bigg)
  dxd\Omega \ .
\end{multline}
Before substituting the Ansatz \eqref{eq:Ansatz_Phi}
into equation \eqref{eq:Physical_energy_expanded},
observe that, using the chain rule,
the coordinate transformation \eqref{eq:HypCoords} and
$\partial_{r_\star}=f\partial_r$, one has
\begin{equation}\label{eq:pxToPtandPr}
  \partial_x = -h'\partial_t + fg'\partial_r \ .
 \end{equation}
Then, using the last expression and the Ansatz \eqref{eq:Ansatz_Phi}
gives
\begin{equation}\label{eq:PartialxPhi}
\partial_x\Phi = \sum_{\ell m} \frac{Y_{\ell m}}{r}\big(\partial_x\phi_{\ell m}-\frac{fg'}{r}\phi_{\ell m}\big) \ .
\end{equation}
Substituting the Ansatz \eqref{eq:Ansatz_Phi} aided by
equation \eqref{eq:PartialxPhi}
into equation \eqref{eq:Physical_energy_expanded},
renders
\begin{multline*}
{}^{[\tau]}E = \sum_{\ell m,\ell',m'} \frac{1}{2}\int_{a}^{b}
\int_{\mathbb{S}^2} \bigg\{ Y_{\ell m}\bar{Y}_{\ell',m'} \bigg(
w\partial_\tau\phi_{\ell m} \partial_\tau \bar{\phi}_{\ell'm'}
 + p \big(\partial_x\phi_{\ell
  m}-\frac{fg'}{r}\phi_{\ell
  m}\big)\\ \times \big(\partial_x\bar{\phi}_{\ell'm'}-\frac{fg'}{r}\bar{\phi}_{\ell'm'}
\big) \bigg)  + \frac{g'f}{r^2} \phi_{\ell
  m}\bar{\phi}_{\ell'm'}\Omega^{AB}\partial_AY_{\ell
  m}\partial_B\bar{Y}_{\ell'm'} \bigg\} dxd\Omega \ ,
\end{multline*}
where $\sum_{\ell m,\ell'm'}$ is a shorthand for $\sum_{\ell m}\sum_{\ell'm'}$.
Expanding and rearranging gives
\begin{multline}\label{eq:Physical_energy_harms}
{}^{[\tau]}E  = \sum_{\ell m,\ell'm'} \frac{1}{2}\int_{a}^{b} \int_{\mathbb{S}^2} 
\bigg\{ Y_{\ell m}\bar{Y}_{\ell'm'}
\bigg( A_{\ell m,\ell'm'}  + B_{\ell m,\ell'm'}\bigg)  +
D_{\ell m,\ell'm'}  \bigg\}
dxd\Omega \ ,
\end{multline}
with
\bea
%\begin{eqnarray*}
&A_{\ell m,\ell'm'} &= w\partial_\tau\phi_{\ell m} \partial_\tau
  \bar{\phi}_{\ell'm'} + p \partial_x\phi_{\ell m} \partial_x
  \bar{\phi}_{\ell'm'} + \frac{fg'}{r^2}\phi_{\ell
    m}\bar{\phi}_{\ell'm'}\nn \\& B_{\ell m,\ell'm'} &=
  -\frac{f}{r}(\bar{\phi}_{\ell'm'}\partial_x\phi_{\ell m} +\phi_{\ell
    m}\partial_x\bar{\phi}_{\ell'm'}) \nn \\& D_{\ell m,\ell'm'} &=
  \frac{fg'}{r^2}\phi_{\ell
    m}\bar{\phi}_{\ell'm'}\Omega^{AB}\partial_AY_{\ell m}\partial_B
  \bar{Y}_{\ell'm'} \ ,
%\end{eqnarray*}
\eea
where we have used that $p=1/g'$ to simplify the expressions.  The
contribution from the $A_{\ell m,\ell'm'}$ part can be
straightforwardly computed exploiting the orthogonality relation
\begin{equation} \label{eq:orthogonality_relation} \int_{\mathbb{S}^2}Y_{\ell
  m}\bar{Y}_{\ell'm'}d\Omega = \delta_{\ell\ell'}\delta_{mm'} \ ,  \end{equation}
as follows
\begin{multline}\label{eq:A_term}
  \sum_{\ell m,\ell'm'} \int_{a}^{b} \int_{\mathbb{S}^2} Y_{\ell
    m}\bar{Y}_{\ell',m'} A_{\ell m,\ell'm'} dxd\Omega = \\ \sum_{\ell
    m} \int_{a}^{b} w\partial_\tau\phi_{\ell m} \partial_\tau
  \bar{\phi}_{\ell m} + p \partial_x\phi_{\ell m} \partial_x
  \bar{\phi}_{\ell m} + \frac{fg'}{r^2}\phi_{\ell m}\bar{\phi}_{\ell
    m} dx \ .
\end{multline}
For determining the contribution of the $D_{\ell m\ell'm'}$ term,
albeit, very cumbersome, one could opt to compute $\partial_A Y_{\ell
  m}$ in a particular coordinate system $x^A$ for
$\mathbb{S}^2$. Instead, it is more convenient to use that
\begin{multline}
\frac{1}{\sqrt{|\Omega|}}\partial_A(\sqrt{|\Omega|}\Omega^{AB}
\bar{Y}_{\ell'm'}\partial_BY_{\ell m}) =
-\ell(\ell+1)\bar{Y}_{\ell'm'}  +
\Omega^{AB}\partial_A\bar{Y}_{\ell'm'}\partial_{B}Y_{\ell m} \ ,
\end{multline}
which follows by using the Leibniz rule and that
$\Delta_{\mathbb{S}^2}Y_{\ell m} = -\ell(\ell+1)Y_{\ell m}$ where
$\Delta_{\mathbb{S}^2}$ is the standard (``round'') Laplacian on
$\mathbb{S}^2$. Then integration over $\mathbb{S}^2$ and an
application of Stokes theorem renders the following expression
\be\label{eq:further_identity_spherical_harmonics}
\int_{\mathbb{S}^2}\Omega^{AB}\partial_AY_{\ell
  m}\partial_{B}\bar{Y}_{\ell'm'} d\Omega =
\ell(\ell+1)\delta_{\ell\ell'}\delta_{mm'} \ . \ee Using the identities
\eqref{eq:orthogonality_relation} and
\eqref{eq:further_identity_spherical_harmonics} one can easily read
the contributions $D_{\ell m\ell'm'}$ as follows
\begin{equation}\label{eq:D_term}
  \sum_{\ell m,\ell' m'} \int_{a}^{b} \int_{\mathbb{S}^2} Y_{\ell
    m}\bar{Y}_{\ell'm'} D_{\ell m,\ell'm'} dxd\Omega =  \sum_{\ell
    m} \int_{a}^{b} \frac{\ell(\ell+1)}{r^2}fg'\phi_{\ell m}
  \bar{\phi}_{\ell m} dx \ .
\end{equation}
To compute the contribution from the $B_{\ell m\ell'm'}$ term we
integrate by parts once to obtain
\begin{multline}\label{eq:B_term}
  \sum_{\ell m,\ell',m'}\int_{a}^{b}\int_{\mathbb{S}^2} Y_{\ell
    m}\bar{Y}_{\ell',m'} B_{\ell m,\ell'm'}dxd\Omega =  \sum_{\ell
    m} \int_{a}^{b}\phi_{\ell
    m}\bar{\phi}_{\ell m}\partial_x\bigg(\frac{f}{r}\bigg)dx
  -\Bigg(\frac{f}{r}\phi_{\ell m}\bar{\phi}_{\ell
    m}\Bigg)\Bigg|_{a}^{b} \ .
\end{multline}
Using equation \eqref{eq:pxToPtandPr} we can rewrite the second term
in \eqref{eq:B_term} in a more convenient form for the eventual
reconstruction of the effective potential inside the final energy
expression,
\begin{equation}
  \partial_x \bigg( \frac{f}{r}\bigg) = -\frac{f^2g'}{r^2}
  +\frac{fg'}{r}\partial_rf \ .
\end{equation}
Altogether, substituting equations \eqref{eq:pxToPtandPr},
\eqref{eq:B_term}, \eqref{eq:D_term} and \eqref{eq:A_term} into
equation \eqref{eq:Physical_energy_harms} gives
\begin{equation}\label{eq:TotalEnergy}
  {}^{[\tau]}E = \sum_{\ell m} {}^{[\tau]}\mathring{E}_{\ell m}
  -\Bigg(\frac{f}{2r}\phi_{\ell m}\bar{\phi}_{\ell
    m}\Bigg)\Bigg|_{a}^{b},
\end{equation}
with,
\begin{equation}\label{eq:mode-energy}
 {}^{[\tau]}\mathring{E}_{\ell m} = \frac{1}{2}\int_{a}^{b}
 w\partial_\tau \phi_{\ell m}\partial_\tau \bar{\phi}_{\ell m} + p
 \partial_x \phi_{\ell m}\partial_x \bar{\phi}_{\ell m}  +
 \tilde{V}\phi_{\ell m}\bar{\phi}_{\ell m} dx \ ,
\end{equation}
where the potential $\tilde{V}$ is given by
\be \label{VtildeDef} \tilde{V} = g'V, \qquad \text{with}\qquad V =
f\bigg(\frac{\ell(\ell+1)}{r^2} + \partial_rf\bigg) \ .  \ee
The boundary term in equation \eqref{eq:TotalEnergy} can be discarded
if one assumes that $f(a)=0$ and $\lim_{x \rightarrow  b}\frac{f|\phi|^2}{r}=0$
as it is in the Schwarzschild metric case
where $a$ and $b$ correspond to the location of the horizon and future null
infinity~\footnote{Actually, $a$ and $b$ are interchanged in \cite{Ansorg:2016ztf,PanossoMacedo:2018hab,Jaramillo:2020tuu}, with $a$ corresponding to null infinity and $b$ to th BH horizon.
  In more general spacetimes $f$ will vanish on horizons, as it is the
  case for asymptotically de Sitter spacetimes.}.
The mode-energy ${}^{[\tau]}E_{\ell m}$ coincides with that
used in \cite{Jaramillo:2020tuu} to compute the pseudospectrum, so the energy norm employed in
\cite{Jaramillo:2020tuu} is correct.

\setcounter{footnote}{0}

\subsubsection{Energy flux at spacetime boundaries.}
\label{s:energy_flux}
Now that the relation between the physical energy ${}^{[\tau]}E$ and
the effective energy ${}^{[\tau]}E_{\ell m}$ has been clarified, we
employ the effective energy momentum tensor of equation
\eqref{eq:effective_EM_tensor} to identify the energy flux at the
boundary.  We drop again momentarily the mode indices $\ell m$, while
discussing the effective approach.  First observe that the divergence
of the effective energy momentum tensor is given by
\begin{align}
  \mathring{\nabla}^a\mathring{T}_{ab} = -\frac{1}{4}\bar{\phi}\phi
  \mathring{\nabla}_bV + \frac{1}{2}(\square_{\mathring{\eta}}\phi -
  V\phi) + c.c. \ .
\end{align}
Notice that, within the effective approach, even if the reduced wave
equation \eqref{eq:reducedeq} is satisfied,
$\mathring{\nabla}^a\mathring{T}_{ab} = 0$ does not hold in general
for a non-constant potential. Contrast this situation with the
4-dimensional case where the wave equation \eqref{eq:Wave_eq_Phi}
implies that $\nabla^aT_{ab}=0$.  Nonetheless, one can still construct
a conserved current in the effective approach since
\begin{align}
  \mathring{\nabla}_a{}^{[t]}\mathring{J}^a =
  -\frac{1}{2}\phi\bar{\phi}\partial_t V \ .
\end{align}
Hence, using that $\partial_tV=0$ one has that
${}^{[t]}\mathring{J}^a$ is a conserved current. Notice that we have
added a ring over the current letter to stress the difference between
the 4-dimensional associated current
${}^{[t]}J^a=(\partial_t)^aT_{ab}$ and the effective 2-dimensional one
${}^{[t]}\mathring{J}^a=(\partial_t)^a\mathring{T}_{ab}$.  Moreover,
recalling that $\partial_t=\partial_\tau$ one has
\begin{equation}
  {}^{[t]}\mathring{J}^a={}^{[\tau]}\mathring{J}^a \ .
\end{equation}
Furthermore, using that $\partial_{\tau} V=0$ then
\begin{equation}\label{eq:effective_cons}
 \mathring{\nabla}_a {}^{[\tau]}\mathring{J}^a = 0 \ .
\end{equation}
From the conservation equation \eqref{eq:effective_cons}, using the
identity $\mathring{\nabla}_a v^a = \partial_i
(\sqrt{|\mathring{\eta}|}v^i)/\sqrt{|\mathring{\eta}|}$ where
$\mathring{\eta}$ is the determinant of $\mathring{\eta}_{ij}$ in
hyperboloidal coordinates $x^i = \{\tau, x\}$, a direct calculation
shows 
\begin{align}\label{eq:effective_energy_flux_relation}
\partial_\tau \mathring{\mathcal{E}} = \partial_x
\mathring{\mathcal{F}} \ ,
\end{align}
where
\begin{align}\label{eq:eff_en_density}
  \mathring{\mathcal{E}} = \frac{1}{2} \Big(w\partial_\tau \phi
  \partial_\tau \bar{\phi} + p \partial_x \phi \partial_x \bar{\phi} +
  \tilde{V}\phi \bar{\phi}\Big) \ ,
\end{align}
  and the effective flux $\mathring{\mathcal{F}}$ is given by
\begin{align}
  \label{eq:effective-flux}
  \mathring{\mathcal{F}} = \gamma \partial_\tau\phi \partial_\tau
  \bar{\phi} + p \;\text{Re} ( \partial_\tau \phi
  \partial_x\bar{\phi}) \ .
\end{align}
In fact, an alternative ---albeit longer--- way to derive equation
\eqref{eq:effective_energy_flux_relation} is to apply directly
$\partial_\tau$ to equation \eqref{eq:mode-energy} and then use the
effective wave equation \eqref{eq:reducedeq} written in hyperboloidal
coordinates to substitute the term $\partial_\tau^2\phi$ ---see
\cite{Jaramillo:2020tuu} for the explicit form of equation
\eqref{eq:reducedeq} in hyperboloidal coordinates. Then, integration
by parts leads to an integrated version of the expression for the
effective flux $\mathring{\mathcal{F}}$.

\medskip

Restoring the mode-indices $\ell m$ in equation
\eqref{eq:effective_energy_flux_relation}, \eqref{eq:eff_en_density}
and \eqref{eq:effective-flux} to make contact with the total physical
energy ${}^{[\tau]}E$ ---as defined in equation
\eqref{eq:Physical_Energy}--- one notices that in fact
$\mathring{\mathcal{E}}_{\ell m} = \frac{d}{dx}
\big({}^{[\tau]}\mathring{E}_{\ell m} \big)$, where
${}^{[\tau]}\mathring{E}_{\ell m}$ is the effective energy given in
equation \eqref{eq:mode-energy}.  To identify the total physical flux,
observe that equation \eqref{eq:TotalEnergy} can be re-expressed as
\begin{equation}\label{eq:TotalEnergyAlt}
  {}^{[\tau]}E = \sum_{\ell m} \int_{a}^{b}
  \bigg(\mathring{\mathcal{E}}_{\ell m} -\partial_x
  \Big(\frac{f}{2r}\phi_{\ell m}\bar{\phi}_{\ell m}\Big)\bigg) dx \ .
\end{equation}
Then applying $\partial_\tau$ to the above equation gives
\begin{equation}
  \partial_\tau \Big({}^{[\tau]}E \Big) = \sum_{\ell m} \int_{a}^{b}
  \bigg(\partial_\tau \mathring{\mathcal{E}}_{\ell m}
  -\partial_x\partial_\tau \Big(\frac{f}{2r}\phi_{\ell
    m}\bar{\phi}_{\ell m}\Big)\bigg) dx \ .
\end{equation}
 Using the effective conservation equation
 \eqref{eq:effective_energy_flux_relation} and formally integrating in
 $x$, we get
\begin{equation}\label{eq:TotalEnergyAlt2}
  \partial_\tau \Big({}^{[\tau]}E\Big) = \sum_{\ell m} \Bigg(
  \mathring{\mathcal{F}}_{\ell m} -
  \frac{f}{2r}\partial_\tau(\phi_{\ell m}\bar{\phi}_{\ell m})
  \Bigg)\Bigg|_{a}^{b}.
\end{equation}
Thus, the total flux $F$ satisfying
\begin{equation}
  \partial_\tau \Big({}^{[\tau]}E\Big) = F \Big|_{a}^{b},
\end{equation}
is given explicitly by
\begin{equation}\label{eq:totalFluxExpression}
  F = \sum_{\ell m} \gamma \partial_\tau\phi_{\ell m} \partial_\tau
  \bar{\phi}_{\ell m} + p \; \text{Re} ( \partial_\tau \phi_{\ell m}
  \partial_x\bar{\phi}_{\ell m}) -
  \frac{f}{2r}\partial_\tau(\phi_{\ell m}\bar{\phi}_{\ell m}).
\end{equation}
For asymptotically flat spacetimes (or for spacetimes regions bounded
by horizons, as de Sitter) the  $\frac{f}{2r}$ term vanishes at the boundaries.
On the other hand, the function $p$ in the hyperboloidal foliation in \cite{Jaramillo:2020tuu}
is devised also to vanish at the boundaries. This is in particular
the case of  Schwarzschild spacetime in \cite{Jaramillo:2020tuu} or Reissner-Nordstr\"om in
\cite{Destounis:2021lum}.
Hence, the flux expression reduces to
\begin{equation}\label{eq:totalFluxExpression_reduced}
  F = \sum_{\ell m} \gamma \partial_\tau\phi_{\ell m} \partial_\tau
  \bar{\phi}_{\ell m} \ .
\end{equation}
This corresponds to equation
\eqref{eq:totalFluxExpression_reduced:intro},
in section \ref{s:intro_QNM_instability} where
our main results have been summarized.

\section{Scalar product in the hyperboloidal approach to QNMs: some applications}
\label{s:applications}
Once the main point of this article is made, namely that the choice
  of scalar product is crucial for the physical assessment of QNM instability
  and that the energy scalar product is our proposal for such assessment,
  in this section we present several applications
  in which the notion of scalar product relevant for the hyperboloidal
  approach to the QNM problem.

\subsection{Energy scalar product and QNM weak formulation}
\label{sec:Weak_Formulation}
%%%%%%%%%%%%
%As briefly explained in the discussion around equation
%\eqref{e:intro_L_eigen},
%%%%%%%%%%%%%%5
QNMs in the present hyperboloidal setting arise by considering the wave
equation \eqref{eq_static_spherically_symetric} in a hyperboloidal
slicing and then solving the eigenvalue problem obtained by
Fourier transform respect in the time coordinate. 
Specifically, introducing a first-order reduction in time by
$\psi_{\ell  m}=\partial_t \phi_{\ell m}=\partial_\tau \phi_{\ell m}$, 
and taking the Fourier transform with respect to $\tau$ ---we simplify the
notation by denoting the Fourier transforms of $\phi_{\ell m}$ and
$\psi_{\ell m}$ simply by $\phi$ and $\psi$--- we are left with the
following eigenvalue problem
\be \label{eq:Eigenvalue_Problem} Lu =
\omega u \ , \ee where $u$ and $L$ are a vector valued function and a
differential operator given respectively by \begin{equation}
\label{eq:L}
u = \begin{pmatrix} \phi \\ \psi \end{pmatrix} , \qquad L
=\frac{1}{i}\!  \left(
  \begin{array}{c|c}
    0 & 1 \\ \hline L_1 & L_2
  \end{array}
  \right) \!  \ , \end{equation} with 
  \begin{equation}
  \label{e:L1L2def}
   L_1=w^{-1}(x)(\partial_x(p(x)\partial_x) - q(x)) \qquad, \qquad
  L_2=w^{-1}(x)(\gamma(x)\partial_x + \partial_x(\gamma(x)
  \cdot{})) \ , \end{equation} where $L_1$ is a Sturm-Liouville operator, $p,w$ are
  functions defined in equations \eqref{wpdefinitions} and
  $q=\tilde{V}$ with $\tilde{V}$ as defined in equation
  \eqref{VtildeDef}. These are essentially equations
   \eqref{e:intro_L_eigen}, \eqref{e:L_operator_intro} and \eqref{e:L_1-L_2_intro}
   in section \ref{s:scalar_product_basics}. The relabeling of the potential $\tilde{V}$
  with $q$ is merely aesthetic as it is intended
   to express the operator $L_1$ with
   standard Sturm-Liouville notation. 
  For further details on the derivation of this
  eigenvalue problem,  see \cite{Jaramillo:2020tuu}.  Let the scalar product
  inherited from the effective energy ${}^{[\tau]}E_{\ell m}$  be
  denoted by $\langle \cdot , \cdot \rangle_{_E}$. From equation
  \eqref{eq:mode-energy} one has that
  \begin{equation}\label{eq_eff_en_inner_product}
   \langle u_1, u_2\rangle_{_E}  =
  \frac{1}{2} \hspace{-1mm}\int_{a}^{b}\hspace{-1mm}
  (w(x)\bar{\psi}_1 \psi_2 +
  p(x)\partial_x\bar{\phi}_1\partial_x\phi_2 +
  \tilde{V}(x)\bar{\phi}_1\phi_2) dx \ .
  \end{equation}
\subsubsection{Weak formulation of the QNM problem.}
The weak problem is obtained by taking a vector valued test function
\be u_T = \begin{pmatrix} \phi_T \\ \psi_T \end{pmatrix} \ , \ee and
considering
\be \langle u_T, Lu \rangle_{_E} \; = \omega \langle u_T,u
\rangle_{_E} \ .\ee In a more explicit notation, the above equation reads
\be \Big\langle \begin{pmatrix} \phi_T \\ \psi_T \end{pmatrix}
, \begin{pmatrix} \psi \\ L_1\phi +L_2\psi \end{pmatrix}
\Big\rangle_{_E} \;=\; i \omega \Big\langle \begin{pmatrix} \phi_T
  \\ \psi_T \end{pmatrix} , \begin{pmatrix} \phi \\ \psi \end{pmatrix}
\Big\rangle_{_E} \ .\ee Using explicitly the form of $\langle \cdot ,
\cdot \rangle_{_E}$ one has
\begin{multline}
\label{eq:WeakFormPreIntegration}
 \int_{a}^{b} \bigg[ w\bar{\psi}_T (L_1\phi + L_2\psi) + p \partial_x
   \bar{\phi}_T \partial_x \psi + \tilde{V}\bar{\phi}_T \psi \bigg]
 \;dx \\ = i \omega \int_{a}^{b} \bigg[w \bar{\psi}_T\psi + p
   \partial_x \bar{\phi}_T \partial_x \phi + \tilde{V}\bar{\phi}_T
   \phi \bigg] \;dx \ .
\end{multline}
One of the advantages of the weak formulation is that one can reduce
the number of derivatives in the equation under the integral sign by
integrating by parts.  In our case, $L_1$ is a second order operator,
so it is convenient to do so. Performing the integrating by parts
of the ``$L_1$ part'' gives \begin{multline}
\label{eq:L1IntegrationByParts}
 \int_{a}^{b} w\bar{\psi_T} L_1\phi dx = \int_{a}^{b} \bar{\psi}_T
 \partial_x (p\partial_x \phi)-q\bar{\psi}_T\phi dx  \\= 
 (p\bar{\psi}_T\partial_x\phi)\big|_{a}^{b} - \int_{a}^{b}
 p\partial_x\phi \partial_x \bar{\psi}_T + q \bar{\psi}_T\phi dx \ .\end{multline}
Since $L_2$ is a first order operator one can opt to leave either it as it is, or 
rather integrate it by parts. Notice that this option is not available for
the third term $p \partial_x \bar{\phi}_T \partial_x \psi$ in equation
\eqref{eq:WeakFormPreIntegration} since integrating by parts would
increase the number of derivatives in the integrand. Integrating 
the ``$L_2$ part'' by parts renders
\begin{multline}
\label{eq:L2IntegrationByParts}
\int_{a}^{b} w\bar{\psi}_TL_2\psi dx = \int_{a}^{b}
\bar{\psi}_T\big(\gamma\partial_x\psi + \partial_x(\gamma \psi)\big)
dx  = 2(\gamma\bar{\psi}_T\psi)\big|_{a}^{b} - \int_{a}^{b} w \psi
L_2 \bar{\psi}_T dx \ .
\end{multline}
Thus, we have two options for the weak formulation of our QNM eigenvalue
problem:
\begin{itemize}
\item[i)] {\em Integrating by parts both $L_1$ and $L_2$}. This 
  gives the following weak problem
  \begin{eqnarray} \label{e:weak_problem_v0}\int_{a}^{b} W_{L}(u_T,u) \; dx \nonumber +
W_B(u_T,u)\bigg{|}_{a}^{b} = i \omega \int_{a}^{b} W_{R}(u_T,u) \;dx,
\nonumber \end{eqnarray} where \begin{eqnarray}\label{eq:Weak_Formulation} & W_L(u_T,u) & =
p (\partial_x \bar{\phi}_T \partial_x \psi -\partial_x\phi \partial_x
\bar{\psi}_T)  +
q(\bar{\phi}_T\psi- \bar{\psi}_T\phi) - w\psi L_2\bar{\psi}_T,
\nonumber \\ & W_R(u_T,u) & = p \partial_x \bar{\phi}_T\partial_x \phi
+ w \bar{\psi}_T\psi + q\bar{\phi}_T \phi \nonumber \\& W_{B}(u_T,u) &
= p\bar{\psi}_T\partial_x \phi + 2\gamma\bar{\psi}_T\psi \ .\end{eqnarray}

\item[ii)]  {\em Integrating by parts only of $L_1$}. 
If one opts not to integrate by parts the $L_2$ term one
is left with the following alternative weak problem
\begin{eqnarray} \int_{a}^{b} W^{*}_{L}(u_T,u) \; dx  +
W^{*}_B(u_T,u)\bigg{|}_{a}^{b} = i \omega \int_{a}^{b} W_{R}(u_T,u)
\;dx \ ,  \end{eqnarray} where \begin{eqnarray}
\label{eq:Weak_Formulation_alternative}
& W_L^{*}(u_T,u) & = p (\partial_x \bar{\phi}_T \partial_x \psi
-\partial_x\phi \partial_x \bar{\psi}_T)  + q(\bar{\phi}_T\psi- \bar{\psi}_T\phi) + w\bar{\psi}_T
L_2 \psi, \nonumber \\& W_{B}^{*}(u_T,u) & = p\bar{\psi}_T\partial_x
\phi .  \end{eqnarray}
\end{itemize}

\subsubsection{Formal adjoint $L^\dagger$ in the weak formulation.}
Equation \eqref{eq:Weak_Formulation} ---or
alternatively \eqref{eq:Weak_Formulation_alternative}--- constitutes
the weak formulation of the problem,  the main focus of this
section. For completeness, notice that this discussion is very close in spirit to
the determination of the formal adjoint. Indeed, if we integrate by parts the
term $ p\partial_x \bar{\phi}_T \partial_x \psi$ and use 
$q=\tilde{V}$ we obtain \begin{multline}  -\int_{a}^{b}\bigg[ w\psi (L_1
  \bar{\phi}_T + L_2 \bar{\psi}_T) + \partial_x\phi
  \partial_x\bar{\psi}_T + \tilde{V}\phi\bar{\psi}_T \bigg] dx +
\bigg(p(\bar{\psi}_T\partial_x\phi + \psi \partial_x \bar{\phi}_T) +
2\gamma \bar{\psi}_T\psi\bigg)\bigg{|}_{a}^{b} \\ =  i
\omega \int_{a}^{b} \bigg[w \bar{\psi}_T\psi + p \partial_x
  \bar{\phi}_T\partial_x \phi + \tilde{V}\bar{\phi}_T \phi \bigg]
\;dx \ ,  \end{multline} which in turn can be rewritten as
\be \Big\langle \begin{pmatrix} \psi_T \\ L_1\phi_T
  +L_2\psi_T \end{pmatrix},
\begin{pmatrix} \phi \\ \psi \end{pmatrix} \Big\rangle_{_E}  + B
\;=\; i \omega \Big\langle \begin{pmatrix} \phi_T
  \\ \psi_T \end{pmatrix} , \begin{pmatrix} \phi \\ \psi \end{pmatrix}
\Big\rangle_{_E},
\ee where
\be B =\bigg(p(\bar{\psi}_T\partial_x \phi + \psi \partial_x
\bar{\phi}_T) + 2\gamma \bar{\psi}_T\psi\bigg)\bigg{|}_{a}^{b} \ .
\ee If one assumes that $p(a)=p(b)=0$ ---as it is indeed the case in the operator $L$
constructed in the compactified hyperboloidal formulation--- then one can absorb the boundary
term and identify the formal adjoint of $L$ as \begin{equation} \label{e:formal_adjoint}L^\dagger
=\frac{1}{i}\!  \left(
  \begin{array}{c|c}
    0 & 1 \\ \hline L_1 & L_2+ L_2^\partial
  \end{array}
  \right) \!  , \end{equation} where \be\label{e:L2_adjoint_pert} L_2^\partial = 2\frac{\gamma}{w} \bigg(
\delta(x-a)-\delta(x-b)\bigg) \ , \ee
which recovers the expression found in \cite{Jaramillo:2020tuu}.

\subsubsection{QNM calculation in the weak formulation: finite elements.}
  \label{sec:FiniteElements}
Equation \eqref{eq:Weak_Formulation} ---or alternatively equation
\eqref{eq:Weak_Formulation_alternative}--- represents the weak
formulation of the eigenvalue problem \eqref{eq:Eigenvalue_Problem}.
From a numerical perspective this formulation of the problem naturally
opens the possibility to use finite elements methods.  This offers an
approach to validate the results obtained with Chebyshev spectral
methods in
\cite{Jaramillo:2020tuu,Jaramillo:2021tmt,Destounis:2021lum}.
To probe these ideas in a
controlled arena, we test the P\"oschl-Teller potential case, 
used as a benchmark in the study of QNMs since it can be solved
exactly, yielding to the following expression for the frequencies
\begin{equation}
\label{e:PT_QNM}
\omega^\pm_n =  \pm \frac{\sqrt{3}}{2} +i\left(n +\frac{1}{2}\right) \ .
\end{equation}

\begin{figure}[h!]
\centering \includegraphics[width=8.5cm]{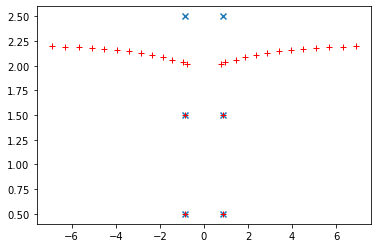}
\caption{P\"oschl-Teller QNM calculation as a variational problem
  solved in FEniCS with a mesh with N=500 points and using the
  Lagrange finite elements function space "CG'' (Continuous Garlekin).
  In blue the exact known eigenvalues of the operator $L$ are shown
  and in red the numerically computed}
\label{fig:Eigenvalues_PT_v1}
\end{figure}

\begin{figure}[h!]
\centering \includegraphics[width=8.5cm]{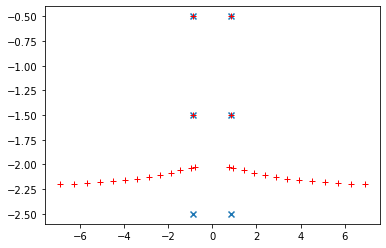}
\caption{P\"oschl-Teller QNM calculation as a variational problem
  solved in FEniCS with a mesh with N=500 points and using the
  Lagrange finite elements function space "CG'' (Continuous Garlekin).
  The eigenvalues of $L^\dagger$ are $\bar{\omega}$ where $\omega$
  are the eigenvalues of $L$.  In blue the exact eigenvalues of the
  $L^\dagger$ operator are shown and in red the numerically computed}
\label{fig:Eigenvalues_PT_v2}
\end{figure}

\noindent For the numerical implementation of finite elements we
use a simple automated implementation using the FEniCS software.
The P\"oschl-Teller potential reads
\begin{equation}
V(r_\star )= {V_o}\;{\mathrm{sech}^2(r_\star)}, \qquad
\ r_\star \in(-\infty,\infty) \ .
\end{equation}
where $V_o$ is a constant.
Using the same hyperboloidal foliation (Bizo\'n-Mach coordinates) as
that discussed for the P\"oschl-Teller potential in \cite{Jaramillo:2020tuu},
corresponds to the following prescription for the height and spatial
compactification functions: $h(x)=\frac{1}{2}\ln(1-x^2)$ and $g(x) =
\text{arctanh}(x)$, on the interval $[a,b]=[-1,1]$.
The latter, in turn, implies
\begin{equation}
\label{e:functions_L1_L2_PT}
 w(x)=1, \quad   p(x) = 1-x^2, \quad
q(x)=\tilde{V}= V_o, \quad \gamma(x)=-x \ .
\end{equation}
As anticipated, in particular it holds $p(a)=p(b)=0$ which 
simplifies the boundary term  $W_B(u,u_T)$ in equations
\eqref{eq:Weak_Formulation} and \eqref{eq:Weak_Formulation_alternative}.
If one assumes that the
test vector valued function is of ``compact support in $[a,b]$''
consistently with the theory of distributions ---see \cite{FriLan98}---
then the boundary term vanishes and we are left with the problem
\be \label{eq:Weak_Formulation_NoBoundary}
 \int_{a}^{b} W_{L}(u_T,u) \; dx  = i \omega \int_{a}^{b} W_{R}(u_T,u) 
 \;dx \ , 
 \ee
 or alternatively
\be \label{eq:Weak_Formulation_alternative_NoBoundary}
 \int_{a}^{b} W^{*}_{L}(u_T,u) \; dx  = i \omega \int_{a}^{b} W_{R}(u_T,u) 
 \;dx \ .
 \ee
 A straightforward implementation in FEniCS using the linear algebra
 PETSc solver renders the numerical results for the first few
 eigenvalues extracted from the weak problem
 \eqref{eq:Weak_Formulation_NoBoundary} and
 \eqref{eq:Weak_Formulation_alternative_NoBoundary} as shown in figure
 \ref{fig:Eigenvalues_PT_v1} and figure \ref{fig:Eigenvalues_PT_v2}
 respectively.

One can readily observe from Figures \ref{fig:Eigenvalues_PT_v1} and
\ref{fig:Eigenvalues_PT_v2} that the first two eigenvalues are
recovered and that from the third eigenvalue onward the analytical
result and numerical one disagree. Additionally, the weak problem
\eqref{eq:Weak_Formulation_NoBoundary} gives direct access to the
eigenvalues of $L$ while the weak problem
\eqref{eq:Weak_Formulation_alternative_NoBoundary} renders the
eigenvalues of $L^{\dagger}$.  As extensively discussed in
\cite{Jaramillo:2020tuu}, in order to increase the number of correct
eigenvalues recovered it is necessary to use enhanced machine
precision and the numerical error can be reinterpreted as a
perturbation of the operator $L$. The branches where the eigenvalues
migrate due to the numerical error are qualitatively similar to those
reported in \cite{Jaramillo:2020tuu} giving confidence that this observed
behavior, suitably interpreted as a random perturbation of the $L$
operator, is not exclusive to the spectral approach used in
\cite{Jaramillo:2020tuu}.  A full numerical analysis is not pursued here as
we consider this discussion as just a proof of concept where the
choice of inner product is brought to the forefront of the numerical
scheme by considering the weak formulation of the equations.

\subsection{Scalar product and QNM resonant expansions: towards BH spectroscopy}
\label{s:resonant_expansions}
As a second application of the use of scalar product in the QNM setting
we discuss resonant expansions of the propagating field in terms of QNMs.
This point has received much attention in the physics literature, going back (at least)
to Gamow's description of $\alpha$ decay \cite{Gamow28}. 
In the gravitational setting it is closely related to the discussion of completeness
of QNMs of BHs and compact objects (cf. e.g. \cite{ChiLeuSue95,Nollert:1998ys,Beyer:1998nu}
and references therein). In particular, it is very interesting the cross-fertilization in this
field between gravitational and optical studies \cite{LeuLiuYou94,ChiLeuMaa98,LalYanVyn17}.
In the present section we revisit this discussion with the emphasis on the normalizability
of QNMs, consequence of the use of a hyperboloidal foliation. This permits to
work on an actual Hilbert space where the scalar product plays a crucial role.

More specifically, our discussion of the expansion of a propagated
(scattered) field in QNMs connects more directly with the treatment of
this problem in Lax-Phillips scattering theory \cite{LaxPhi89} (see
also, for detailed reviews,
e.g. \cite{TanZwo00,zworski2017mathematical,dyatlov2019mathematical}).
The latter stands as an approach to resonances in terms of poles of
the meromorphic extension of the resolvent. As commented above, here
we revisit the problem in an approach that uses critically the Hilbert
space structure in the hyperbolic approach to resonances, in
particular the scalar product we are here discussing.  Specifically,
in the setting of the non-selfadjoint spectral problem defined in the
hyperbolic approach to QNMs, we make use of an asymptotic expansion by
Keldysh of the resolvent of a non-selfadjoint
operator~\cite{Keldy51,Keldy71,MenMol03,BeyLatRot12,Beyn12}.

This approach allows us to recover the results in
Lax-Phillips.  But, beyond that and in contrast with the standard (Cauchy) approach to
resonances, the normalizability of QNM eigenfunctions in the
hyperboloidal setting allows us to introduce expansion coefficients,
for which we provide explicit expressions that reduce, in the
selfadjoint case, to the standard normal modes expressions. This could provide
some insight into the quantification of the strength of QNM ringing discussed
and related ``excitation coefficients'' in \cite{Nollert:1998ys}.

\subsubsection{Resolvent of non-selfadjoint operators: Keldysh's expansion.} 
\label{s:Keldysh_expansion}

Let us consider a non-selfadjoint operator $L$ in a Hilbert (more
generally, Banach) space, and its adjoint $L^\dagger$ with respect to
the given scalar product $\langle \cdot, \cdot \rangle$. Then, as introduced in
equation \eqref{e:intro_LR_eigen}, right- $v_n$ and left- $w_n$ (proper) eigenvectors
are defined, respectively, as the eigenvectors of $L$ and $L^\dagger$
\begin{equation}
\label{e:left-right_vectors}
L v_n = \lambda_n v_n \ \ , \ \ L^\dagger w_n = \overline{\lambda}_n
w_n \ .
\end{equation} Notice that $v_n$ and $w_n$ are normalizable
vectors (in contrast to equation \eqref{e:intro_LR_eigen}, we have not yet normalised them).
Let us assume, for simplicity, that $L$ and $L^\dagger$ are
diagonalizable and that their eigenvalues are simple (for the general case,
see \cite{MenMol03,BeyLatRot12}).  In
this context, instead of a standard normalization in terms of the norm
of individual vectors, let us adopt the condition~\footnote{This
  condition can be generalized, on the one hand, to degenerate
  eigenvalues and operators with non-trivial Jordan blocks and, on
  the other hand, to spectral problems with a non-linear dependence on
  the spectral parameter, namely pencil operators (see
  \cite{Keldy51,Keldy71,MenMol03,BeyLatRot12,Beyn12}).}
\begin{equation} \langle
w_n , v_n\rangle = -1 \ .
\end{equation} We can then normalize one of them, but not both.
That is, in general $v_n$ and $w_n$ are not
normalized: $||v_n|| \neq 1 \neq ||w_n||$.  In this setting, we consider~\footnote{We
  use the notation $\lambda\text{-}\mathbb{C}$ to refer to the complex plane in
  the complex variable $\lambda$. This is to distinguish the different spectral parameters
appearing in the discussion.} a bounded domain
$\Omega\in\lambda\text{-}\mathbb{C}$. Under appropriate hypothesis (namely
the discreteness and isolation of $\lambda_n$ in the spectrum
$\sigma(L)$, guaranteed if $L$ is Fredholm), there is a finite number
$N$ of eigenvalues $\lambda_n\in \Omega$.

With these elements, we can introduce Keldysh's expansion
\cite{Keldy51,Keldy71,MenMol03,BeyLatRot12,Beyn12} of the resolvent of
$L$, namely $(L-\lambda)^{-1}$, where $\lambda\in \Omega\setminus\sigma(L)$.
Specifically, the Green function (the integral Kernel of the
resolvent) for $\lambda\in \Omega\setminus\sigma(L)$ can be written as
\begin{equation} G_\lambda(x',x) = \sum_{\lambda_j\in\Omega}
\frac{w^\dagger_j(x')v_j(x)}{\lambda - \lambda_j} + H(x,x';\lambda) \ ,
\end{equation} where $H(x,x';\lambda)$ is analytic in $\Omega$ (see full
technical details of this case in \cite{BeyLatRot12}).  Then, we can
formally write the action of the resolvent on a given source $S=S(x)$ (see below in equation
\eqref{e:source_S_u}) in terms of the scalar product, namely taking care of the
integration in the $x'$ variable \begin{equation}
\big((L- \lambda)^{-1} S\big)(\lambda,x) = \langle
G_\lambda(\cdot,x), S\rangle + H(\lambda)(S) =
\sum_{\lambda_j\in\Omega} \frac{v_j(x)}{\lambda - \lambda_j} \langle
w_j, S\rangle + \ H(\lambda)(S) \ ,  \end{equation} where $H(\lambda)$
is an operator, analytic in $\lambda\in \Omega$.  This notation is a
bit cumbersome and can be made more transparent in the terms of familiar (formal)
``bra's''s and ``ket's'', that take care of the adequate application
of the scalar product \begin{equation}
\label{e:Keldysh_expansion_normalized}
\!\!\!\!\!\!\!\!\!\!\!\!(L-\lambda)^{-1} = \sum_{\lambda_j\in\Omega}
\frac{|v_j\rangle\langle w_j|}{\lambda - \lambda_j} + H(\lambda) \ \ ,
\ \ \lambda\in\Omega\setminus\sigma(L) \ , \end{equation} where $\langle \;\cdot\;
|$ and $|\;\cdot \;\rangle$ must be understood with respect to the
given $\langle \;\cdot\;,\;\cdot \;\rangle$ in the Hilbert space.  Let
us denote the relation (\ref{e:Keldysh_expansion_normalized}) formally as \begin{equation}
\label{e:Keldysh_expansion_normalized_formal}
(L-\lambda)^{-1} \sim \sum_{\lambda_n\in\Omega}
\frac{|v_n\rangle\langle w_n|}{\lambda - \lambda_n} \ \ ,
\ \ \lambda\in\Omega\setminus\sigma(L) \ ,\end{equation} where the resolvent in
a (bounded) $\Omega$ region is written as a finite sum of poles plus an
(omitted) analytical function.  Note that the sum is a finite one and not a
series. This is a key point for the later interpretation of the
asymptotic nature of the resonant expansion, in contrast with
convergent series of selfadjoint operators.  That is, in spite of the
formal similitude to the expression of the resolvent for selfadjoint
operators in terms of eigenfunctions and eigenvalues,
the ``sum'' symbol has a completely different content (see
\cite{SheJar20} for details).
       
We can use condition $\langle w_n , v_n\rangle = -1$ to
recast equation (\ref{e:Keldysh_expansion_normalized_formal}) as \begin{equation}
\label{e:Keldysh_expansion_general}
(L-\lambda)^{-1} \sim \sum_{\lambda_n\in\Omega}
\frac{|v_n\rangle\langle w_n|}{\langle w_n , v_n\rangle}
\frac{-1}{\lambda-\lambda_n} = \sum_{\lambda_n\in\Omega}
\frac{|v_n\rangle\langle w_n|}{\langle w_n , v_n\rangle}
\frac{1}{\lambda_n - \lambda} \ .  \end{equation} Introducing now the condition
number $\kappa_n$ associated with the eigenvalue $\lambda_n$ \begin{equation}
\label{e:condition_number}
\kappa_n := \frac{||w_n||\;||v_n||}{\langle w_n , v_n\rangle} \ , \end{equation}
we can write \begin{equation}
\label{e:Keldysh_expansion_kappa}
(L-\lambda)^{-1} \sim \sum_{\lambda_n\in\Omega} \kappa_n
\frac{|v_n\rangle\langle w_n|}{||w_n||\;||v_n||}
\frac{1}{\lambda_n-\lambda} \ , \end{equation} and, in terms of the normalized
left- and right-eigenvectors \begin{equation} \hat{w}_n = \frac{w_n}{||w_n||} \ \ ,
  \ \ \hat{v}_n = \frac{v_n}{||v_n||} \ , \end{equation}
we can finally write (note that expression
  \eqref{e:condition_number_intro} for $\kappa_n$ is here recovered, when using in
  particular the  energy scalar product)
  \begin{equation}
\label{e:Keldysh_expansion_kappa_normalized}
(L-\lambda)^{-1} \sim \sum_{\lambda_n\in\Omega} \kappa_n
\frac{|\hat{v}_n\rangle\langle \hat{w}_n|}{\lambda_n-\lambda} \ \ ,
\ \ \lambda\in\Omega\setminus\sigma(L) \ .  \end{equation} Note that this
expression formally recovers the expression of the resolvent of a
selfadjoint (more generally, 'normal') operator, in which
$\hat{w}_n=\hat{v}_n$ and $\kappa_n=1$.  However, the statement in the
selfadjoint case is much stronger, since the sum there indicates a
series convergence. In our case, if $L$ is actually selfadjoint case,
expression (\ref{e:Keldysh_expansion_kappa_normalized}) can indeed be
extended to the whole $\lambda\text{-}\mathbb{C}\setminus \sigma(L)$ in a convergent
sense, but this requires other (Hilbert space) techniques.

\subsubsection{QNM resonant expansions in the hyperboloidal approach.}
\label{s:QNM_hyperboloidal_expansions}
Let us apply the previous discussion in the context of
the hyperboloidal initial data problem of a wave
equation for a field $\phi$ on a domain $\Sigma\times\mathbb{R}^+$, formulated in a
first-order in time setting and, for concreteness, employing the
energy scalar product $\langle\cdot,\cdot\rangle_{_E}$. We start, as in equation
\eqref{e:wave_eq_1storder_intro}, by reducing the wave equation to a
first-order (in time) formulation
\begin{equation}
\label{e:psi_u_def}
\psi=\partial_\tau \phi \ \ , \ \ u =
\begin{pmatrix}
  \phi \\ \psi
\end{pmatrix}  \ .
\end{equation}
Then, the wave equation can be written as
\begin{equation}
\label{e:wave_eq_1storder_u}
\displaystyle\left\{
\begin{array}{l}
 \partial_\tau u  = i L  u \ , \\
 u(\tau=0,x)=u_0(x) \ \ , \ \ ||u_0||_{_E}<0
 \end{array}
 \right. 
\end{equation}
where the operator $L$ and its adjoint $L^\dagger$ with respect to
the scalar product $\langle\cdot ,\cdot\rangle_{_E}$---we repeat here the
equations \eqref{e:L_operator_intro} and \eqref{eq:L} for $L$
and equation \eqref{e:formal_adjoint} for $L^\dagger$---
have the structure
\begin{equation}
\label{e:L_operator}
L =\frac{1}{i}\!
\left(
  \begin{array}{c|c}
    0 & 1 \\
    \hline 
   L_1 & L_2
  \end{array}
  \right) \qquad , \qquad L^\dagger =\frac{1}{i}\!
\left(
  \begin{array}{c|c}
    0 & 1 \\
    \hline 
   L_1 & L_2+L_2^{\partial}
  \end{array}
  \right) \ .
  \end{equation}
%The adjoint $L^\dagger$, with respect to the energy scalar product
%$\langle\;\cdot\;,\;\cdot\;\rangle_{_E}$, is written as
%% \begin{equation}
%% \label{e:L_adjoint}
%% L^\dagger =\frac{1}{i}\!
%% \left(
%%   \begin{array}{c|c}
%%     0 & 1 \\
%%     \hline 
%%    L_1 & L_2+L_2^{\partial}
%%   \end{array}
%%   \right) \ .
%%   \end{equation}
  Explicit expressions for $L_1$, $L_2$ and $L_2^{\partial}$
  in the 1+1-dimensional case are given in equations  \eqref{e:L1L2def} ---see also
  \eqref{e:L_1-L_2_intro}--- and
  \eqref{e:L2_adjoint_pert}.

  \medskip
  
  We address the resolution of the time evolution problem
  \eqref{e:wave_eq_1storder_u} in a spectral setup, adopting a
  Laplace transform approach.  Specifically, we consider for
  $\mathrm{Re}(s)>0$
\begin{equation}
 \label{e:Laplace_transform_1st_u}
       u(s;x) := \big({\cal L} u\big)(s;x) =\int_0^\infty e^{-s\tau}
       u(\tau,x)d\tau \ .
\end{equation}
Applying this transformation to equation  (\ref{e:wave_eq_1storder_u}),
we get \begin{equation} s\;u(s;x) - u(\tau=0,   x) = \frac{1}{i} L u(s; x) \ .
\end{equation} Dropping the explicit       $s$-dependence and using (\ref{e:wave_eq_1storder_u})
for the initial data, we can write
\begin{equation} \label{e:non-homogenous_s}
\big(L + is\big)u(s;x) = i S(x) \ ,
\end{equation} where the source $S$ is  defined in terms of the initial data
\begin{equation}  \label{e:source_S_u}
S(x) := u_0(x) \ .
\end{equation} This is a linear non-homogeneous equation.
For its resolution in a Green's function approach, we need an expression for the resolvent of $L$.
This is the point in which the above-discussed Keldysh's expansion
of the resolvent enters.  We need then first to consider the eigenfunctions
of the homogeneous part of the equation.  Writing, for notation purposes,
equation \eqref{e:non-homogenous_s} in terms of the Fourier parameter $\omega$
by using the relation $s=i\omega$, we can write
\begin{equation}  \label{e:non-homogeneous_L-S_Fourier}
(L - \omega) u(\omega; x) = i S(x) \ .
\end{equation} The spectral problem of $L$ associated with the homogeneous part is
given in \eqref{e:intro_LR_eigen}
\begin{equation} \label{e:eigenvalue_problem_L_Fourier_v1}
 L \hat{v}_n = \omega_n \hat{v}_n \ \ , \ \ L^\dagger
\hat{w}_n = \overline{\omega}_n \hat{w}_n \ ,
\end{equation}
namely the spectral problem \eqref{e:intro_LR_eigen} in the
$\omega$-spectral parameter and for normalized vectors $\hat{v}_n$ and
$\hat{w}_n$ with respect to the energy scalar product
$\langle\cdot,\cdot\rangle_{_E}$.  From this spectral problem, the
resolvent $(L-\omega)^{-1}$ is constructed in a bounded
$\Omega_\omega\in\omega\text{-}\mathbb{C}\setminus \sigma(L)$, so
$u(\omega;x)$ writes
\begin{equation}
  u(\omega;x) = i(L- \omega)^{-1} S  \ \ , \ \ \omega\in\Omega_\omega\setminus \sigma(L) \ ,
  \end{equation}
  and using equation (\ref{e:Keldysh_expansion_kappa_normalized})
  \begin{equation}
  u(\omega;x) = i(L- \omega)^{-1} S \sim i\sum_{\omega_j\in\Omega_\omega}
  \kappa_j\frac{\langle \hat{w}_j|S\rangle_{_{E}}}
  {\omega_j - \omega}|\hat{v}_j\rangle \ .
  \end{equation}  
  Writing it again in terms of the Laplace spectral parameter
  $s=i\omega$, with $s\in\Omega_s=i\Omega_\omega$ and reinserting the
  analytic part of the resolvent in
  (\ref{e:Keldysh_expansion_normalized}), we have, for
  $s\in\Omega_s\setminus\sigma(L)$ \begin{equation}
  \label{e:u(s,x)_bounded}
  u(s;x) =\sum_{s_j\in\Omega_s} \kappa_j\frac{\langle
    \hat{w}_j|S\rangle_{_{E}}}{s - s_j}|\hat{v}_j\rangle + iH(s)(S)
  \ .  \end{equation} In order to construct the time evolution solution $u(\tau,x)$ we 
  take the inverse Laplace transform %of  equation \eqref{e:u(s,x)_bounded},
 \begin{equation}  \label{e:Laplace_inverse}
 u(\tau,x) = \frac{1}{2\pi i }\int_{c-i\infty}^{c+i\infty} e^{st}u(s;x) ds \ ,
 \end{equation}
 with $c\in \mathbb{R}^+$. But expression (\ref{e:u(s,x)_bounded}) only provides
 $u(s;x)$ in a bounded $\Omega_s\setminus\sigma(L)$.
 In this setting, taking a finite $R\in \mathbb{R}$, we can write (assuming the limit exists)
 %%%%%%
 %\begin{widetext}
 %%%%%
      \bea
         \label{e:u_Keldysh_v1}
       u(\tau,x) &=& \lim_{R\to\infty}\frac{1}{2\pi i }\int_{c-iR}^{c+iR}  e^{is} u(s;x) ds  
       \\&=&
       \lim_{R\to\infty}\frac{1}{2\pi i }\int_{c-iR}^{c+iR} e^{is}
       \Big(\sum_{s_j\in\Omega_s} \kappa_j\frac{\langle \hat{w}_j|
         S\rangle_{_{E}}}{s - s_j}|\hat{v}_j\rangle + iH(s)(S)\Big) ds \ .\nn
       %      \end{multline}
       \eea
       Let us consider now a contour $C$ in $s\text{-}\mathbb{C}$ formed by the interval
       $[c-iR, c+iR]$  closed on the left half-plane by a circle (of radius $R$,
       centered at $c+i0$) and denote  by $\Omega_R$ the bounded domain in $s\text{-}\mathbb{C}$
       delimited by $C$.  In the context of the (Fredholm) operators we are considering,
       the number of $s_j\in\Omega_R$ is finite,
       so we can safely interchange the (finite) sum and the integral 
       \begin{equation}
       \label{e:u_Keldysh_v2}
       u(\tau,x) =  \lim_{R\to\infty} \Bigg(\sum_{\omega_j\in\Omega} \frac{1}{2\pi i }
       \oint_C e^{s\tau} \kappa_j\frac{\langle \hat{w}_j|S\rangle_{_{E}}}{s - s_j}|\hat{v}_j\rangle ds +
       \frac{1}{2\pi  }
       \oint_C e^{s\tau} H(s)(S) ds + 
      \hbox{(circle part)}\Bigg) \ .
       \end{equation}
%%%%%%
%\end{widetext}
%%%%%
The contour $C$ integral involving the analytic expression $e^{s\tau}
H(s)(S)$ vanishes, but nothing guarantees that its integral along the
semi-circle vanishes, this depending on the specific dependence of the
function $H(s)$, that is not fixed by Keldysh expansion. In
general, the last term ``$\hbox{(circle part)}$'' gives a term $C_R(\tau; S)$.  Using now Cauchy
theorem, we can write
 \begin{equation}
 \label{e:u_Keldysh_v3}
 u(\tau,x)  = 
 \lim_{R\to\infty}\Big(  \sum_{\omega_j\in\Omega} 
 e^{s_jt}\kappa_j \langle \hat{w}_j|S\rangle_{_{E}} \hat{v}_j + C_R(\tau; S)\Big) \ .
   \end{equation}
   In contrast with the selfadjoint case, nothing guarantees that this limits exists.
   First, for strongly non-selfadjoint (more generally non-normal) operators
   the condition number $\kappa_n$ can present a strong growth and, on the other hand, the terms
   $C_R(\tau; S)$ does not need to converge as $R\to\infty$, in particular not
   needing to vanish.
   In this context we cannot write $u(\tau,x)$ as an actual convergent series, but we
   can still write an asymptotic QNM resonant expansion
   \begin{equation}
   \label{e:u_Keldysh_v4}
   u(\tau,x) \sim \sum_n e^{s_{j}\tau}\kappa_j \langle \hat{w}_j|S\rangle_{_{E}} \hat{v}_j \ ,
   \end{equation}
   or, in the Fourier spectral-parameter
   \begin{equation}
   \label{e:u_Keldysh_v5}
   u(\tau,x) \sim \sum_n e^{i\omega_{j}\tau}\kappa_j \langle \hat{w}_j|S\rangle_{_{E}} \hat{v}_j \ ,
   \end{equation}
   this meaning that in a bounded domain $\Omega$ 
   the number of QNMs is finite and we can write
   \begin{equation}
   \label{e:u_Keldysh_v6}
   u(\tau,x) = \sum_{\omega_j\in\Omega} e^{i\omega_{j}\tau}\kappa_j \langle \hat{w}_j|S\rangle_{_{E}}
   \hat{v}_j + E_\Omega(\tau;S) \ ,
   \end{equation}
   where the structure of the term $e^{s\tau}H(s)(S)$ in the ``$\hbox{(circle part)}$''
   of (\ref{e:u_Keldysh_v2})
   permits to bound the ``error'' $E_\Omega(\tau;S)$ in the estimation of $u(\tau,x)$
   by the (finite) resonant
   expansion. Explicitly, defining $a= \max\{\mathrm{Im}(\omega), \omega\in \Omega\}$,
   there exists a constant $C_\Omega(a,L)$ depending on $a$ and the operator $L$ (but not on
   the initial data), such that
   \begin{eqnarray}
   \label{e:bound_E}
   ||E_\Omega(\tau;S)||_{_E} \leq C_\Omega(a,L) e^{-a\tau}||S||_{_E} \ .
   \end{eqnarray}
   On behalf of clarity, denoting by $N$ the (finite) number of QNMs in $\Omega$ and numbering
   $\omega_n$ from $1$ to $N$, we can rewrite equation \eqref{e:u_Keldysh_v6} as
   \begin{eqnarray}
   \label{e:u_Keldysh_v7}
   u(\tau,x) &=& \sum_{j=1}^N e^{i\omega_{j}\tau}\kappa_j \langle \hat{w}_j|S\rangle_{_{E}}
   \hat{v}_j + E_N(\tau;S) \nn \\
   \hbox{with } &&||E_N(\tau;S)||_{_E} \leq C_N(a,L) e^{-a\tau}||S||_{_E} \ .
   \end{eqnarray}
   This is essentially the content of Lax-Phillips resonant
   expansion \cite{LaxPhi89,TanZwo00,zworski2017mathematical,dyatlov2019mathematical},
   here expressed in terms of normalizable QNM
   functions $\hat{v}_j$  and providing an explicit prescription for the evaluation
   of the associated coefficient. In sum, we can write 
   \bea
   \label{e:Keldysh_QNM_expansion_u}
   u(\tau,x) \sim \sum_j e^{i\omega_{j}\tau} a_j \hat{v}_j(x) \ \ \ , \ \ \ \hbox{with} \
   a_j = \kappa_j \langle \hat{w}_j|u_0\rangle_{_{E}} \ ,
   \eea
   where we have rewritten $S$ in terms of initial data $u_0$.
   This provides the QNM resonant expansion of the propagating field in terms
   of its initial data, in particular entailing a prescription for the
   quantification of the QNM ringing in the signal \cite{Nollert:1998ys}.

 \subsubsection{Keldysh QNM expansion for the scattered field.}
 \label{s:Keldysh_second_order}
 In order to facilitate the comparison of expression (\ref{e:Keldysh_QNM_expansion_u})
 with the natural quantities in QNM resonant expansions  in (the second-order formulation of)
 Lax-Phillips theory, let us rewrite the main steps above explicitly
 in terms of $\phi$ and $\psi$ fields in equation
 (\ref{e:psi_u_def}). First, the wave equation (\ref{e:wave_eq_1storder_u}) writes
\begin{eqnarray}
\label{e:wave_eq_1storder_two_components}
\displaystyle\left\{
\begin{array}{l}
  \partial_\tau \begin{pmatrix} \phi \\  \psi\end{pmatrix} = iL
\begin{pmatrix} \phi \\  \psi\end{pmatrix} \ , \\
 \begin{pmatrix} \phi \\  \psi\end{pmatrix}(t=0,x)=\begin{pmatrix}
  \varphi_0(x) \\
  \varphi_1(x)
\end{pmatrix}  \ \ , \ \ \varphi_0\in H^1_V,  \varphi_1\in L^2
 \end{array}
 \right. 
\end{eqnarray}
where the operator $L$ and its adjoint have the structure
in equations \eqref{e:L_operator}.
 %% Explicit expressions for $L_1$, $L_2$ and
%%   $L_2^{\partial}$ and the energy scalar product $\langle\cdot,
%%   \cdot\rangle_{_E}$ in in the $1+1$-dimensional case are given in
%%   equations \eqref{e:L1L2def}.
Regarding the scalar product, expression \eqref{eq_eff_en_inner_product} is generalised to
(odd) dimension $n$ in the form
  \begin{equation}
\label{e:scalar_product_1st}
\!\!\!\!\Big\langle\begin{pmatrix}
  \phi_1 \\
  \psi_1
\end{pmatrix}, \begin{pmatrix}
  \phi_2 \\
  \psi_2
\end{pmatrix}\Big\rangle_{_{E}}  \\
= \frac{1}{2} \int_\Sigma \!\!\!\left(\! w(x)\overline{\psi_1}
\psi_2 +p(x)\overline{\nabla{\phi}_1}\cdot \nabla\phi_2 + \tilde{V}
\overline{\phi_1} \phi_2 \!\right) d\Sigma \ ,  \end{equation}
Under Laplace
transform, the source of equation (\ref{e:non-homogenous_s}) is
explicitly written in terms of the initial data as  
\begin{equation}
\label{e:source_S}
S = \begin{pmatrix} \varphi_0\\ \varphi_1\end{pmatrix} \ .
\end{equation}
Then, the right and left spectral problem \ref{e:eigenvalue_problem_L_Fourier_v1}
writes, by using
\begin{equation}
  \label{e:w_v_normalized}
  \hat{v}_n=\begin{pmatrix} \hat{\phi}^{_{\mathrm{R}}}_n \\  \hat{\psi}^{_{\mathrm{R}}}_n\end{pmatrix} \ \ , \ \
  \hat{w}_n=\begin{pmatrix} \hat{\phi}^{_{\mathrm{L}}}_n \\  \hat{\psi}^{_{\mathrm{L}}}_n\end{pmatrix} 
  \end{equation}
  as
  \begin{equation}
  \label{e:left-right_vectors_phi-psi}
\left(
  \begin{array}{c|c}
    0 & 1 \\
    \hline 
   L_1 & L_2
  \end{array}
  \right)
  \begin{pmatrix} \hat{\phi}^{_{\mathrm{R}}}_n \\  \hat{\psi}^{_{\mathrm{R}}}_n\end{pmatrix}
    =
    i\omega_n
    \begin{pmatrix} \hat{\phi}^{_{\mathrm{R}}}_n \\  \hat{\psi}^{_{\mathrm{R}}}_n\end{pmatrix}, \qquad 
 \left(
  \begin{array}{c|c}
    0 & 1 \\
    \hline 
   L_1 & L_2+L_2^{\partial}
  \end{array}
  \right)
  \begin{pmatrix} \hat{\phi}^{_{\mathrm{L}}}_n \\  \hat{\psi}^{_{\mathrm{L}}}_n\end{pmatrix}
   =
    i\overline{\omega}_n
    \begin{pmatrix} \hat{\phi}^{_{\mathrm{L}}}_n \\  \hat{\psi}^{_{\mathrm{L}}}_n\end{pmatrix} \ .
      \end{equation}
      \smallskip
      Notice that here the functions ${\phi}^{_{\mathrm{L,R}}}_n$ and
      $\hat{\psi}^{_{\mathrm{L,R}}}_n$ are, respectively, the first
      and second components of the left and right $\hat{w}_n$ and
      $\hat{v}_n$ vectors, and the ``hat'' indicate (``simultaneous'')
      normalization of the 2-component vector with the energy norm
      $||\cdot||_{_E}$
\begin{equation}
       \label{e:normalization_E}
\!\!\!\!\!\Big\langle\begin{pmatrix}
  \hat{\phi}^{_{\mathrm{L,R}}}_n \\
  \hat{\psi}^{_{\mathrm{L,R}}}_n
\end{pmatrix}, \begin{pmatrix}
  \hat{\phi}^{_{\mathrm{L,R}}}_m \\
  \hat{\phi}^{_{\mathrm{L,R}}}_m
       \end{pmatrix}\Big\rangle_{_{E}}
       =\delta_{nm} \ .
       \end{equation}
       Now, we can express the QNM resonant expansion (\ref{e:Keldysh_QNM_expansion_u})
       in terms of $\phi$ and $\psi$ as
   \begin{equation}
    \label{e:Kel_phi_psi}
   \begin{pmatrix}
  \phi \\
  \psi
\end{pmatrix} \sim \sum_n e^{i\omega_{j}\tau} a_n \begin{pmatrix}
  \hat{\phi}^{_R}_j \\
  \hat{\psi}^{_R}_j
\end{pmatrix} \ ,
   \end{equation}
   with the coefficient $a_n$ given by 
   \begin{equation}
       \label{e:a_j_v1_Kel}
       a_j = \kappa_j \langle \hat{w}_j|S\rangle_{_{E}} = \kappa_j\Big\langle\begin{pmatrix}
  \hat{\phi}^{_L}_j \\
  \hat{\psi}^{_L}_j
\end{pmatrix}, \begin{pmatrix}
  \varphi_0 \\
  \varphi_1
\end{pmatrix}\Big\rangle_{_{E}} \ .
       \end{equation}
   This can be made more explicit by inserting the expression for the scalar product
   in \eqref{e:scalar_product_1st}
 \begin{equation}
       \label{e:a_j_ve_Kel}
       a_j = \frac{\kappa_j}{2}\int_\Sigma \!\!\!\left(\!w(x)\overline{\hat{\psi}^{_L}_j} \varphi_1 +
       p(x)\overline{\nabla\hat{\phi}^{_L}_j}\cdot
       \nabla\varphi_0
+  \tilde{V} \overline{\hat{\phi}^{_L}_j} \varphi_0 \!\right) d\Sigma  \ , 
\end{equation}
 and using now, from the first component in the spectral problem for left eigenvectors in
 equation (\ref{e:left-right_vectors_phi-psi}), the relation
$\hat{\psi}^{_L}_j = i\overline{\omega}\hat{\phi}^{_L}_j$, we have
%%%
%\begin{widetext}
%%
\begin{flalign}
       \label{e:a_j_v3_Kel}
       a_j &= \frac{\kappa_j}{2}\Big(\int_\Sigma \!\!\!\!
       w(x)\overline{i\overline{\omega}_j\hat{\phi}^{_L}_j} \varphi_1 d\Sigma 
       + \int_\Sigma \!\!\!\left(p(x)\overline{\nabla{\hat{\phi}}^{_L}_j}\cdot \nabla\varphi_0
       +  \tilde{V} \overline{\hat{\phi}^{_L}_j} \varphi_0 \!\right) d\Sigma \Big) \nn \\
       &= \frac{\kappa_j}{2}\Big(-i\omega_j\int_\Sigma \!\!\!\! w(x)
       \overline{\hat{\phi}^{_L}_j} \varphi_1 d\Sigma 
       + \int_\Sigma \!\!\!\left(p(x)\overline{\nabla{\hat{\phi}}^{_L}_j}\cdot \nabla\varphi_0
       +  \tilde{V} \overline{\hat{\phi}^{_L}_j} \varphi_0 \!\right) d\Sigma \Big) \nn \\
       &=  \frac{\kappa_j}{2}\Big(-i \omega_j\langle \hat{\phi}^{_L}_j, \varphi_1\rangle_{_{(2,w)}}
       + \langle \hat{\phi}^{_L}_j,  \varphi_0\rangle_{H_{(V,p)}^1} \Big) \ .
\end{flalign}
%%%%
%\end{widetext}
%%%
Then, putting together the first component of equation (\ref{e:Kel_phi_psi})
and the expression of $a_j$ in (\ref{e:a_j_v3_Kel})
       \bea
       \label{e:Keldysh_QNM_expansion_phi}
       &&\phi(\tau,x) \sim \sum_n e^{i\omega_{j}\tau} a_n \hat{\phi}^{_R}_j(x)
       \ \ \ , \ \ \ \phi(t,x) \sim \sum_n e^{i\omega_{j}t}  a_n e^{i\omega_{j}h(x)}\hat{\phi}^{_R}_j(x)\nn \\
       && \text{with} \quad
       a_j = \frac{\kappa_j}{2}\Big(\langle \hat{\phi}^{_L}_j,
       \varphi_0\rangle_{H_{(V,p)}^1} -i \omega_j\langle \hat{\phi}^{_L}_j,
       \varphi_1\rangle_{_{(2,w)}}\Big) \ ,
       \eea
       where the factors $e^{i\omega_{j}h(x)}$ in the ``Cauchy'' expression
       $\phi(t,x)$ follow from the height function $h(x)$ in the transformation
       \eqref{eq:HypCoords} from Cauchy to hyperboloidal slices~\footnote{Note that, given
         $\mathrm{Im}(\omega_{j})>0$ and the structure of $h(x)$ at $\pm\infty$,  the factors
         $e^{i\omega_{j}h(x)}$ explode at infinity. This is the signature of the lack of normalizability
         of QNMs in the Cauchy description. It is when using the hyperboloidal slice,
         and therefore a time $\tau$ that asymptotes to ``retarded'' and ``advanced'' times,
         that we have finite spatial functions.}.
       These expressions provide a QNM resonant expansion for the scattered field $\phi(\tau,x)$
       whose form reduces to that of normal modes in the selfadjoint case.
       Specifically, in the selfadjoint case we have $\hat{\phi}^{_L}_j=\hat{\phi}^{_R}_j
       = \hat{\phi}_j$
       and $\kappa_j\to 1$ 
       and we find in a compact domain $D$ in a ``Cauchy slice''
       (the factor $e^{i\omega_{j}h(x)}$ is absent in the already normalizable 
       Cauchy selfadjoint case) 
       \begin{equation}
       \label{e:normal_modes_1}
       \!\!\!\phi(t,x) =\!\!\! \sum_{j\in\mathbb{Z^*}} \!\!\!e^{i\omega_{j}t}
       \frac{1}{2}\Big(\langle \hat{\phi}_j,  \varphi_0\rangle_{H_V^1}
       \!-i \omega_j\langle \hat{\phi}_j, \varphi_1\rangle_{_2}\Big) \hat{\phi}_j(x)   \ ,
       \end{equation}
       where the $\hat{\phi}_j$ are still normalized with respect to the energy
       scalar product $\langle\cdot,\cdot\rangle_{_E}$.
       This last expression reduces indeed to the standard expression (e.g. equation (2.52)
       in \cite{zworski2017mathematical})), in terms of eigenfunctions
       $\phi_j$ of $P_V:=-\Delta + V$ (with homogeneous Dirichlet conditions)
       normalized with respect to the $L^2$ norm (see \cite{SheJar20}) for details)
\begin{equation}       
\label{e:normal_modes_2}
 \phi(t,x) 
  = \sum_{j\in\mathbb{Z^*}} \!\!e^{i\omega_{j}t} \frac{1}{2}
       \Big(\langle \phi_j, \varphi_0\rangle_{_2}
       \!-\frac{i}{\omega_j}\langle \phi_j, \varphi_1\rangle_{_2}\Big) 
  \phi_j(x)   
  \end{equation}
  The crucial difference between equations (\ref{e:normal_modes_1})
  and (\ref{e:normal_modes_2}), on the one hand, and equation
  (\ref{e:Keldysh_QNM_expansion_phi}) is the sense of the ``sum''
  symbol: whereas in  equations (\ref{e:normal_modes_1}) and
  (\ref{e:normal_modes_2}) this corresponds to a convergent series and
  the set of normal modes  $\{\phi_j\}_{j\in\mathbb{N}}$ is an actual
  Hilbert basis, in the case of equation (\ref{e:Keldysh_QNM_expansion_phi})
  this is just an asymptotic expansion in the sense of equations
  (\ref{e:u_Keldysh_v5}), (\ref{e:u_Keldysh_v6}) and  (\ref{e:bound_E}).
  More generally, coming back to the non-selfadjoint case,
  equation (\ref{e:Keldysh_QNM_expansion_phi})
  provides exactly the version in the Keldysh setting of
  the Lax-Phillips asymptotic resonant expansion.
  Indeed, equation (\ref{e:Keldysh_QNM_expansion_phi}) recovers
  the structure of Lax-Phillips expansion  (cf. e.g. equation (2.50)
  in \cite{zworski2017mathematical}; note the sign change in the convention
  for the spectral parameter), in particular (crucially)
  recovering the correct dependence in $\omega_j$, and not
  in the conjugated $\overline{\omega_j}$, a non-trivial check result
  of a subtle intertwining of the Hilbert space elements in the Keldysh's  setting.
  Finally, in contrast with the selfadjoint case, the expression of $a_n$
  cannot be further reduced to purely $L^2$ scalar products, since $\hat{\phi}^{_L}_j$
  is not  the eigenfunction of the operator $P_V$.

%\section{QNM instability high-frequency limit: Regge QNM branches and Burnett's conjecture}
%\label{s:Burnett}

\section{QNM spectral instability: an emerging general picture}
\label{e:QNM_instability_general_picture}
The BH instability phenomenon identified by Nollert and Price in
\cite{Nollert:1996rf,Nollert:1998ys} has been recently revisited in
\cite{Jaramillo:2020tuu,Daghigh:2020jyk,Qian:2020cnz,Liu:2021aqh,Jaramillo:2021tmt,Destounis:2021lum}
---see also results in section \ref{sec:FiniteElements}.
In particular, from the work in
\cite{Jaramillo:2020tuu,Jaramillo:2021tmt} a consistent picture of
this phenomenon emerges, in which the fundamental QNMs are stable under
arbitrary perturbations (respecting the asymptotic structure) whereas
overtones are unstable under high-frequency perturbations. In
this section we discuss first the main ingredients building our current understanding
of the problem and then detail a description of the emerging picture.
We close the section by considering some possible implications of such general
picture for BH spectroscopy in gravitational wave astronomy.

\subsection{Elements in the QNM spectral instability problem}

\subsubsection{QNM-free regions: logarithmic boundaries and pseudospectra.}
\label{s:QNM-free_regions}
As discussed in sections \ref{s:scalar_product_basics} and \ref{sec:Weak_Formulation},
the calculation of QNMs in the hyperbolic approach is cast in terms of
the eigenvalue problem of a non-selfadjoint operator $L$. In this setting,
QNM instability is reduced to the study of the spectral instability of
such an operator. Crucially, this can be studied at the level of the
original unperturbed operator $L$, without implementing the actual perturbations
$\delta L$.
Specifically, as presented in section \ref{s:spectral_inst_intro}
and the applied in the example in section \ref{s:pseudospectru_ABC},
the notion of pseudospectrum (see \cite{trefethen2005spectra,Sjo19Book}
and references in \cite{Jaramillo:2020tuu}) provides an approach to assess
such instability, in terms of the analytical properties of the resolvent of $L$.

When the operator $L$ is perturbed, its eigenvalues can migrate inside
specific regions in the complex plane around the original spectrum $\sigma(L)$,
depending on the size (norm) of the enforced perturbations. If such possible regions
are concentrated around $\sigma(L)$, the operator is spectrally stable but if, on the
contrary, such regions are well extended in $\mathbb{C}$, then large instabilities
can show up.
More specifically ---applying the first characterization in equation \eqref{e:pseudospectrum_def}
to QNM frequencies $\omega$---
the QNM $\epsilon$-pseudospectrum $\sigma^\epsilon(L)$ region is
the set of complex QNM frequencies that are actual resonances of some perturbed 
operator ---defining the perturbed scattering problem---
whose difference $\delta L$ with the original one is $O(\epsilon)$, that is
 \begin{equation}
   \sigma^\epsilon(L)=\{\omega\in\mathbb{C}, \exists \; \delta L\!\in\!
   M_n(\mathbb{C}), ||\delta L||<\epsilon:
\lambda\!\in\!\sigma(L+\delta L) \}  \ .
\end{equation}
The $\epsilon$-pseudospectrum $\sigma^\epsilon(L)$ is then the maximal
region in the complex plane that QNMs can reach under perturbations of
norm $\epsilon$.  Then, for a given $\epsilon$, the regions beyond the
$\epsilon$-pseudospectrum boundaries are QNM-free, or resonance-free,
regions in the complex plane.

\medskip

In the setting of Schwarzschild's potential, the results in
\cite{Jaramillo:2021tmt} (see also \cite{Destounis:2021lum} in
Reissner-Nordstr\"om) permit to identify the resonance-free region
from the numerically calculated pseudospectra, concluding that such
QNM-free regions are asymptotically bounded below by logarithmic
curves of the form (here $C_1$, $C_2$ and $C_3$ are
constants)
\begin{equation}
\label{e:log_pseudospectrum_lines}
\mathrm{Im}(\omega) \sim C_1 + C_2 \ln \big(\mathrm{Re}(\omega) +
C_3\big) \ \ , \ \ |\mathrm{Re}(\omega)|\gg 1 \ .
\end{equation} This
is consistent with theoretical analytical results. Indeed, logarithmic
lower bounds $\mathrm{Im}(\omega) \sim M\ln\mathrm{Re}(\omega)$ of
resonant-free regions appear in the setting of scattering by obstacles
(where the field does not penetrate the scatterer) and smooth compact
support potentials, as shown by Regge \cite{Regge58}, Lax \& Phillips
\cite{LaxPhi71,LaxPhi89} and Vainberg \cite{Vainb73} (see
\cite{zworski2017mathematical,dyatlov2019mathematical} for a detailed
discussion and references). This generalizes to (non-trapping) smooth
non-compact support potentials \cite{Sjoes90,Marti02,SjoZwo07}. The
interesting result is that in the Schwarzschild (also
Reissner-Nordstr\"om and P\"oschl-Teller) case, such asymptotics
extend far from the asymptotic region into the ``inner'' region around
non-perturbed QNMs, if we allow for a translation in the imaginary and
real parts of $\omega$ ---respectively $C_1$ and $C_3$--- that
disappear at large asymptotics.  This asymptotically logarithmic shape
of the QNM-free region, determined from the pseudospectrum, is the
first element in our general QNM spectral instability picture.

\subsubsection{Regular high-frequency QNM perturbations: Nollert-Price-like branches and ``inner'' QNMs.} 
\label{s:regular_perturbations}
The second element building a general picture of QNM
instability concerns the migration of QNMs to perturbed QNM branches.
Consistently with the pseudospectrum analysis discussed above,
perturbed QNMs corresponding to $||\delta L|| < \epsilon$ must lay
above the $\epsilon$-pseudospectrum contour lines in
$\mathbb{C}$. Given the large size of pseudospectra logarithmic
regions in our problem, this is not a tight constraint and, indeed,
low frequency perturbed QNMs stay far from pseudospectra
boundaries. Instability is triggered only by high-frequency
perturbations, that leave stable the fundamental QNM but push overtones
to open branches that we generically refer as Nollert-Price branches,
to acknowledge the first identification of this phenomenon in
\cite{Nollert:1996rf,Nollert:1998ys} (although different types of
branches can and do appear \cite{Jaramillo:2021tmt}).

Interestingly, as discussed in \cite{Jaramillo:2021tmt}, several new
``Nollert-Price'' branches can appear under perturbations, depending on
the size and frequency of the latter. The passage from one such branch
to another in the complex plane is marked by the presence of new QNMs
in the ``interior'' of the new branches, typically less damped and with
smaller oscillation frequency.  Such ``inner QNMs'' do not place
themselves along structured branches, but rather seem to populate the
region in the interior of the open branches.  These inner QNMs are
indeed eigenvalues of the operator $L+\delta L$ (they pass a convergence
test) and are very sensitive to the applied perturbation. Their role
in the QNM instability picture remains to be elucidated.

In generic terms, the relation between the pseudospectrum and the
perturbation of QNMs frequencies can be understood in terms of the
Bauer-Fike theorem.  Given an operator $A$, the `tubular neighborhood' $\Delta_\epsilon(A)$
of radius $\epsilon$ around the spectrum $\sigma(A)$, defined as
\begin{equation}
\label{e:tubular_epsilon}
\Delta_\epsilon(A) = \{\lambda\in \mathbb{C}: \mathrm{dist}\left(\lambda,\sigma(A)\right)<\epsilon\}  \  ,
\end{equation}
is, by definition of the pseudospectrum, always contained in the $\epsilon$-pseudospectrum $\sigma^\epsilon(A)$~\cite{trefethen2005spectra}
\begin{equation}
\Delta_\epsilon(A)\subseteq \sigma^\epsilon(A) \ .
\end{equation}
Therefore, the question is how far the tubular neighborhood of radius $\epsilon$ stays away
from the pseudospectrum boundary. Selfadjoint (more generally, normal operators) indeed satisfy
\cite{trefethen2005spectra}
\begin{equation}
\label{e:tubular_normal}
\sigma_2^\epsilon(A) = \Delta_\epsilon(A) \ ,
\end{equation}
where $\sigma_2^\epsilon(A)$ indicates the use of a $||\cdot||_2$
norm. As a consequence, pseudospectra of normal operators show
characteristic concentric 'circles' tightly packed around the eigenvalues.
When moving away from normality, the instability of a given eigenvalue
$\lambda_n$ is controlled by the so-called condition
number $\kappa_n$ introduced in equation \eqref{e:condition_number}, through the
inequality \eqref{e:intro_eigenvalue_perturbation}.
%% . Specifically, considering the right and left-eigenvectors of
%% $\lambda_n$, respectively denoted by $v_n$ and $w_n$, ---as already
%% defined in equation \eqref{e:left-right_vectors}--- satisfying
%% \begin{equation}
%% L v_n = \lambda_n v_n \ \ , \ \ L^\dagger w_n = \overline{\lambda}_n w_n \ .
%% \end{equation}
%% the condition number $\kappa_n$ of $\lambda_n$ is defined as
%% \begin{equation}
%% %\label{e:condition_number}
%% \kappa_n := \frac{||w_n||\;||v_n||}{\langle w_n , v_n\rangle} \ ,
%% \end{equation}
%%namely the inverse of the cosines between $v_n$ and $w_n$.
In the normal case, %left and right eigenvectors are colinear and
$\kappa_n =1$.  However, in the non-normal case $\kappa(\lambda_i)> 1$, and can
grow unbounded if left- and right- eigenvectors in \eqref{e:condition_number}
tend towards orthogonality.  The essential content of Bauer-Fike theorem ---cf.
\cite{trefethen2005spectra}--- is the identification of such $\kappa_n$ numbers
as the key element in the relation between pseudospectra and eigenvalue
perturbations, by determining the effective radii in which perturbed
QNMs can move around the spectrum
\begin{equation}
\label{e:tubular_error}
\sigma^\epsilon(A)\subseteq \Delta_{\epsilon\kappa}(A):=\bigcup_{\lambda_i\in\sigma(A)} \Delta_{\epsilon\kappa(\lambda_i)+O(\epsilon^2)}(\{\lambda_i\})
\ .
\end{equation}
In sum, QNM branches whose eigenvalues have large condition
numbers $\kappa_n$ are subject to potential instabilities. In contrast,
branches with low $\kappa_n$ remain stable under
perturbations.

\medskip

\setcounter{footnote}{0}

In our problem, non-perturbed BH QNM overtones have large $\kappa_n$
\cite{Jaramillo:2020tuu}, consistently with the large logarithmic
pseudospectra regions. They are therefore potentially unstable under
physical perturbations. However, this does not mean that {\em any}
perturbation triggers the instability. In fact, low frequency
perturbations do not modify QNMs in an unstable way, but rather
spectrum modifications are of the same scale as that perturbation.  On
the contrary, high energy perturbations lead to perturbation patterns
(mainly ~\footnote{Notice also the mentioned presence of
``inner QNMs'', in the interior region of these branches and not
  along them. This different pattern may respond to a distinct
  underlying instability mechanism.}) structured in the new branches we have referred to as
Nollert-Price branches, sharing the following qualitative points:
\begin{itemize}
\item[i)] Perturbed QNMs became less damped, lasting longer in time,
  therefore $\mathrm{Im}(\omega_n)$ reduces.
\item[ii)] Oscillations probe smaller scales, specifically
  $\mathrm{Re}(\omega_n)$ is not bounded above.
\item[iii)] As a result, perturbed QNM
  branches show a characteristic opening in the complex plane, growing
  from a first perturbed QNM overtone in the non-perturbed QNM
  spectrum.
\item[iv)] The first perturbed overtone gets closer to the real axis
  as the frequency and/or the size of the perturbation increases. The
  fundamental QNM always stay stable under this type of low-amplitude and high-frequency
  perturbation~\footnote{The high-frequency perturbations respect a ``non-trapping
    condition'' for the perturbed potential, namely there are not closed trajectories in the
    phase space ---apart from the ``point trajectory'' corresponding to the
    unstable equilibrium at the (single) maximum of the potential. If we allow for
    closed trajectories in the unperturbed potential, such as the ``well in an island''
    potential  \cite{BinZwo} with two maxima, then perturbations of the fundamental
    QNM can occur even for very tiny double potentials (``flea on the elephant''
    effect \cite{JonMarSco81,GraGreJon84,Simon85}).
    We thank V. Cardoso for signalling this point.}.
\item[v)] The resulting perturbed QNM branches are stable under further
  perturbations, accordingly perturbed QNMs have lower condition numbers $\kappa_n$.
\end{itemize}
Such qualitative behaviour of QNMs is compatible with the generic one found in
the scattering by matter compact objects, namely $w$-modes ---or
curvature QNMs \cite{Kokkotas:1999bd}--- or in QNMs of compact
obstacles. This
`universality', however, is not very constraining concerning the 
shapes of the perturbed branches, specifically in the asymptotic large
frequency regime.  To illustrate this, we mention the case of convex
obstacles, where the asymptotic form of QNM branches (under a `pinched
curvature assumption') can be explicitly established
\cite{SjoZwo99,Zworski99}
  \begin{equation}
  \label{e:QNM_Gevrey_s=3}
\mathrm{Im}(\omega_n) \sim K |\mathrm{Re}(\omega_n)|^{\frac{1}{3}} + C \ \ , \ \ n\gg 1 \ .
\end{equation}
Such QNM power-law behavior follows the boundary of resonant-free regions
for scatterers whose regularity is of Gevrey type. Specifically
$s$-Gevrey classes interpolate between analyticity, corresponding to $s=1$, and
$C^\infty$ regularity, for $s=\infty$. The asymptotic
behavior of the resonant free regions for scatterers of $s$-Gevrey class
is given by $\mathrm{Im}(\omega) > K\omega^{1/s}$
\cite{Goodh73}, consistent with the actual Gevrey $s=3$ regularity in the example in
\eqref{e:QNM_Gevrey_s=3}.
In our BH setting, the analysis in
\cite{Gajic:2019oem,Gajic:2019qdd,galkowski2020outgoing} suggests a
special role for Gevrey $s=2$ classes~\footnote{Note that in our QNM
  {\em stability} discussion, a special role is played by the energy
  scalar product and the associated energy norm (namely a $H^1$-like
  norm).  This is because we privilege ``energy'' as as measure of the
  size of the perturbations. If we are rather interested in the QNN
  {\em definition} ---a problem of different nature to QNM
  stability--- $H^1$-like norms are not sufficient to characterize the
  spectrum. Specifically, a stronger ``control'' of the regularity of
  eigenfunctions associated with QNMs is needed, as the one provided
  by Gevrey classes (controlling higher derivatives with appropriate
  weights).  In pseudospectrum terms, this would amount to much
  tighter $\epsilon$-pseudospectrum lines, defining the spectrum in
  the $\epsilon\to 0$ limit.  In other words, one would use first
  Gevrey classes to define QNMs in both the non-perturbed and the
  perturbed potentials and, in a second step, one would use the energy
  norm to assess if the associated operator perturbations ---possibly
  inducing QNM instabilities--- are actually big or small (in energy
  terms).  The relation between the two problems above is fundamental
  and requires further specific investigation.}  and, therefore,
branches of type $\mathrm{Im}(\omega_n) \sim K
|\mathrm{Re}(\omega_n)|^{\frac{1}{2}}$. However, the numerical
methodology adopted in \cite{Jaramillo:2020tuu,Jaramillo:2021tmt}
makes very difficult to determine the precise asymptotics and,
actually, branches with different detailed behaviors are observed
\cite{Jaramillo:2021tmt} ---together with the ``inner QNMs'' marking the
passage between such different branches.  In spite of the resulting
complexity and the numerical difficulties for the study of large-$n$
asymptotics, a remarkable regularity shows up: QNMs on Nollert-Price
branches tend to place themselves uniformly along the branch, this
ultimately leading to a Weyl law for the asymptotic counting of
QNMs~\footnote{This suggests the possibility of extracting fine-structure
  information from fluctuations (with respect to the 'regular
  unfolding' of the spectrum, using Weyl law's), by adopting a
  'spectrum statistics' approach to the QNM spectrum properties
  (cf. e.g. \cite{Pook-Kolb:2020jlr} and references therein).  It is
  also interesting to note that, together with the different `open branch'
  Nollert-Price regimes, other 'non-branch' behavior associated with 'inner QNMs'
  show up. Such inner QNMs seem to follow well-defined probabilistic distributions
  in $\mathbb{C}$ (the latter entering in the spectral  analysis
  of perturbation of non-selfadjoint operators~\cite{Sjo19Book}),
  suggesting the need to import stochastic
  tools to the study of QNMs.}. Inner QNMs do not change such Weyl
law, though they induce a transition in the relevant length
scale~\cite{Jaramillo:2021tmt}.

  Beyond the complexity of QNM Nollert-Price branches, the (Gevrey)
  example in equation (\ref{e:QNM_Gevrey_s=3}) illustrates a
  fundamental point: in setting we are studying, (a part of) perturbed
  QNMs tend to migrate towards the pseudospectra boundaries. This is
  systematically observed in the numerical studies in
  \cite{Jaramillo:2020tuu,Jaramillo:2021tmt}. In spite of the
  difficulties to study numerically QNM asymptotics, on can soundly
  conclude: as long as perturbations are regular (at least,
  $C^\infty$), they stay distinctly above the corresponding
  $\epsilon$-pseudospectrum asymptotically logarithmic lines.  That
  is, perturbed QNM tend to track pseudospectrum lines, getting closer
  to them as frequency increases but  always staying above
  as long as perturbations are at least $C^\infty$.
  This is the second element in our general picture of high-frequency QNM spectral instability.

\subsubsection{Non-regular $C^p$ QNM perturbations: logarithm Regge branches.}
\label{s:Cp_perturbations}
When $C^\infty$ regularity is lost, perturbed QNMs reach the
logarithmic boundaries of the QNM-free regions. This is the third key
ingredient in our picture of QNM instability
picture. Perturbations with $C^p$ regularity, i.e.  with a
discontinuous $p$-th derivative (for any $p<\infty$) make QNM migrate
to branches that saturate the pseudospectrum boundaries, following
logarithmic lines (\ref{e:log_pseudospectrum_lines}).  Specifically,
$C^{p<\infty}$-perturbed QNMs follow Regge QNM branches
\cite{Regge58,BinZwo,zworski2017mathematical,dyatlov2019mathematical}
\begin{eqnarray}
\label{e:Regge_branches}
\mathrm{Re}(\omega_n) &\sim& \pm\Big(\frac{\pi}{L} n + \frac{\pi
  \gamma_p}{2L}\Big) \\ \mathrm{Im}(\omega_n) &\sim& \frac{1}{L}
\bigg[\gamma\ln\bigg(\Big(\frac{\pi}{L} n + \frac{\pi
    \gamma_p}{2L}\Big) + \frac{\pi \gamma_\Delta}{2L}\bigg) - \ln
  S\bigg] \ , \nn \end{eqnarray} for $n\gg 1$. This asymptotic result, first
found by Regge \cite{Regge58} has been independently recovered later
in different settings. Note that such branches place themselves on the top
of asymptotic pseudospectrum lines (\ref{e:log_pseudospectrum_lines}),
with $C_1=-\ln(S)/L$, $C_2=\gamma/L$ and $C_3= \pi
\gamma_\Delta/(2L)$.

In the setting of compact-support potentials (actually the context in
Regge's work \cite{Regge58}), this result has been discussed in full detail by
Zworski \cite{Zwors87}.  Specifically, considering a potential with
compact support in the interval $[a,b]$ and with behavior
\begin{flalign}
\label{e:compact_support_V_boundaries}
\begin{array}{rclcl}
V(x) &\sim& V_1 (x-a)^m \ \ &,& \ \ \hbox{as} \ x\to a^+  \\
V(x) &\sim& V_2 (x-b)^k \ \ &,& \ \ \hbox{as} \ x\to b^-
\end{array}
\end{flalign}
with $m,k$ non-negative integers, the QNMs follow Regge branches (\ref{e:Regge_branches})
with
\bea
\label{e:parameters_Zworski}
L &=& (b-a) \ \ , \ \ \gamma = \frac{m+k+4}{2}\nn \\
\gamma_p &=& \gamma + \big(1 + \mathrm{sign}(V_1V_2)\big) \ \ , \ \
\gamma_\Delta = -\gamma_p \nn \\
S &=& \Big(\frac{|V_1V_2|k!m!(k+m+2)}{2^{m+k+1}}\Big)^{\frac{1}{2}} \ .
\eea
This rigorous result fully accounts for the branches found by Nollert
in \cite{Nollert:1996rf}, since the stepwise potential there employed
to approach Schwarzschild is, in particular, a compact-support $C^0$
potential. Interestingly, the same theorem states that QNMs of smooth
($C^\infty$) class must lay strictly above the logarithmic lines, consistently with
the pseudospectrum analysis in section \ref{s:QNM-free_regions}.  The
result does not directly apply to the `spiked truncated dipole'
potential in \cite{Nollert:1998ys}, since this is a non-compact
support potential with a finite jump at $x=x_0$ and a Dirac-delta
perturbation at $x=x_\delta$, namely $\delta V = V_o
\delta(x_\delta)$. In spite of the non-compactness, the asymptotic
analysis by Nollert and Price leads exactly to Regge branches
(\ref{e:Regge_branches}), now with
\begin{eqnarray}
\label{e:parameters_NP}
L &=& (x_\delta-x_0) \ \ , \ \
\gamma = \gamma_p = \frac{3}{2}\nn \ \ , \ \ \gamma_\Delta = 0 \ \ , \ \
S = \Big(\frac{V_\delta}{4x_0^2}\Big)^{\frac{1}{3}} \ .
\end{eqnarray}
More generally, a higher-order WKB approach \cite{Berry82,BerMou72} can be
envisaged to capture the QNM asymptotics of general (non-compact)
$C^{p<\infty}$ potential barriers.  This approach boils down to the
determination by Berry \cite{Berry82} of the ``reflection coefficient'' $R$ at a
point $x=x_0$, where the potential $V$ is discontinuous in the
$p$th-derivative (cf. equation (27) in \cite{Berry82}), namely
\begin{equation}
\label{e:Delta_V}
R \sim \Big(\lim_{x\to x_0^+} \frac{d^p V}{dx^p} - \lim_{x\to x_0^-} \frac{d^p V}{dx^p} \Big)
  := \Delta_p V \ .
\end{equation}
This WKB approach has been applied in Zhang et al. \cite{ZhaWuLeu11}
to recover Regge branch asymptotics of polytropic neutron stars (with $L$
the ``tortoise'' radius of the star, $\gamma = p+2$
fixed by the polytropic index,
$\gamma_p= \theta_b$ depending on the wave parity,
$\gamma_\Delta=0$ and $S$ proportional to $\Delta_p V$) and in Qian et
al. \cite{Qian:2020cnz,Liu:2021aqh} for BH-like potentials with a
``jump'' or ``cut'' ($L$ is there given in terms of the position of
the jump $x_{\mathrm{cut}}$ and $S$ is again controlled by $\Delta_pV$).

In sum, QNMs of $C^p$ potentials follow logarithm
Regge branches that saturate the pseudospectrum boundary lines. In
particular, QNMs in \cite{Nollert:1996rf} are indeed Regge
branches~\footnote{We prefer in any case use the term Nollert-Price
  branches to all perturbed branches (even not logarithmic as those in
  section \ref{s:regular_perturbations}), to stress the first authors
  that identified the QNM instability branch opening phenomenon.}. The explicit models behind 
Regge branches lead to an interpretation of the coefficients \cite{Jaramillo:2021tmt}:
\begin{itemize}
\item[i)] {\em ``Length scale'' $L$}: straightforwardly obtained as
  $L = \pi \big(\mathrm{Re}(\omega_{n+1})-\mathrm{Re}(\omega_{n})\big)^{-1}$, it is
  related to a 'bouncing' scale \cite{Regge58}, as illustrated by $L=(b-a)$ in  \cite{Zwors87}
  or $L=x_\delta-x_0$ in \cite{Nollert:1998ys} (in \cite{Liu:2021aqh} $T=2L$ has been proposed
  as a bouncing time in an ``echoes'' mechanism).

\item[ii)] {\em ``Small structure'' coefficient $\gamma$}: this coefficient is explicit identified
  in \cite{Zwors87} with the regularity of the perturbations. This raises the question:
  {\em can one read the (effective) regularity of spacetime from QNM observational data?}

  More soberly, we refer to this parameter as a ``small structure'' coefficient, something
  further  supported  in a matter setting, where $\gamma$ 
  is ---as mentioned above--- directly related in \cite{ZhaWuLeu11} to the polytropic index
  in the equation of state of polytropic (neutron) stars. This parameter should strongly
  depend on the frequency of the underlying perturbation.

\item[iii)] {\em ``Strength coefficient'' $S$} (or {\em ``imaginary shift''}): this parameter is
  directly related to the ``size'' of the perturbation, namely $V_1$ and $V_2$
  in equation (\ref{e:compact_support_V_boundaries})
  in \cite{Zwors87}, $V_\delta$ in \cite{Nollert:1998ys} or $\Delta_p$ in equation (\ref{e:Delta_V})
  used in  \cite{ZhaWuLeu11,Qian:2020cnz}. However, it also depends on the regularity/frequency
  of the perturbation and the size and frequency effects can be difficult to disentangle.
  In this sense, it should be determined after $\gamma$. The term $\ln(S)/L$ provides a 
  translation in the imaginary part in the $\mathrm{Im}(\omega_n) =
  f(\mathrm{Re}(\omega_n))$
  functional dependence.

\item[iv)] {\em ``Parity term'' $\gamma_p$}: it shifts the location of Regge QNMs along
  the logarithmic branches according to sign/parity properties of the potential.
  In \cite{Zwors87} it depends on the relative sign of $V_1$ and $V_2$,
  whereas in \cite{ZhaWuLeu11}
  it depends on the axial/polar character of the perturbation. It seems to
  play a role in the ``polarity-alternant'' pattern in $w$-like modes. 

\item[v)] {\em ``Real shift'' $\gamma_\Delta$}: it induces a translation in the real part of the
  relation  $\mathrm{Im}(\omega_n) = f(\mathrm{Re}(\omega_n))$ determining,
 with  $\ln(S)/L$,  the first overtone
  affected by a perturbation.

\end{itemize}

The $C^p$ potentials here considered provide an instance of (infinite)
ultraviolet limit ``high-frequency'' spacetime perturbation.
They are not only logarithmic, but they also provide a very specific form
of the QNM distribution along that logarithmic branches.
A natural question is posed: {\em how generic is such Regge QNM distribution in the
  ultraviolet perturbation limit?}

\subsection{Emerging picture: universality in QNMs of compact objects}
\label{s:emerging picture}
The previous points build a consistent picture for the BH QNM
instability pattern:
\begin{itemize}
\item[i)] The fundamental BH QNM (of each $\ell$-fixed branch) is
  stable under (generic ``non-trapping'') perturbations respecting the
  asymptotic structure at null infinity.

\item[ii)] QNM-free regions are bounded-above by logarithmic curves,
  namely the pseudospectra contour line boundaries.
  
\item[iii)] Generic regular ($C^\infty$) high-frequency perturbations
  make QNM overtones migrate to either a) Nollert-Price-like branches
  whose opening is associated with a ``bouncing/reverberance''
  mechanism, or b) ``inner QNMs'' marking the transition between
  different perturbed branches.  As long as the perturbation frequency
  stays finite, perturbed QNMs lay (strictly) above pseudospectrum
  logarithmic curves.

\item[iv)] Low-regularity $C^p$ perturbations, in particular involving
  an (ultraviolet) infinite high-energy limit, are 'optimal' in
  reaching the logarithmic QNM-free region boundaries.

\item[v)] Perturbed QNMs migrated to Nollert-Price-like branches are
  stable under further perturbations, whereas ``inner QNMs'' are
  highly unstable, distributing in ``transition regions'' in the interior of
  Nollert-Price-like branches.

\item[vi)] The resulting  perturbed  BH Nollert-Price QNM
  branches %, with their different regimes,
  share some of the
  qualitative features of QNM branch patterns of matter compact objects:
  high-frequency perturbations induce a transition from the non-perturbed BH
  (structurally unstable) QNM branches to new pertubed (structurally stable) QNM
  branches qualitatively similar for all (vacuum or matter)
  compact objects, pointing towards a possible
  universality~\cite{Jaramillo:2020tuu}.
\end{itemize}
Some remarks concerning the previous heuristic picture:
\begin{itemize}
\item[a)] The (in)stability of QNMs is claimed under {\em generic}
  perturbations. This does not preclude that certain fine-tuned
  perturbations can induce specific non-generic behaviors. This may
  be relevant in specific models.
  
\item[b)] The asymptotics of {\em generic} ultraviolet perturbations,
  when the perturbation frequency tends to infinity in an arbitrary
  way, is a point one would like to elucidate.  As stated in point iv)
  above, the enforcement of $C^p$ perturbations realizes an instance
  of such ultraviolet situation, but not as a limiting process. This issue will be
  revisited in section \ref{s:Burnett}, after in introducing in section
  \ref{sec:physical_interpretation_energy} an additional piece of information,
  leading to the conjecture that arbitrary infinite high-frequency limits
  lead to a $C^p$ state, so  Regge QNMs are the precise and
  generic QNM pattern of generic ultraviolet perturbations.

\item[c)] The nature of the ``inner QNMs'' and the associated Nollert-Price-like branch
  transitions observed in \cite{Jaramillo:2021tmt} remains quite elusive. An understanding of this problem
  could shed light on the different scales involved in the QNM
  instability problem.
\end{itemize}

\subsection{BH spectroscopy and $\epsilon$-dual QNM resonant expansions.}
\label{s:epsilon_dual_QNMexpansions}
Let us consider a non-perturbed BH with non-perturbed QNMs $\omega_n$.
On the one hand, section \ref{e:QNM_instability_general_picture} provides a
description of the QNM spectral instability phenomenon,
according to which high-frequency perturbations with $||\delta L||=\epsilon$
perturb the QNMs to very different new values $\omega^\epsilon_n$.
Beyond the ``frequency-domain'' spectral computations in \cite{Jaramillo:2020tuu,Jaramillo:2021tmt},
the ``time-domain'' results in \cite{Jaramillo:2021tmt} crucially confirm the 
presence of such perturbed QNMs $\omega^\epsilon_n$  in the time signal.

On the other hand, in section \ref{s:resonant_expansions} we have constructed
resonant expansions of the time-dependent field scattered by a given
potential in terms of the QNMs of {\em that} potential. 
Given i) a non-perturbed potential $V$ with  $\hat{v}_n(x)$ and $\omega_n$ QNM
functions and frequencies, and ii) a perturbed potential $V^\epsilon = V + \epsilon \delta V$ (with $||\delta V||_{_E}=1$)
with $\hat{v}^\epsilon_n(x)$ and $\omega^\epsilon_n$ QNMs, we can write the respective scattered
fields $u(\tau,x)$ and $u^\epsilon(\tau,x)$ in QNM resonant expansions as
\begin{equation}
\label{e:resonant_expansions}
u(\tau,x) \sim  \sum_n e^{i\omega_n \tau} a_n \hat{v}_n(x) \quad, \quad
u^\epsilon(\tau,x) \sim 
\sum_n e^{i\omega^\epsilon_n \tau} a^\epsilon_n \hat{v}^\epsilon_n(x) \ ,
\end{equation}
in accordance with the resonant expansion structure in general scattering theory \cite{LaxPhi89,Vainb73,dyatlov2019mathematical}.

\subsubsection{GW signal time domain stability versus QNM spectral instability.} 
In this setting, let us note the stark contrast in the stability ---based on the same
energy norm $||\cdot||_{_E}$--- of the
``frequency domain'' and ``time domain'' problems associated with $L$:
\begin{itemize}
\item[i)] Spectral QNM problem of $L$, equation \eqref{e:eigenvalue_problem_L_Fourier_v1}: instability in the energy norm.
  As reviewed in section \ref{e:QNM_instability_general_picture}, this problem is strongly unstable under
  high-frequency perturbations.
\item[ii)] Time evolution problem  of $L$, equation \eqref{e:wave_eq_1storder_u}:  stability in the energy norm.
   As  discussed  in \cite{Nollert:1996rf,Nollert:1998ys} and later revisited in
   \cite{Daghigh:2020jyk,Qian:2020cnz},  time evolution signals $u(\tau,x)$ and $u^\epsilon(\tau,x)$ ``look very similar''
   so the dynamical problem seems stable in spite of the strong QNM
   spectral instability. This can be understood on general grounds
   from (finite) energy considerations and from the absence of
   unstable QNMs. In particular, it can be better assessed in terms of
   the stability of this dynamical problem in the energy
   norm \footnote{Note that we refer here to stability upon perturbations $\delta L$
     of the operator $L$ defining the  dynamics.
     This notion of stability is different from that of the
     usual PDE perspective, focused on analysing the stability of solutions $u(t,x)$ of a given fixed equation upon perturbations $\delta u_{0}(x)$ of the initial data with respect to reference initial data $u_0(x)$.} From
   the Cauchy problem \eqref{e:wave_eq_1storder_u} for the
   non-perturbed potential, one can write
  \bea
  \label{e:dyn_energy_stab}
||u - u^\epsilon||_{_E} \lesssim \bar{C} ||u||_{_E} \epsilon \lesssim C \epsilon \ ,
\eea
that provides a crude (energy) estimate of the difference between the unperturbed and perturbed
time signal. Certainly this is not the pointlike estimate for the time series
at null infinity (namely, the idealized  position of the detector) that one would need
to compare the observed signals, but it gives an idea that for sufficiently regular time signals
their difference is proportional to the perturbation size, then providing
a notion of stability.
%To control the regularity of the signal,
%one would need to use Sobolev norms.
%in $H^k=W^{k,2}$).
%Since the 
\end{itemize}
There is no actual contradiction in such radically different stability
behaviour (in the energy norm~\footnote{If one wants
  to better control the regularity of the signal to give a better estimate than 
  the one in equation \eqref{e:dyn_energy_stab}, one would need to use rather  norms
  controlling  higher derivatives, namely Sobolev norms in $W^{k,p}$. In the setting
  of this work where the scalar product plays a key role is natural to choose
  Hilbert spaces $H^k=W^{k,2}$.
  Interestingly, given that the QNM instability is a high-frequency phenomenon,
  we can expect significant differences in the signal at high-wave numbers.
  Then, in terms of a high-$k$ Sobolev norm the difference $||u - u^\epsilon||_{_{H^k}}$
  could be much larger than $\epsilon$, consistently with an instability
  in the time-domain signal. So both the spectral and time domain problem would
  be unstable under high-frequency perturbations, when measuring the operator pertubation
  in the energy norm.
  
  One could also think the problem alternatively, by changing the manner of measuring
  the size of the operaror perturbation $\delta L$: % in the frequency-domain problem,
  if we choose the operator norm induced from $||\cdot||_{_{H^k}}$, a high-frequency perturbation $\delta L$
  that we would interpret as small in the energy norm ($H^2$-like norm) would be actually
  very large in the Sobolev norm (i.e. $||\delta L||_{_{H^k}}>>||\delta L||_{_{E}}$ for high-frequency
  perturbations and sufficiently high $k$), so both the large perturbations in the eigenvalues
  and the large $||u - u^\epsilon||_{_{H^k}}$ would be consistent with spectral stability.
  But such reconciliation of frequency- and time-domain stability in the $H^k$ norm is not very
  interesting in our setting, since it is only a rephrasing of the known stability under
  low frequency perturbations. These issues will be addressed in more detail elsewhere.}) of these two related, but different, problems associated with $L$.
But this stark contrast suggests the following question: writing formally, from the crude energy estimation
(\ref{e:dyn_energy_stab}) and for sufficiently regular time signals, the approximation
\begin{eqnarray}
\label{e:u_epsilon_perturb}
u^\epsilon(\tau,x) \sim u(\tau,x) + O(\epsilon) \ ,
\eea
then, writing the respective QNM resonant exoansions \eqref{e:resonant_expansions},
we have
\begin{eqnarray}
\label{e:dual-epsilon_QNM_expansions}
%\hspace{-18mm}u^\epsilon(\tau,x) \sim u(\tau,x) + O(\epsilon) \quad , \quad
\sum_ne^{i\omega^\epsilon_n \tau} a^\epsilon_n  \hat{v}^\epsilon_n(x)
\sim\sum_n
e^{i\omega_n \tau} a_n \hat{v}_n(x)+O(\epsilon) \ .
\end{eqnarray}
In other words, in a physical situation with perturbations, the perturbed field $u^\epsilon(\tau,x)$
admits an ``exact'' resonant expansion in terms of the perturbed QNMs $\omega^\epsilon_n$, but also 
an expansion in terms of non-perturbed QNMs $\omega_n$ that is indistinguishable (and therefore valid)
as long as energies involved are above or of the order of the energy $\epsilon$ of the triggering
perturbation.

Then, which expansion should one use? The message we would like
to convey is that both can and must be employed: whereas non-perturbed
QNMs encode information about the large scale properties of the BH,
the perturbed QNM expansion encodes information on the perturbation
physics. We refer to QNM-expansions in
(\ref{e:dual-epsilon_QNM_expansions}) as '$\epsilon$-dual QNM
expansions'.

\subsubsection{Data analysis inverse problem: perturbation-enhanced degeneracy.}
If such a notion of $\epsilon$-dual QNM expansion is pertinent in the BH spectroscopy
problem, then unfortunately it comes with a very non-trivial challenge data analysis:
if very similar GW waveforms can actually correspond to very different sets of QNM frequencies,
triggered by very tiny perturbations characterised by a large parameter space, then
the data analysis inverse problem seems extremely degenerate~\footnote{Actually,
  there would be not only one pair of $\epsilon$-dual QNM expansions, but a whole
  (continuous) family of them parametrized by all possible perturbations of size $\epsilon$.
  This looks like a formidable data analysis problem.}. Developing 
specifically suited strategies would be required.

Given an observed GW signal, the QNM expansion based on non-perturbed QNMs has
a special feature: if it can actually be constructed,
then it should minimize the number of excited modes as compared with other
$\epsilon$-dual QNM expansions. The reason stems from the instability of non-perturbed
QNM overtones $\omega_n$ --- with large condition numbers are large $\kappa_n\gg 1$ ---
and the stability of already perturbed QNMs $\omega^\epsilon_n$ --- with $\kappa^\epsilon_n\sim 1$.
From equation \eqref{e:Keldysh_QNM_expansion_u} we can write for the expansion
coefficients in equation \eqref{e:resonant_expansions}
\bea
a_n = \kappa_n\langle \hat{w}_n|u_0\rangle_{_{E}} \quad , \quad  
a^\epsilon_n =  \kappa^\epsilon_n\langle \hat{w}^\epsilon_n|u_0\rangle_{_{E}} \ .
\eea
Then, given the growth of $\kappa_n$ with $n$ \cite{Jaramillo:2020tuu}, the initial
data $u_0$ should project only on the first non-perturbed modes $\hat{w}_n$, 
to preclude the series from exploding. In contrast, since $\kappa^\epsilon$ are of order unity,
more $\hat{w}^\epsilon_n$ can be involved. Typically, a QNM expansion in
perturbed QNMs will need more terms than one in non-perturbed ones.
%~\footnote{Note  that if a $H^k$ scalar product is used, the instability of would overtones decrease.}
This is consistent with results in \cite{Liu:2021aqh,Jaramillo:2021tmt}.

If one aims at constructing the QNM expansion of the observed GW waveform built on the
perturbed QNM frequencies corresponding to the underlying perturbation, one
lacks (in contrast with non-perturbed QNMs) an analytical parametrization
of the exact possible solutions to be used as an ``a priori'' input in the data analysis
extraction. In this context, by enforcing as Ans\"atze educated guesses
for perturbed QNM branch patterns
(e.g. the Regge QNM branches \eqref{e:Regge_branches}, as suggested in \cite{Jaramillo:2021tmt})
could render feasible the data analysis process.

Finally, if such an ``$\epsilon$-dual resonant expansion'' strategy can actually be implemented,
it would be informative (up to precision $\epsilon$ in energy) about two
complementary physics regimes: i) non-perturbed BH QNM expansions would retrieve
the physical parameters of the underlying 'averaged BH' in a standard Kerr uniqueness paradigm,
whereas ii) Nollert-Price-like BH QNM expansions would probe the small scales of the
environmental perturbations or, if Planck scales are involved, the
effective regularity/granularity/stochasticity of spacetime.

%Coefficients in Hilbert space $H^k = W^{k,2}$.

\section{Geometry and QNM instability: heuristics on further structural questions}
\label{s:Geometry_QNMs}
In this section we initiate the exploration, at a heuristic level, of some of
  the stuctural geometric aspects underlying the BH QNM instability. In this sense,
  it must be understood as an invitation to a more detailed research, rather than
  an account of actual results.

\subsection{Physical interpretation of the energy norm: role of spacetime density perturbations}
%\label{sec:Heuristic}
%\subsubsection{Physical interpretation of the energy norm}
\label{sec:physical_interpretation_energy}

First, we discuss the relation between the characteristics of
spacetime metric perturbations and the norms of operator perturbations
$\delta L$ in the eiganvalue problem \eqref{e:intro_L_eigen} and
\eqref{e:intro_perturbed_spectral_problem}.
%  inducing perturbations in the potential,
and the associated norms in the QNM problem of equation
%\eqref{eq:Eigenvalue_Problem}.
Since this is a heuristic argument we will use the symbol $\sim$ to
denote equality where subleading terms are ignored.

\setcounter{footnote}{0}

The potential $V$ ---and hence its rescaled version $\tilde{V}$---
appearing in equation \eqref{eq:reducedeq} can be put in
correspondence with the Ricci scalar of a conformally rescaled metric
$\tilde{g}_{ab} = \Xi^2 g_{ab}$ where $g_{ab}$ is the physical metric
solving the Einstein field equations and $\Xi$ is a conformal factor
vanishing at $\scri^+$ (see e.g. \cite{PanossoMacedo:2018hab}), so that formally we can write
\begin{equation}
\tilde{V} \sim C+\tilde{R} \ ,
\end{equation}
where $C$ is a term associated with the  angular part of the propagating field and the particular
coordinates used ---in the case discussed in section
\ref{sec:Energy_associated_inner_products}, this corresponds to
the constant term $\ell(\ell+1)$--- which will not enter at a first order 
in our discussion, since we will be interested in variations
of the potential related to the subjacent geometric background, so $\delta C \sim 0$ and
\begin{equation}
\delta \tilde{V} \sim \delta\tilde{R} \ .
\end{equation}
On the other hand, from a variational point of view, one can relate
variations in the metric with variations in the Ricci scalar via
\begin{equation}\label{eq:variationR}
  \delta \tilde R = \delta (\tilde{g}^{ab}\tilde{R}_{ab})=
  \tilde{R}_{ab} \delta \tilde{g}^{ab} + \tilde{g}^{ab}\delta
  \tilde{R}_{ab} \ ,
\end{equation}
where the conformally rescaled
Ricci tensor $\tilde{R}_{ab}$ and the physical one
 $R_{ab}$ are related through
\begin{equation}\label{eq:ConfTransRicci}  
  \tilde{R}_{ab} = R_{ab} - 2 \Xi^{-1}\tilde{\nabla}_a\tilde{\nabla}_b\Xi
  -\tilde{g}_{ab}\tilde{g}^{cd}(\Xi^{-1}\tilde{\nabla}_c
  \tilde{\nabla}_d\Xi 
  -3 \Xi^{-2}\tilde{\nabla}_c\Xi\tilde{\nabla}_d\Xi) \ ,
\end{equation}
(cf. for instance \cite{CFEBook}).
Observe that $\tilde{R}_{ab}$ is formally singular at $\scri$.
Additionally, notice that
\begin{equation}
  \delta \tilde{g}^{ab} = \Xi^{-2} \delta g^{ab},
\end{equation}
where we have assumed that $\delta \Xi=0$, consistent with the ``scri
fixing'' philosophy of the hyperboloidal approach. Thus overall, the
first term in equation \eqref{eq:variationR} is formally singular at
$\scri^+$ unless $\delta g^{ab} \sim O(\Xi^3)$ close to $\scri$ or
compactly supported perturbations $\delta g^{ab}$ are considered.  For
the rest of this section, this latter assumption will be made so the main
contribution in equation \eqref{eq:variationR} comes from the
second term~\footnote{Another approach to justify
  $\delta \tilde R \sim  \tilde{g}^{ab}\delta  \tilde{R}_{ab}$ is a reminiscent
  of a virial-like reasoning. Indeed, the first term in equation \eqref{eq:variationR}
  can be understood as a kinetic energy term, whereas the second one plays the role of a
  potential energy term. Under the assumption that the considered spacetime perturbations
  are {\em virialised} (in some appropriate sense to be better seized), the two terms are commensurate
  $\tilde{R}_{ab} \delta \tilde{g}^{ab} \sim \tilde{g}^{ab}\delta \tilde{R}_{ab}$ and \eqref{eq:variationR_v2} holds.
  This would be consistent (in an averaged ergodic sense) with the requirement of stationarity in the characterization and treatment of
  QNMs.} and we can write
\begin{equation}\label{eq:variationR_v2}
  \delta \tilde R \sim \tilde{g}^{ab}\delta \tilde{R}_{ab} \ ,
\end{equation}
where the right hand side can be cast in the form
%The second term in equation  %\eqref{eq:variationR} is given by
\begin{equation}
\tilde{g}^{ab} \delta \tilde{R}_{ab} = \tilde{\nabla}^av_a \ ,
\end{equation}
where (see e.g. \cite{Wald84})
\begin{equation}
 v_a= \tilde{\nabla}^b(\delta \tilde{g}_{ab})-\tilde{g}^{cd}\tilde{\nabla}_a\delta \tilde{g}_{cd} \ .
\end{equation}
Hence,  we can formally relate the variations in the potential 
to variations in the metric 
\begin{equation}\label{eq_deltaV_metricPert}
  \delta \tilde{V} \sim \delta \tilde R \sim |\partial^2 \delta \tilde{g}_{ab}|
  \sim  \Xi^{-2} |\partial^2 \delta {g}_{ab}| \ ,
\end{equation}
so that we can write
\be
\delta g \; \delta \tilde{V} \sim \Xi^{-2} |\partial \delta{g}_{ab}|^2 \ .
\ee
%that, in Fourier space, would write
%\begin{equation}\label{eq_deltaV_metricPert_Fourier}
%\delta \tilde{V} \sim \delta \tilde R \sim k^2 \delta \tilde{g} \sim k^2 \Xi^{-2} \delta g \ ,
%\end{equation}
where $k$ is the wave number.
Therefore, given different metric
perturbations, those with higher frequencies will have
a stronger impact in $\delta \tilde{V}$
and, consequently, in the QNM spectrum.
This is consistent with the findings in \cite{Jaramillo:2020tuu,Jaramillo:2021tmt}
that identify the QNM instability phenomenon as a high-frequency perturbation effect. It also
matches the observation that in linearized gravity, if one considers the Isaacson stress-energy tensor
---see \cite{Alc08}, the energy density of the perturbed gravitational field is associated to terms quadratic
in derivatives of the metric perturbation
\begin{equation}\label{eq_Isaacson_energy}
\rho_E(\delta g; x) \sim |\partial \delta g|^2  .
\end{equation}

Now, let us relate this to the energy norm of the perturbations of the operator $L$.
Let us consider a perturbation $\delta L$ associated with a perturbation
  of the potential $\tilde{V}$. In other words
\begin{equation}
\label{eq:deltaL}
\delta L = \left(
  \begin{array}{c|c}
    0 & 0 \\ \hline \delta \tilde{V} & 0
  \end{array}
  \right) \!  ,
\end{equation} whose operator norm induced by the energy norm is given (cf. equation (B.15) in \cite{Jaramillo:2020tuu}) by
\begin{equation}\label{eq:matrix_norm_def}
    || \delta L ||_{_E} = \Big(\rho\big((\delta L)^\dagger\delta L\big)\Big)^{\frac{1}{2}} = 
    \max\{\sqrt{\lambda}\; |\; \lambda \in \sigma\big((\delta L)^\dagger\delta L\big)\} \ ,
\end{equation}
  where $\dagger$ represents the adjoint respect to the inner
  product $\langle \cdot, \cdot\rangle_{_E}$.
  Although $L$ and $\delta L$,
  are in general differential operators,
  in the current heuristic assessment we will make use of discrete approximates,
  to cast the discussion into matricial terms (a proper differential operator treatment
  is needed and
  will be developed elsewhere). 
   Let   $[L] $ and  
   $[\delta L] $ denote the discrete version of $L$ and
   $\delta L$, respectively, and let $[L]$ and $ [\delta L]\in \mathbb{C}_{2n,2n}$.
   In this discrete setup, the
  scalar product can be encoded in the Gram matrix $[G^E] \in
  \mathbb{C}_{2n,2n}$ and the adjoint operator is given by $[\delta
    L]^\dagger = [G^{E}]^{-1}\cdot [\delta L]^{*}\cdot [G^E]$ where
  the expression is understood in the matricial sense and the symbol ${}^{*}$
  denotes again the standard transposition-conjugation operation for
  matrices.    
  Let the Gram matrix and its inverse be schematically
  represented as~\footnote{Observe that although in general $(G^{\textbf{i}})^{-1} \neq G_{\textbf{i}}$,
  ---here $\textbf{i}$ are label-indices
  with $\textbf{i}=A,B,C,D$--- the relation between
  $G_{\textbf{i}}$ and $G^{\textbf{i}}$ can be found explicitly under
  suitable assumptions using Schur's complement method for the treatment of block matrices ---see \cite{zha05}.
  These relations  will not be needed for the subsequent discussion.}
 \begin{eqnarray}
 \label{eq:gram}
 [G^E]^{-1} = \left(
  \begin{array}{c|c}
    G_A & G_B \\ \hline  G_C & G_D
  \end{array}
  \right) \quad  , \quad
 [G^E] = \left(
  \begin{array}{c|c}
    G^A & G^B \\ \hline  G^C & G^D
  \end{array}
  \right) \!  \ .
  \end{eqnarray}
 Using this notation, a direct calculation renders 
\begin{equation}
     [\delta L]^\dagger[\delta L] =
     \left(
\begin{array}{c|c}
  G_A[\delta \tilde{V}]^{*}G^D[\delta \tilde{V}] & 0 \\
  \hline  G_C[\delta \tilde{V}]^{*}G^D[\delta \tilde{V}] & 0
\end{array}
\right) \ .
\end{equation}
For the upcoming discussion, we note that given a block matrix $M$ of the form
\begin{equation}
\label{eq_blockMatrix}
    M:=
     \left(
\begin{array}{c|c}
   M_A & M_B \\ \hline M_C & M_D
\end{array}
\right)  ,
\end{equation}
one can exploit the Schur's complement of the matrix \eqref{eq_blockMatrix} under the 
assumption that $M_D$ is invertible ---see \cite{zha05}---  to express
the determinant of $M$ in terms of the block matrices as %and $M_{\textbf{i}}$ via
\begin{eqnarray} \label{eq_det_schur}
\det M = \det M_D \det(M_A -M_B (M_D)^{-1}M_C).
\end{eqnarray}
In order to address the estimation of the norm $||\delta L||_{_E}$ given by
equation \eqref{eq:matrix_norm_def}, we set
\begin{eqnarray}
\label{eq:Mblock_spe}
&M_A = G_A[\delta \tilde{V}]^{*}G^D[\delta \tilde{V}] -\lambda I,
\qquad &  M_B=0, \nn \\
 &M_C=G_C[\delta \tilde{V}]^{*}G^D[\delta \tilde{V}],  & M_D=-\lambda I \ ,
\end{eqnarray}
where $I$ is the identity matrix of size $n$.
Substituting then equations \eqref{eq:Mblock_spe} in 
\eqref{eq_det_schur}, one obtains
\begin{equation}
\det ([\delta L]^\dagger[\delta L]-\lambda I)
= (-\lambda)^n \det(G_A[\delta \tilde{V}]^{*}G^D[\delta \tilde{V}] -\lambda I) \ .
\end{equation}
Observe that the assumption that $M_D$ is invertible
is equivalent in the present case to $\lambda \neq 0$. Thus, 
as long as $0 \notin \sigma([\delta L]^\dagger[\delta L])$
one has that
\begin{equation}\label{eq_deltaL_general_gram}
\sigma([\delta L]^\dagger[\delta L])
= \sigma (G_A[\delta \tilde{V}]^{*}G^D[\delta \tilde{V}]) \ .
\end{equation}
The discussion up to this point has been independent of
the form of the matrix $[\delta \tilde{V}]$ and the chosen Gram matrix,
with the only assumptions $0 \notin \sigma([\delta L]^\dagger[\delta L])$.
We impose now $[\delta \tilde{V}]$
to be a diagonal matrix, as it corresponds to a perturbation of the potential $\tilde{V}$, namely
\begin{equation} \label{eq_deltaV_diag}
  [\delta \tilde{V}] = \text{diag} (\delta \tilde{V}_{1},...,\delta \tilde{V}_{n}) \ ,
\end{equation}
with $\delta \tilde{V}_{i}, i\in\{1, \ldots, n\}$ the discretized values of the
function $\delta\tilde{V}$ perturbing $\tilde{V}$.

Before applying the previous discussion to the Gram matrix
associated with the energy scalar inner product \eqref{eq_eff_en_inner_product},
it is illustrative
to consider the simpler case of the standard $L^2$-scalar product for
which the Gram matrix reduces to the identity; $[G_{_2}]=I_{2n}$.
In other words, $G_A=G^A=G_D=G^D=I_n$
and $G_B=G^B=G_C=G^C=0$. Then,
using equation \eqref{eq_deltaL_general_gram}
along with \eqref{eq_deltaV_diag} and
\eqref{eq:matrix_norm_def} one concludes
that
\begin{equation}\label{eq_deltaL_id}
|| \delta L ||_{_2} = \max\big\{|\delta \tilde{V}_{i}|, i\in\{1, \ldots, n\}\big\} \ ,
\end{equation}
that is, $|| \delta L ||_{_2}$ is given by the largest (in absolute value) of the
discretised values of the potential perturbation $\delta\tilde{V}$. This 
discretised reasoning leads to the point-like expression 
\bea ||\delta L||_{_E} = \max_{\mathrm{supp}(\delta \tilde{V})}{|\delta
  \tilde{V}|} \ ,
\eea
for the norm of $||\delta L||_{_E}$,
where $\mathrm{supp}(\delta \tilde{V})$ is the support of the perturbation $\delta \tilde{V}$.
We conclude that large perturbations of the $L$ operator are associated with ``peaked'' distributions
of the perturbation of the potential. This conclusion, with ``modulations'' in the case of the
energy norm (see below), is the main message to retain from this discussion.

For the energy scalar product,
observe that equation \eqref{eq_eff_en_inner_product}
can be written as
\begin{eqnarray}
\langle u_1, u_2\rangle_{_E}
=
\int_{a}^{b}
 \begin{pmatrix} \bar{\phi}_1, \bar{\psi}_1 \end{pmatrix}
 \begin{pmatrix}
 \begin{array}{c|c} -\partial_x( p\; \partial_x) + \tilde{V}  & 0 \\
   \hline
   0 & w \end{array}
 \end{pmatrix} 
 \begin{pmatrix} \phi_2 \\ \psi_2 \end{pmatrix}
 dx \ .
\nonumber
\end{eqnarray}
where $u=(\phi,\psi)$, we have used integration by parts and
exploited that $p(a)=p(b)=0$.
Thus, one can make the following identifications
 $G_B=G^B=G_C=G^C=0$ and
\begin{align}\label{eq_energy_gram_matrix}
 & G^D = w I, && G^A = -\partial_x( p\; \partial_x) + \tilde{V}, \nonumber \\
 & G_D = w^{-1}I, && G_A = (-\partial_x( p\; \partial_x) + \tilde{V})^{-1} \ .
\end{align}
Hence using equations \eqref{eq_deltaL_general_gram}, \eqref{eq_deltaV_diag},
and  \eqref{eq_energy_gram_matrix} renders 
\begin{equation}\label{eq_deltaL_Energy_Gram}
\sigma([\delta L]^\dagger[\delta L])
= \sigma ((-\partial_x( p\; \partial_x) + \tilde{V})^{-1} w|\delta \tilde{V}|^2) \ .
\end{equation}
Using definition of the operator $L_1$ in equations \eqref{e:L_1-L_2_intro} (or \eqref{e:L1L2def}
taking into account the notation $q=\tilde{V}$),
one has that $L_1^{-1}= (\partial_x( p\; \partial_x) - \tilde{V})^{-1}w $
and thus, one can rewrite equation \eqref{eq_deltaL_Energy_Gram} as
\begin{equation}\label{eq_deltaL_Energy_Gram_improved}
\sigma([\delta L]^\dagger[\delta L])
= \sigma (-L_1^{-1}|\delta \tilde{V}|^2) \ .
\end{equation}
This last expression can be read as a relation between the spectrum of
operators rather than matrices.
Thus, using expression \eqref{eq:matrix_norm_def}, we have
\begin{equation}\label{eq:matrix_norm_L1deltaV}
|| \delta L||_{_E}^2 = \max \{|\lambda|, \lambda\in \sigma (L_1^{-1}|\delta \tilde{V}|^2)\}.
\end{equation}
An estimate of this quantity, appropriate in the setting of the present
heuristic discussion, is given by the notion of ``numerical
abscisa'' $\omega(A)$ of an operator $A$, defined as the supremum of the real
part of its numerical range $W(A)$ and that
can be characterised \cite{trefethen2005spectra} as
\be
\omega(A) = \sup \sigma\Big(\frac{1}{2}(A+A^\dagger)\Big) \ .
\ee
Since $\frac{1}{2}(A+A^\dagger)$ is selfadjoint, we can use Rayleigh-Ritz theorem (e.g. \cite{Zet11,Berger03})
to characterize~\footnote{We could have just written this from the starting definition
  of the numerical abscissa $\omega(A)$ in terms if the numerical range of $A$, namely
  $W(A) = \{ \langle u, A u \rangle, ||u||=1 \}$, but we have preferred to pass
through the Rayleigh-Ritz, familiar for physicists in the quantum setting.
See \cite{trefethen2005spectra} for details and properties of numerical range and abscissa.
}
\begin{equation}
\omega(A) = \sup_{||u||=1} \{ \langle u, A u \rangle \} \ .
\end{equation}
In our case,  although $L_1$ ---and
hence also $L_1^{-1}$--- and $|\delta \tilde{V}|^2$ are selfadjoint
operators, the composite operator $L_1^{-1}|\delta \tilde{V}|^2$ is
not, since
\begin{equation}\label{eq_L1deltaVDagger}
 (L_1^{-1}|\delta \tilde{V}|^2)^\dagger =   (|\delta \tilde{V}|^2)^\dagger (L_1^{-1})^\dagger   = 
|\delta \tilde{V}|^2 L_1^{-1} \neq L_1^{-1} |\delta \tilde{V}|^2 \ .
\end{equation}
We can put together the previous discussion to estimate $|| \delta L||_{_E}^2$
in equation \eqref{eq:matrix_norm_L1deltaV} as
\bea
\label{eq_deltaL_Energy_ToPinv}
|| \delta L||_{_E}^2 &\sim& \sup \sigma\Big(\frac{1}{2}\big(L_1^{-1}|\delta \tilde{V}|^2
+(L_1^{-1}|\delta \tilde{V}|^2\big)^\dagger)\Big) \nn \\
&=& \sup_{||u||=1} \Big\{ \langle u, \frac{1}{2}\big(L_1^{-1}|\delta \tilde{V}|^2
+  |\delta \tilde{V}|^2L_1^{-1}|\delta \tilde{V}|^2\big)u \rangle \Big\}
\eea
%For the subsequent discussion we will use the usual bra-ket notation
%as the calculations that follow are clearer in this format.
%Using this notation, equation \eqref{eq:matrix_norm_L1deltaV} and
%expression \eqref{eq:estimation_with_selfadjoint_version} yield
%\begin{equation}\label{eq_deltaL_Energy_ToPinv}
%|| \delta L ||_{E}^2  \sim \max_{||u||=1}\{ \langle u| Q | u\rangle\},
%\end{equation}
%where $Q= \big(\frac{1}{2} L_1^{-1}|\delta \tilde{V}|^2
%+ \frac{1}{2}  (L_1^{-1}|\delta \tilde{V}|^2)^\dagger \big)$.
Exploiting the fact that  $L_1$ is a Sturm-Liouville operator,
namely, selfadjoint in the space $(L^2,w(x)dx)$,
one can express its inverse (adopting a usual bra-ket notation) as
\be
%\begin{eqnarray}
  \label{eq_PinvExpansion} L_{1}^{-1}=
\sum_{n}\frac{1}{\lambda_n}|\ell_n\rangle \langle \ell_n| \ ,
%\end{eqnarray}
\ee
where $\lambda^1_n$, and $| \ell_n\rangle$ denote now respectively, the eigenvalues
and normalized eigenvectors of $L_1$, using that the former constitute
an orthonormal basis for the function space where $L_1$ is defined.
In order to justify this expression, in  this discussion $L_1$ is thought
as an operator defined on the compact, 3-dimensional manifold corresponding to the
compactified hyperboloid $\Sigma$ with regularity imposed on
the corresponding cut of null infinity $\Sigma \cap
\scri^{+}$. This guarantees the discrete character of the spectrum and,
for generic potentials, the trivial kernel of $L_1$ and therefore
the absence of zero eigenvalues. Therefore expression \eqref{eq_PinvExpansion}
makes sense.

Similarly, % to equation \eqref{eq_PinvExpansion},
for the operator $\delta
\tilde{V}$ we can write \begin{equation}\label{eq_deltaVExpansion} |\delta
\tilde{V}|^2 = \int dx |\delta \tilde{V}|^2|x\rangle \langle x| \ , \end{equation}
where $| x \rangle$ denotes the  eigenvectors
of the position operator $X$, namely $X|x\rangle = x |x \rangle$.
Using  expansions
\eqref{eq_PinvExpansion} and \eqref{eq_deltaVExpansion}
to estimate the right hand side of equation \eqref{eq_deltaL_Energy_ToPinv},
we get
\begin{flalign}\label{eq_uQu}
2|| \delta L||_{_E}^2 & \sim
\langle u| \sum_{n}\frac{1}{\lambda_n}
\ell_n\rangle \langle \ell_n|\int dx |\delta
V|^2|x\rangle \langle x|  u\rangle  + 
\langle u|\int dx |\delta
V|^2|x\rangle \langle x| \sum_{n}\frac{1}{\lambda_n}
\ell_n\rangle \langle \ell_n|  u\rangle \nonumber \\
 \lesssim  &
\frac{\sup{|\delta \tilde{V}|^2}}{\min{\lambda_n}}  \langle u| \sum_{n}
\ell_n\rangle \langle \ell_n|\int dx
|x\rangle \langle x|  u\rangle 
+ 
\frac{\sup{|\delta \tilde{V}|^2}}{\min{\lambda_n}}
\langle u|\int dx |x\rangle \langle x| \sum_{n}
\ell_n\rangle \langle \ell_n|  u\rangle \nonumber
 \\ & = 2\frac{\sup{|\delta \tilde{V}|^2}}{\inf{\lambda_n}}\langle u| u\rangle 
= 2\frac{\sup{|\delta \tilde{V}|^2}}{\inf{\lambda_n}} \ ,
\end{flalign}
where we have used the expression of the identity operator in the Hilbert
bases $\{|\ell_n\rangle\}$ and $\{|x\rangle\}$ and, in the last line, the normalization
$||u||=1$. Finally, using the order of magnitude estimate 
$\inf{\lambda_n} \sim \max_{\mathrm{supp}(\tilde{V})}\tilde{V}>0$, one concludes
that~\footnote{Note that the expression is consistent  from the perspective of
  physical dimensions, since the operator $L$ (and therefore $\delta L$) has the physical
dimensions of a frequency, as follows from equation \eqref{e:intro_L_eigen}.}
\begin{equation}\label{eq_deltaL_energy_heuristic}
  || \delta L ||_{E} \sim \alpha \max_{\mathrm{supp}(\delta \tilde{V})}|\delta \tilde{V}| \quad ,
  \quad \alpha =  \frac{1}{\displaystyle\Big(\max_{\mathrm{supp}(\tilde{V})}\tilde{V}\Big)^{\frac{1}{2}}} \ .
\end{equation}
Contrast this result with the estimation in equation \eqref{eq_deltaL_id} from
the standard $L^2$ scalar product: we still have an estimation such that large
norms $|| \delta L ||_{_E}$ correspond to ``peaked'' distributions of the perturbation potential
$\delta\tilde{V}$, but the estimation is not really pointwise, but rather non-local,
since maxima of $\delta \tilde{V}$
and $\tilde{V}$ occur generically at different locations.

Finally, to make contact with spacetime perturbations $\delta g_{ab}$,
we consider $\epsilon$-sized perturbations $\delta L$ (i.e. $|| \delta L ||_{_E}=\epsilon$),
so  putting together \eqref{eq_deltaL_energy_heuristic} with the estimation \eqref{eq_deltaV_metricPert}
for $\delta \tilde{V}$
\begin{equation}
\epsilon \sim \alpha  \max_{\mathrm{supp}(\delta g)} |\partial^{2}(\delta g_{ab})| \ ,
\end{equation}
that,  combined with \eqref{eq_Isaacson_energy}, relates $\epsilon$ to
the density of energy of the spacetime perturbation
\begin{equation}
  \epsilon \; \delta g \sim \alpha \max_{\mathrm{supp}(\delta g)}  |\partial (\delta g_{ab})|^2 \sim
  \alpha \max_{\mathrm{supp}(\delta g)} \rho_E(\delta g; x) \ ,
\end{equation}
Therefore,  although modulated by a factor $\alpha$ ---consequence of
the energy-inner product $\langle \cdot, \cdot \rangle_{_E}$ and controlled
by the ``curvature maximum'' of the non-perturbed geometry--- spacetime perturbations
involving larger $\epsilon$ perturbations of the operator $L$ (and therefore triggering stronger
QNM instabilities of the overtones), are associated with strong peaks
in the ``energy distribution'' of the perturbation,
rather than to the  integrated energy distribution: to efficiently perturb with a
given available energy, use peaked distributions of perturbations.

\subsection{High-frequency limit of general relativity and spacetime regularity: Burnett's conjecture
and QNM high-frequency perturbations}
\label{s:Burnett}
We have closed section \ref{s:Cp_perturbations} and the discussion of the
emerging picture of BH QNM instability in section \ref{s:emerging picture}
by posing the question about the possible universality 
of BH perturbed QNM branches asymptotics, in particular the relation to pseudospectra
boundaries, when the perturbation frequency
tends to infinity. In this context, the precedent section
\ref{sec:physical_interpretation_energy} provides an additional key ingredient, namely
the fact that perturbations with a ``spiked structure'' are more efficient
in triggering BH QNM overtone instabilities. This points towards a
low-regularity phenomenon underlying the genericity of perturbed BH QNM asymptotics.

In addition, as commented in section \ref{s:emerging picture},
perturbed BH QNM branches resulting from
high-frequency perturbations share some qualitative features with QNMs
of matter compact objects, namely with so-called $w$-modes or curvature modes
of neutron stars. In this
setting relating matter and high-frequency vacuum  spacetimes,
it may be of interest to bring attention to the so-called
Burnett's conjecture \cite{Burne89}, namely stating that in the limit
of infinite high-frequency a vacuum spacetime can be understood in
effective terms as a spacetime filled with matter, namely with a
stress-energy tensor given by Vlasov massless matter.

The high-frequency limit of general relativity has been the subject of
systematic studies \cite{Isaac68a,Isaac68b,Choqu69,Penro69b,MacCalTau73}, in particular aiming at
elucidating the properties of rapidly oscillating gravitational wave spacetimes. Burnett's
approach considers a sequence of vacuum spacetimes $\{(M, g_n)\}_{n=1}^\infty$ (we follow the
presentation in \cite{Luk:2020pyn}) 
satisfying $R_{ab}=0$, (weakly) converging to $g_\infty$ in the limit
$n\to\infty$. However, the resulting metric
$g_\infty=\lim_{n\to\infty} g_n$ is generically not a vacuum
spacetime, but satisfies $R_{ab}-\frac{1}{2}R g_{ab}=8\pi T_{ab}$ for
an appropriate stress-energy tensor.  For this result, specific
conditions are required on $\{(M, g_n)\}_{n=1}^\infty$ that encode the
desired properties of the high-frequency limit, in particular the
vanishing of perturbations amplitudes as the frequency goes to
infinity (see \cite{Burne89} for details).  Specifically, it assumes the
existence of constants $C$ and $\lambda_n$ (with $\lim_{n\to\infty}
\lambda_n = 0$) such that
\begin{eqnarray}
\label{e:conditions_Burnett}
|g_n - g_\infty|\leq \lambda_n \ \ , \ \ |\partial g_n|\leq C \ \ ,
\ \ |\partial^2 g_n|\leq C\lambda_n^{-1} \ .  \end{eqnarray} Under these
conditions, and building on a theorem on the convergence of the
sequence $\{(M, g_n)\}_{n=1}^\infty$, Burnett conjectures that the
limiting metric $g_\infty$ is isometric to a solution of the massless
Vlasov-Einstein system, namely
\begin{equation} G_{ab} = 8\pi T_{ab} \ \ ,
\ \ \nabla_aT^{ab}= 0 \ , \end{equation} with \begin{equation} T_{ab} = \int a^2(x,k)
k_ak_b dV_k \ \ , \ \ k^c\nabla_c a^2(x,k) = 0 \ ,
\end{equation}
where $a^2$ is a ``particle'' distribution function in the cotangent
bundle $T^*M$, with $k_a$ constrained to be null ($k^ak_a=0$) and, therefore, the
integration ---with measure $dV_k$--- is performed on the null cone in the cotangent space $T_x^*M$
(at each $x\in M$).

Although no proof is available in the fully general case, Burnett's
conjecture has been proved under the further condition of $\mathbb{U}(1)$ symmetry
\cite{Huneau:2019jcx} (see also \cite{Huneau:2017led,Huneau:2017tfc}).
Burnett's conjecture captures the idea that a vacuum spacetime, under
(gravitational wave) high-frequency perturbations, behaves effectively
as a matter system, a tantalizing picture when comparing perturbed BH
QNM overtones with $w$-modes in matter compact stars.

Although Burnett's conditions (\ref{e:conditions_Burnett})
allow for infinitely rapid oscillations, they do not permit
``concentration'' of the oscillations.  However, as seen in section \ref{sec:physical_interpretation_energy},
concentration of perturbations is a crucial element in the analysis of
BH QNM perturbations. Such capability to deal with concentrations
(in addition to oscillations) requires the relaxation of conditions
(\ref{e:conditions_Burnett}). This has been addressed by Luk \&
Rodnianski in their study of ``null shells'' in the more general setting of
low-regularity problems of Einstein equations
\cite{Luk:2020pyn} (cf. also \cite{LukRod15,LukRod17} and references therein).  In particular, by
relaxing convergence conditions of $\{(M, g_n)\}_{n=1}^\infty$ to \begin{equation}
\label{e:conditions_Luk-Rodnianski}
g_n\to g_\infty \ \ \hbox{in } C^0 \ \ , \ \ \partial g_n\to \partial
g_\infty \ \ \hbox{in } L^2 \ , \end{equation}
Luk \& Rodnianski have proven \cite{LukRod17} that the limiting metric is
$C^0\cap H^1$ (with $H^1=W^{1,2}$ the Sobolev space controlling the energy,
precisely the kind of norm in our QNM instability assessment)
and, appropriately formulated, it satisfies Einstein equation \cite{Luk:2020pyn}
in an Einstein-null dust system. This provides a resolution of a relaxed
version of Burnett's conjecture.

The key point for the assessment of BH QNM asymptotics is that the
limit of infinite high-frequency oscillations allowing
from ``perturbation concentrations'' is a ``low-regularity'' limit, in particular
the limiting metric is generically in the $C^0$ class. Putting this together with our discussion
of Regge QNMs for $C^p$ potentials in section \ref{s:Cp_perturbations}, the involved
low-regularity limit directly leads to refining 
the conjecture in \cite{Jaramillo:2021tmt} about the logarithmic asymptotics of BH QNM branches in the
infinite high-frequency limit under generic ultraviolet oscillations.

\subsubsection{Ultraviolet QNM perturbations: Regge QNM branches conjecture.}
\label{e:ReggeQNM_conjecture}
Let us recall the behaviour of perturbed BH QNM overtones under perturbations
of increasing frequency. As described in section \ref{s:emerging picture},
increasing the frequency of the perturbations push the
perturbed BH QNMs towards the logarithm pseudospectrum lines.
Such perturbed QNMs stay strictly above the logarithmic lines
as long as the frequency is high but finite: there is no guarantee that
the logarithmic lines are attained in the high-frequency limit and, in case
QNMs indeed reach the pseudospectrum boundary in this a limit,
no information about the specific distribution of QNMs along such logarithmic lines
is available. On the the other hand, for the instance
of ``infinite frequency'' provided by $C^p$
potentials, the resulting Regge QNM branches are indeed logarithmic
branches with a very specific distribution of QNMs given by expressions \eqref{e:Regge_branches}.

Putting together the $C^p$ potential case with Luk \& Rodnianski version
of Burnett's conjecture allowing for oscillations and concentration,
it seems natural to expect that Regge QNMs are not just an instance
but they describe the generic infinite high-frequency limit of {\em all}
perturbations. Specifically, we propose the following Regge QNM conjecture:

{\em In the limit of infinite frequency, generic ultraviolet perturbations
  push BH QNMs in Nollert-Price branches to asymptotically logarithmic branches
  along the boundary of the QNM-free region, following precisely
  the Regge QNM  asymptotic pattern in (\ref{e:Regge_branches}).
}

\subsection{Towards a geometric description of QNMs}
\label{s:geometric_approach}
Following \cite{Jaramillo:2020tuu}, our discussion of BH QNM instability explicitly relies on a formulation of the hyperboloidal
approach to QNMs  depending on an appropriate choice of coordinates, in particular tied to 
spherical symmetry.  However, the discussions in the previous sections strongly suggest the naturality of 
promoting the whole discussion to a genuine geometric formulation, in arbitrary dimensions
and to generic spacetimes without symmetries~\footnote{We acknowledge A. Ashtekar for stressing 
the relevance and need of addressing this geometric aspect.}.  
This section is simply meant as a heuristic exploration, putting
in order some of the elements to be incorporated in a systematic geometric formulation to
be discussed in a future work~\cite{GasJarMac22}.

Arguably, the only crucial features in the construction are i) that the spacetime
foliation intersects null-infinity  and the BH horizon at, respectively,
cuts $\mathcal{S}^\infty_\tau$ and ${\cal S}^{\cal H}_\tau$
and ii) that the key
analytic structures admit a lifting to a geometric set up, where the coordinates
and  particular foliation do not play a fundamental role.
Methodologically, we focus on a hyperboloidal approach, even if this is not the only
option~\footnote{Null slicings of spacetime could also be considered here. This will be addressed
elsewhere.}. We list the main steps in the construction:
\begin{itemize}
\item[1.] {\em Compactified hyperboloidal slicing: geometric data}. Spacetime is described in
  a compactified hyperboloidal slicing $\{\Sigma_\tau\}$, where slices are compactified
  manifolds with boundary. Boundaries  ${\cal S}^\infty_\tau$ correspond to the
  intersection of $\Sigma_\tau$ with future null infinity (together with horizon slices ${\cal S}^{\cal H}_\tau$
  in BH spacetimes). Data are given by: i) the slicings  $\{\Sigma_\tau\}$, $\{\cal S^\infty_\tau\}$
  (and $\{\cal S^{\cal H}_\tau\}$ in BH spacetimes), ii) the intrinsic and extrinsic geometries
  of $\Sigma_\tau$ (respectively $\gamma_{ab}$ and $K_{ab}$, possibly including an effective potential $V$),
  iii) a choice of transverse null normals $k^a$ at
  $\scri^+$, normal to $\{\cal S\}^\infty_\tau$ and the null generator $\ell^a$ at $\scri^+$
  (analogously at the horizon $\cal H$),
  iv) a function $\gamma$ on  $\{\cal S\}^\infty_\tau$
  (similarly $\{\cal S\}^{\cal H}_\tau$),
  and iv) an integration measure $d\mu = w(x) d\Sigma_\tau$ on $\Sigma_\tau$.
  Constraints exist among these data.

\item[2.] {\em Geometric formulation of dynamics and scalar
  product}. A first order formulation of the dynamics is formulated in
  terms of an evolution operator $L$ constructed from operators $L_1$
  and $L_2$ on $\Sigma_\tau$.  $L_1$ is elliptic and depending
  only on the intrinsic geometry (and on $V$) of $\Sigma_\tau$, whereas $L_2$ is a first order
  operator depending on the extrinsic geometry.  A scalar product is
  built from the data, $L_2$ accounting for the loss of
  selfadjointness.

\item[3.] {\em A geometric approach to QNM
  characterization}. QNMs are proposed to arise from an
  eigenvalue problem for $L$. Outgoing boundary conditions are encoded
  in $L_2$ and a geometric condition of regularity for
  eigenfunctions, namely enforcing $L_1$ to ``live'' on a smooth
  closed (compact without boundary) version $\mathring{\Sigma}_\tau$
  of $\Sigma_\tau$.  In particular, smoothness of
  $\mathring{\Sigma}_\tau$ would entail the elimination of the
  continuous (``branch cut'') spectrum of $L$.
  
\item[4.] {\em Towards a notion of normal modes for isolated systems:
  additional degrees of freedom and BMS symmetry}.  Lack of
  selfadjointness in our scattering problem follows from the flow of
  degrees of freedom through null infinity (and the
  horizon). Restituting selfadjointness requires enlarging the system
  to account for the loss of such degrees of freedom at the
  boundaries, so that the full system is conservative.  We claim that
  such degrees of freedom can be encoded in terms of the asymptotic
  (dynamical) BMS symmetry at null infinity (and an analogous
  structure at the horizon). Resulting eigenvalues would provide a
  notion of (real) normal modes of isolated object spacetimes.

\end{itemize}
In the rest of this section, we provide some further details in the scheme above,
having the scattering of a scalar field as the main reference problem in mind.

\subsubsection{Compactified hyperboloidal slicing: geometric data.}
Let us start with a stationary (with Killing vector $t^a$) asymptotically simple physical spacetime $(\mathcal{M},g)$,
with null infinity $\scri^+$ and a horizon ${\cal H}$, in case of dealing
with a BH spacetime. Topology of $\scri^+$ and ${\cal H}$ are $\mathbb{S}^2\times\mathbb{R}$.
Let us consider a conformal extension of the
physical spacetime $(\tilde{\mathcal{M}},\tilde{g})$.
Let us denote by $\ell^a$ the null generators along $\scri^+$ and ${\cal H}$.
We proceed then in the following stages:

\begin{itemize}

\item[i)] Start by choosing  a slicing $\{\cal S^\infty_\tau\}$ of $\scri^+$.
  Given the generator $\ell^a$ along $\scri^+$ consider, at each
  slice $\cal S^\infty_\tau$, the only null normal $k^a$ to $\cal S^\infty_\tau$
  satisfying $\ell^a k_a = -1$.
  In a BH spacetime, repeat the procedure at the BH (Killing) horizon ${\cal H}$, i.e.
  choose a slicing  $\{\cal S^{\cal H}_\tau\}$  fixing
  the associated transverse  $k^a$ null normal, $\ell^a k_a = -1$, at horizon sections $\cal S^{\cal H}_\tau$.
  
\item[ii)] Choose a function $\gamma$ on null infinity sections $\cal S^\infty_\tau$ (respectively, also a function $\gamma$ on
  $\cal S^{\cal H}_\tau$ in BH spacetimes) and consider the null vector $\gamma^a = \gamma k^a$, outgoing from $\mathcal{M}$.

\item[iii)] Extend the vector $\gamma^a$ arbitrarily to the bulk, subject to the following
  constraints: a) $\gamma^a$ is tangent to a spacelike slicing $\{\Sigma_\tau\}$ of  ${\cal M}$
  adapted to stationary (namely, the Killing vector $t^a$ transports slices of $\{\Sigma_\tau\}$
  into slices of $\{\Sigma_\tau\}$),
  b) $\gamma^a$ is spacelike everywhere, except at the boundaries, where it is null.
  The slicing $\{\Sigma_\tau\}$ is a hyperboloidal one.

\item[iv] Slices  $\Sigma_\tau$ in $\tilde{\mathcal{M}}$ are compact manifolds with boundary,
  with boundaries given by spheres. Let us consider the intrinsic and extrinsic geometry of $\Sigma_\tau$,
  namely the induced metric $\gamma_{ab}$
  and the extrinsic curvature $K_{ab}={\cal L}_n \gamma_{ab}$, with $n^a$ the unit timelike normal to $\Sigma_\tau$.
  Let us introduce a measure $d\mu = wd\Sigma_\tau$, with $w$ a weight
  function on $\Sigma_\tau.$

\item[v)] The resulting data~\footnote{These data are redundant,
  namely they admit constraints. They ultimately should be reduced
  just to the choice of slicings of the boundaries smoothly extending to
  the bulk in a hyperboloidal slicing.  Given the slicings $\{\cal
  S^\infty_\tau\}$ of $\scri^+$ and $\{\cal S^{\cal H}_\tau\}$ of
  ${\cal H}$, extend and join them by a a hyperboloidal slicing
  $\{\Sigma_\tau\}$.  Denote the induced metric on $\Sigma_\tau$ by
  $\gamma_{ab}$ and the extrinsic metric by $K_{ab}={\cal L}_n
  \gamma_{ab}$. Then, defining $\gamma = K_{ab}\ell^ak^a$ at the
  boundaries, consider the outgoing null vector at the boundaries
  $\gamma^a = \gamma k^a$. Considering the timelike vector $t^a$, we
  can decompose it as $t^a = N n^a + \beta^a$, in terms of lapse
  function $N$ and shift vector $\beta^a$. Then $K_{ab} =
  \frac{1}{2N}(\partial_t\gamma_{ab} + D_a\beta_b + D_b\beta_a)=
  \frac{1}{2N}(D_a\beta_b + D_b\beta_a)$.  Also, note $\sqrt{g}d^n x =
  N d\tau d\Sigma_t$.}  in the problem are:
  $(\{\Sigma_\tau\},  \gamma_{ab}, K_{ab}, w; \{\cal S^\infty_\tau\}, \{\cal S^{\cal   H}_\tau\},\gamma)$.
  
\end{itemize}

\subsubsection{Geometric formulation of dynamics and scalar product.}
One of the most interesting results in \cite{Jaramillo:2020tuu}, is
the identification of analytic structure of operators $L_1$ and $L_2$,
building the evolution generator $L$, as appearing in equation
(\ref{e:L_1-L_2_intro}). In particular, the singular Sturm-Liouville
structure of $L_1$ is closely related to a Laplacian-like operator,
whereas the structure of $L_2$ also calls for a natural geometric
generalization in terms of a vector field $\gamma^a$.

For concreteness, denoting by $\phi$ the degrees of
freedom whose hyperbolic dynamics we are considering, we write
their first-order (in time) evolution in terms of %the geometric
data above
as
\begin{eqnarray}
\label{e:wave_eq_1storder_gen}
\partial_\tau u  = i L  u \ \ , \ \  u =
\begin{pmatrix}
  \phi \\
  \psi = \partial_\tau \phi
\end{pmatrix}  \ \ , \ \ L =\frac{1}{i}\!
\left(
  \begin{array}{c|c}
    0 & 1 \\
    \hline 
   L_1 & L_2
  \end{array}
  \right) \ ,
\end{eqnarray}
where
\begin{equation}
\label{e:L_1-L_2}
L_1 = \frac{1}{w}\big(\tilde{\Delta}- \tilde{V}(x)\big), \qquad
L_2 = \frac{1}{w}\big(2\gamma\cdot \tilde{D} + \tilde{D}\cdot \gamma\big) \ ,
\end{equation}
Here $L_1$ and $L_2$ are operators living on $\Sigma_\tau$ (with
boundaries), $\tilde{\Delta}$ denotes the Laplacian with respect to
the compactified metric $\tilde{\gamma}_{ab}$, namely
$\tilde{\Delta}=\tilde{\gamma}^{ab}\tilde{D}_a\tilde{D}_b$, and
$\tilde{V}$ is a possible effective potential, depending on the nature
of the dynamical field $\phi$.

The scalar product is then proposed to be
\begin{eqnarray}
\label{e:mode-energy_gen}
   \hspace{-1mm}\langle u_1, u_2\rangle_{_E}  \hspace{-0.5mm}=\hspace{-0.5mm}
  \frac{1}{2} \hspace{-1mm}\int_{\Sigma_\tau}\hspace{-2mm}
  (w\bar{\psi}_1 \psi_2 \hspace{-0.5mm}+\hspace{-0.5mm}
  \tilde{D}\bar{\phi}_1\hspace{-0.5mm}\cdot\hspace{-0.5mm}\tilde{D}\phi_2 \hspace{-0.5mm}+\hspace{-0.5mm}
  \tilde{V} \bar{\phi}_1\phi_2\hspace{-0.5mm}) d\tilde{\Sigma} \ ,
\end{eqnarray}
where the $\cdot$ scalar product is calculated with respect to $\tilde{\gamma}_{ab}$ and
$d\tilde{\Sigma}$ denotes the volume element in ($\Sigma_\tau, \tilde{\gamma}_{ab}$).
Then, the formal adjoint operator would be
\begin{equation}
\label{e:formal_adjoint_pert}
L^\dagger = L + L^{\partial} \ \ , \ \ 
L^{\partial} =\frac{1}{i}\!
\left(
  \begin{array}{c|c}
    0 & 0 \\
    \hline 
   0 & L^\partial_2
  \end{array}
  \right) \ ,
  \end{equation}
  with $L^\partial_2$ given by the expression
  \begin{equation}
  \label{e:L2_boundary}
  L^\partial_2 = -2\frac{\gamma^a \ell_a}{w}\left(\delta_{S^{\cal H}}-\delta_{S^\infty}\right) =
  -2\frac{(\gamma \ell^a)k_a}{w}\left(\delta_{S^{\cal H}}-\delta_{S^\infty}\right)
  \end{equation}
  where $\gamma^a \ell_a = \gamma k^a \ell_a = - \gamma$.
  The reason of rewriting it as $(\gamma \ell^a)k_a$,
  with $\ell^a$ the generator of the null boundaries, will be apparent
  below.

  In sum, expressions in this subsection provide a (purely formal) geometric extension
  of the relevant equations in \cite{Jaramillo:2020tuu}, upon the
  choice of the boundary foliations.
  
  \subsubsection{A geometric approach to QNM characterization.}
  Taking Fourier transform in time of equation (\ref{e:wave_eq_1storder_gen}), we obtain
  the eigenvalue problem characterization QNMs. Outgoing boundary conditions
  are enforced by the geometric frame, given that the boundaries are null
  hypersurfaces and the light cones point outwards. However, such boundary
  conditions must be supplemented with appropriate regularity conditions
  on the eigenfunctions. This defines a challenging  problem in analysis, namely
  the identification of the appropriate functional space of eigenfunctions
  \cite{Warnick:2013hba,Gajic:2019oem,Gajic:2019qdd,galkowski2020outgoing}.
  In particular, the treatment in \cite{Jaramillo:2020tuu} suffers from the severe
  problem that the spectrum of $L$ contains a continuous part.

  The approach we propose here is to {\em define} QNMs in a
  ``radically'' geometric manner, by imposing $L_1$ to be
  defined/regularized in a closed manifold
  $\mathring{\Sigma}_\tau$. This would be obtained by:
  \begin{itemize}
  \item[i)] First, shrinking the ${\mathbb S}^2$ boundaries  of $\Sigma_\tau$ to
    respective points $i_\infty$ and $i_{\cal S}$. The manifold
    $\Sigma_\tau$ is then a 3-sphere $\mathbb{S}^3$ with two
    pinched-holes corresponding to its boundaries.

  \item[ii)] Second, the operator $L_1$ is regularized to an
    operator $\mathring{L}_1$, so that it is well defined on a full
    smooth $3$-sphere $\mathbb{S}^3$ obtained by adding ``two points'' to complete
    $\Sigma_\tau$.
  \end{itemize}
  Certainly, problems will appear in general when trying to obtain a regular $\mathring{L}_1$.
  For instance, whereas the problem seems straightforward in the P\"oschl-Teller case,
  in Schwarzschild one should expect obstructions due to the non-vanishing mass, leading
  to the existing branches. Regularizing the operator on a smooth $3$-sphere can then
  be seen as a tantamount of ``cutting'' the continuous part of the spectrum, leaving only
  eigenvalues representing the ``geometric counterpart'' of the identification of the
  proper functional space in the analytical approach. Assuming this can
  be achieved, QNMs are defined from the (pure)-eigenvalue problem
   \begin{equation}
  \left(
  \begin{array}{c|c}
    0 & 1 \\
    \hline 
   \mathring{L}_1 & L_2
  \end{array}
  \right)
  \begin{pmatrix}
  \phi_n \\
  \psi_n
  \end{pmatrix}
  =
  \omega_n
  \begin{pmatrix}
  \phi_n \\
  \psi_n
  \end{pmatrix}  \ ,
  \end{equation}
  where $\mathring{L}_1$ lives on a smooth $\mathbb{S}^3$, whereas $L_2$ lives on a
  ``pinched'' $\mathbb{S}^3$, with holes at   $i_\infty$ and $i_{\cal S}$
  where the outgoing ``degrees of freedom'' are encoded.
  Note that, given the absence of boundaries,
  the scalar product (\ref{e:mode-energy_gen}) can then be written in terms
  of the $\mathring{L}_1$ as
\begin{equation}
  \label{e:mode-energy_gen_L1}
   \langle u_1, u_2\rangle_{_E}  =
  \frac{1}{2} \hspace{-1mm}\int_{\Sigma_\tau}\hspace{-1mm}
  (w\bar{\psi}_1 \psi_2 - \bar{\phi}_1 \mathring{L}_1\phi_2)
  d\tilde{\Sigma}
  \end{equation}
 Note that $L_2$ plays no role in this (energy) scalar product.

\subsubsection{Missing degrees of freedom and BMS symmetry: towards normal modes in isolated objects.}
\label{s:BMS}
Non-selfadjointness in dynamical problems generically reflects 
that some degrees of freedom are missing or are being lost: the system is not isolated or is
part of a larger system, leading to non-conservative dynamics. In order
to cast such systems in terms of a selfadjoint problem, one must
``complete'' the system by adding an appropriate set of degrees of
freedom. In our setting, how could we render our BH perturbation problem
selfadjoint?

The compactified picture provides a geometric frame to address such a
question.  In our scattering case, degrees of freedom flow to infinity and
through the back hole horizon.  If we consider adding formal degrees
of freedom at future null infinity $\scri^+$ and at the BH horizon
${\cal H}$, in such a way that they account for the degrees of freedom and energy leaving
the system, then we could consider formulating the dynamics in an enlarged system
where energy is actually conserved. It would be as placing ``{\em
  Geiger} counters'' at the infinite and horizon boundaries, such that
the total number of degrees of freedom is conserved.

Being anchored to the boundary, such boundary degrees of freedom would
be non-propagating ones (in the bulk).  In order to grasp their nature and
structure, let us consider the expression of the flux in equation
(\ref{eq:totalFluxExpression_reduced}), integrating it in the boundary
cuts
\begin{equation}\label{eq:totalFluxExpression_reduced_onemode}
  F(\tau)  = \int_{{\cal S}^{\cal{H}}} \gamma |\partial_\tau\phi|^2 dS
  + \int_{{\cal S}^{\infty}} \gamma |\partial_\tau\phi|^2dS \ .
\end{equation}
This is the relevant equation in our line of thought towards the
missing degrees of freedom. Indeed, the second term in the
right-hand-side has the structure of the Bondi-Sachs flux of
energy-momentum at null infinity if: i) $\gamma$ is built from the
$\ell=0,1$ spherical harmonics and ii) we identify $\partial_\tau\phi$
as a news function ${\cal N}$ (see
e.g. \cite{Ashtekar:1981bq,Ashtekar:2014zsa}).  However, in general,
$\gamma$ will be an arbitrary function on ${\cal S}^\infty_\tau$.
In this general situation, the vector field $\xi^a$ on $\scri^+$
\begin{equation}
\label{e:xi_BMS}
\xi^a = \gamma \ell^a \ \ , \ \ {\cal L}_\ell \gamma = 0 \ ,
\end{equation}
corresponds to a supertranslation in the asymptotic BMS group (if
$\gamma$ is spanned by the $\ell=0,1$ spherical harmonics, this is then a
BMS translation, namely in the Poincar\'e group).  In the spirit of the
discussion in \cite{Ashtekar:1981bq,Ashtekar:2014zsa}, the expression (integrated in time)
corresponding to \eqref{eq:totalFluxExpression_reduced_onemode}
in the generic $\gamma$ case would be
the ``Hamiltonian'' $H_\xi$ (compare with equation (3.18) in \cite{Ashtekar:2014zsa})

\begin{equation}
  \label{e:Ham_xi}
  H_\xi = \int_{\scri^+} \left(  \gamma \partial_\tau\phi \partial_\tau\phi +  \gamma \partial_\tau\phi
  + \partial_\tau\phi \Delta_{_{{\cal S}^\infty}} \gamma\right) 
d\tau dS \ ,
\end{equation}
corresponding to the BMS supertranslation $\xi^a$ and generating symplectomorphisms in the (appropriate) phase
space of degrees of freedom that ``live on the boundary''.
  
The important point is that such expression couples the (missing)
degrees of freedom $\phi$, propagating through the bulk and flowing
away through the asymptotic boundary, with the BMS elements (namely
supertranslations $\xi^a$) living on the boundary.  Moreover, such
$\xi^a$ supertranslations appear explicitly in the term accounting for
the lack of selfadjointness, namely in equation (\ref{e:L2_boundary}), that is the
signature of missing degrees of freedom.  In the same heuristic spirit, an analogous discussion
could be developed at the event horizon (more generally, isolated horizon), in
terms of the $\mathbb{R}^+-$ MOTS-gauge symmetry described in
\cite{Jaramillo:2014oha,Jaramillo:2015twa} in the MOTS-stability
setting, and precisely presented in terms of such $\gamma \ell^a$
vectors.  In this perspective, $L_1$ is a bulk operator
accounting for the energy and therefore the scalar product, whereas the
operator $L_2$ is essentially a BMS object living on the null
boundaries.

In this setting, our proposal is that missing degrees of freedom can
be encoded in BMS supertranslation degrees of freedom:
\begin{itemize}
\item[i)] BMS supertranslations $\xi^a$ would stand as a dynamical
  symmetry, acting on (and generating) the phase space of degrees of
  freedom on the boundary (the ``clicks'' in the Geiger counter
  analogy), with $\gamma$ a degree of freedom
  living on the null boundaries.
\item[ii)] The whole geometric construction starts from the choice of
  slicing of null boundaries. This is arbitrary, but all choices
  are related by an appropriate supertranslation.
\item[iii)] The set of degrees of freedom $(\phi, \gamma)$ would be
  complete. By this we mean that, as $\phi$ flows away, degrees of
  freedom $\gamma$ are ``activated'' through a coupling controlled by
  the Hamiltonian (\ref{e:Ham_xi}), so a total energy of the type
  $E_o = E(\phi,\psi)  + H_\xi$ would be conserved.
\end{itemize}

Endowing this heuristic (and, admittedly, bold) picture with a sound
foundation will be the subject of dedicated research
\cite{GasJarMac22}.  The hope would be that, from the perspective of
the extended space of degrees of freedom $(\phi, \gamma)$, one could
gain insight into the observed instability of BH QNMs. Paraphrasing
this in terms of inverse scattering, the additional BMS degrees of
freedom could provide {\em necessary} additional data at the
boundaries (complementary to transmission and reflection amplitudes),
needed to determine the scattering potential ~\footnote{This approach
  in the spirit of ``inverse scattering'' theory, connects with the
  cross-correlation approach to strong-field spacetime dynamics in
  \cite{Jaramillo:2011re,Jaramillo:2011rf,Jaramillo:2012rr} (see also
  \cite{Prasad:2020xgr,Iozzo:2021vnq}). This also resonates with
  recent approaches in celestial holography studying bulk dynamics in
  terms of asymptotic (holographic) data (see e.g.
  \cite{raclariu2021lectures} and references therein).}.

On the other hand, if a conservative formulation of the dynamics can
be constructed in the extended space of degrees of freedom $(\phi,
\gamma)$, then a notion of global spacetime {\em normal modes}, as
eigenfunctions of the time evolution generator, could be envisaged.
This could be of interest both from a phenomenological and from a fundamental
(quantization) perspective.

\section{Conclusions and perspectives}\label{sec:Conclusions}
\label{s:conclusions}

\subsection{Conclusions}
In this work we have pushed forward the research program on BH QNM
instability proposed in \cite{Jaramillo:2020tuu}, with a focus on the
role of the scalar product in the pseudospectrum analysis and its 
specific structural aspects in the hyperboloidal approach to QNMs.

We have obtained the following main results:
\begin{itemize}
\item[i)] Derivation of the energy scalar product in the hyperboloidal scheme
  from a spacetime perspective, recovering the effective expression
  in \cite{Jaramillo:2020tuu} and extending it to incorporate more general
  boundary/asymptotic conditions. Likewise, an explicit expression for the
  energy flux across spacetime null boundaries has been derived, permitting the identification
  of the specific element in the formalism accounting for loss of selfadjointness,
  namely a function $\gamma$ living on null infinity and the horizon sections.

\item[ii)] Development of a weak formulation of the BH QNM problem, specially
  well suited to address low regularity issues in this problem. As an application,
  demonstration of the genericity of the BH QNM overtone instability as
  regards the employed numerical scheme. This is obtained by implementing a
  finite elements scheme that recovers results from the Chebyshev spectral 
  approach. This addresses one of the caveats in \cite{Jaramillo:2020tuu},
  regarding possible artifacts related to Chebyshev's derivation matrices.

\item[iii)] Derivation of QNM resonant expansions of the scattered field (e.g.
  the GW waveform) based on QNM normalizability in the hyperboloidal
  scheme and Keldysh's asymptotic expansions of the resolvent in a Hilbert (more generally,
  Banach) space. This recovers Lax-Phillips resonant expansions, adapted to the
  case of normalizable eigenfunctions and therefore permitting the identification of
  expansion coefficients. In particular, we provide explicit expressions for the expansion
  coefficients obtained through projection on left-eigenvectors, valid
  in spite of the generic loss of orthogonality of eigenfunctions, and reducing
  to the known expressions for normal modes in the selfadjoint case.

\item[iv)] More heuristically, from estimations of the norm of the operator $L$ (the
  generator of the dynamics) in terms of the energy density of spacetime perturbations,
  we find evidence that spiked (low-regularity) distributions of
  BH perturbations are more efficient in triggering
  BH QNM instabilities than smooth distributions. On the one hand, this can be of astrophysical
  interest, suggesting sharp energy/matter concentrations as the preferred candidates
  among BH environmental disturbances
  to impact on BH spectroscopy through QNM instabilities. For instance,
  distributions of compact objects around a supermassive BH
  would impact BH QNMs more than smooth gas distributions. On the other hand, at a fundamental
  physics level, an effective low-regularity of spacetime (e.g. granularity, stochasticity)
  emerging from (quantum) first principles, could enhance QNM instability.

\end{itemize}
Together with these results we have made the following
proposals/conjectures:
\begin{itemize}
\item[a)] A notion of {\em $\epsilon$-dual QNM
  expansions} for BH spectroscopy has been introduced, providing equally valid
  expressions for a GW waveform, up to errors of order
  $\epsilon$.  Specifically, one of the expansions is built on
  non-perturbed BH QNMs, whereas the other is constructed on
  ``Nollert-Price'' QNMs under perturbations of order $\epsilon$. Both
  expansions are meant to be applied to the same observed waveform 
  to provide complementary (``dual'') information on large average scales
  (non-perturbed QNMs) and small scales (perturbed QNMs).

\item[b)] On the basis of the ongoing work on this research program
  (namely
  \cite{Jaramillo:2020tuu,Jaramillo:2021tmt,Destounis:2021lum,SheJar20}
  and the present work), we have proposed a {\em general picture of the
    high-frequency BH QNM instability phenomenon}.  This provides an
  effort towards a self-consistent comprehensive framework to encompass the
  ensemble of different  related results found in the literature.

\item[c)] As a specific aspect of the latter, intimately related to the
  concentration  mechanism underlying the singularized ``spiked-perturbations''
  for BH QNM instability,
  we have refined the conjecture proposed in \cite{Jaramillo:2021tmt} on
  the limit of QNMs under high-frequency perturbations:
  {\em in the limit of infinite high-frequency,
    the asymptotics of Nollert-Price BH QNMs branches converge
    universally to Regge QNM branches}. In such ultraviolet
  limit, BH QNMs would migrate to asymptotically logarithmic pseudospectra
  contours, distributing along them in the precise pattern determined
  by Regge broad resonances. This supports the further conjecture 
  suggesting the universality of (curvature) QNMs of all compact objects,
  either high-frequency perturbed BHs or matter compact stars.

\end{itemize}

\medskip 

\subsubsection{Complementary developments in the QNM instability problem.}
Together with the results and proposals above, the present work
provides some additional detailed discussions with either pedagogical or
perspective-enlarging value.

In particular, we have illustrated the effect of the choice of the
scalar product in eigenvalue instability, by using an elliptic one-dimensional
second-order differential operator with
constant coefficients. This operator provides a
rich illustration of the different subtle issues concerning spectral
instability. If use is made of the standard $L^2$ scalar product, the operator
is non-selfadjoint but it is still ``formally normal''.  However,
spectral stability is not realized since the operator is actually
non-normal, but this can only be understood at the level of the
domains of the  operator and its adjoint.  We conclude that, if using the associated $L^2$-norm, the
operator is spectrally unstable, as confirmed by its
pseudospectrum. However, using an alternative scalar product
determined by the Sturm-Liouville form of the operator, the latter is not
only normal but actually selfadjoint.  In the associated
``Sturm-Liouville'' norm the operator is spectrally stable, again
confirmed by the calculation of the pseudospectrum. This example
illustrates the care required to conclude either spectral stability or
instability.

Finally, some elements towards a genuine geometric setting for QNMs in
a compactified hyperboloidal approach have been discussed. In particular,
the identification of the $L_2$ block-operator as an object essentially
depending on the geometry at the conformal null boundaries, and oblivious
to the bulk, hints to the plausible role of BMS-supertranslations as
the objects controlling non-selfadjointness of the evolution generator
$L$ and, therefore, accounting for non-trivial BH pseudospectra and QNM
instability.  In particular, BMS-charges could be part of the
scattering data needed to reconstruct the potential in an inverse scattering approach.

\subsection{Perspectives}
The present work addresses some of the caveats in \cite{Jaramillo:2020tuu},
but many points remain open. In this sense, the present work
shares the perspectives in the research program presented in \cite{Jaramillo:2020tuu}.
More specifically related to the present work, we can list the following points:
\begin{itemize}
\item[a)] Construction of the energy scalar product for tensorial fields,
  extending the current scalar product discussion to arbitrary electromagnetic
  and gravitational (GW) perturbations.

\item[b)] Exploration of a scalar product for BH-like
  (e.g. P\"oschl-Teller) potentials rendering a ``stable pseudospectrum'',
  with an appropriately weight-modified scalar product.
  Even if keeping the energy scalar product
  on the basis of physical considerations,  such an alternative
 scalar product would be informative on the analytical
 aspects of the problem.

\item[c)] Formalization of the $O(\epsilon)$-equivalence of
  $\epsilon$-dual QNM expansions and assessment of their impact
  on BH spectroscopy, in particular the possible degeneracy of the
  associated inverse data analysis problem for retrieving QNM overtones from
  GW waveforms.

\item[d)] Proof of the ``Nollert-Price to Regge'' BH QNM ultraviolet instability conjecture.

\item[e)] Development of the geometric hyperboloidal framework for QNMs, in particular
  assessing the role of BMS-supertranslations and their relation to inverse scattering.

\end{itemize}

\bigskip

\section*{Acknowledgments}

The authors would like to thank Lamis Al Sheikh, Oscar Meneses-Rojas,
Rodrigo Panosso-Macedo and Johannes Sj\"ostrand for numerous and
continuous discussions. We also thank Abhay Ashtekar, Piotr Bizon,
Graham Cox, Oscar Reula and Jeffrey Winicour for raising specific points to be
developed in this work, as well as Emanuele Berti, Vitor Cardoso and
Kyriakos Destounis for the intense and enriching scientific
interaction in the late stage of the project.  The numerical
implementation discussed in Section \ref{sec:FiniteElements} is based
on an adaptation of the FEniCs notebooks and related material of the
PDE course of Oscar Reula and Manuel Tiglio.  We acknowledge support
from The European Union’s H2020 ERC Consolidator Grant ``Matter and
Strong-Field Gravity: New Frontiers in Einstein’s Theory'', Grant
Agreement No. MaGRaTh-646597, the PO FEDER-FSE Bourgogne 2014/2020
program and the EIPHI Graduate School (contract ANR-17-EURE-0002) as
part of the ISA 2019 project. We also thank the ``Investissements
d'Avenir'' program through project ISITE-BFC (ANR-15-IDEX-03), the ANR
``Quantum Fields interacting with Geometry'' (QFG) project
(ANR-20-CE40-0018-02), and the Spanish FIS2017-86497-C2-1 project
(with FEDER contribution).

\bigskip 
%%%%%%%%%%%%%%%%%%%%%%
%\bibliographystyle{spmpsci}
%\bibliographystyle{iopart-num}
%\bibliographystyle{reporthack}
%\bibliography{ThesisGRbib}
%\bibliography{Oct2021Filebib.bib}
%%%%%%%%%%%%%%%%%%%%%%

\end{document}